\documentclass[fleqn,usenatbib]{rasti}

\usepackage{newtxtext,newtxmath}

\usepackage[T1]{fontenc}

\DeclareRobustCommand{\VAN}[3]{#2}
\let\VANthebibliography\thebibliography
\def\thebibliography{\DeclareRobustCommand{\VAN}[3]{##3}\VANthebibliography}

\usepackage{graphicx}	
\usepackage{amsmath}	
\usepackage{bm}         
\usepackage{subcaption}
\usepackage{enumitem}
\usepackage{hyperref} 
\usepackage{fontawesome} 
\usepackage[dvipsnames]{xcolor} 

\newcommand\rb{\bm{r}} 
\newcommand\thetab{\bm{\theta}} 

\newcommand\githubcodelink{\href{https://github.com/isaackanowski/ROHSA_SNAPD}{\faGithubSquare}}

\newcommand{\revonechanged}[1]{{\color{black} #1}}

\newcommand{\revtwochanged}[1]{{\color{black} #1}}

\title[Spatially non-parametric recovery of intrinsic kinematic maps]{Spatially non-parametric recovery of intrinsic kinematic maps in pre- to post-merger galaxies}

\author[I. Kanowski et al.]{
Isaac Kanowski,$^{1,2}$\thanks{E-mail: isaac.kanowski@anu.edu.au}
Emily Wisnioski,$^{1,2}$
J. Trevor Mendel$^{1,2}$
Antoine Marchal$^{1,2}$
and Takafumi Tsukui$^{1,2}$
\\
$^{1}$Research School of Astronomy and Astrophysics, Australian National University, Canberra, ACT 2611, Australia\\
$^{2}$ARC Centre of Excellence for All Sky Astrophysics in 3 Dimensions (ASTRO 3D), Australia
}

\date{Accepted XXX. Received YYY; in original form ZZZ}

\pubyear{2025}

\begin{document}
\label{firstpage}
\pagerange{\pageref{firstpage}--\pageref{lastpage}}
\maketitle

\begin{abstract}
We introduce an adaptable kinematic modelling tool called {\tt ROHSA-SNAPD}, "Spatially Non-parametric Approach to PSF Deconvolution using {\tt ROHSA}". {\tt ROHSA-SNAPD} utilises kinematic regularisation to forward model the intrinsic emission-line flux and kinematics (velocity and \revonechanged{linewidth}) of 3D data cubes. Kinematic regularisation removes the need to assume an underlying rotation model (eg. exponential disc, \revonechanged{tilted-ring}) to deconvolve kinematic data. We evaluate the code on mock observations of simulated galaxies: one idealised disc model and three more complex galaxies from a cosmological simulation with varying levels of kinematic disturbance, from pre-merger to post-merger state.
The mock observations are designed to approximate published results at $z\sim 1-2$ from 8-metre class near-infrared spectroscopic facilities, using realistic observational parameters including spatial and spectral resolution, noise and point spread function. We demonstrate that {\tt ROHSA-SNAPD} can effectively recover the intrinsic kinematics of complex systems whilst accounting for observational effects.
\revonechanged{{\tt ROHSA-SNAPD} is publicly released on Github \githubcodelink.}
\end{abstract}



\section{Introduction}
\label{sec:Introduction}
\revonechanged{Galaxies predominantly grow over cosmic time through star formation fuelled by gas accretion and minor mergers \citep{Tacconi_2020, Forster_Schreiber_2020}. Integral field spectroscopy (IFS) and interferometry have been key to sharpening this picture, through the analysis of the kinematics and interstellar medium of galaxies across cosmic time \citep[eg.][]{Arribas_2014,Wisnioski_2015,Wisnioski_2019,Stott_2016,Harrison_2017,Johnson_2018, Tacconi_2018,Ubler_2019,Forster_Schreiber_2019,Tiley_2021,Genzel_2020,Genzel_2023,Mai_2024}. A major result from these studies is that massive disc populations were clumpy and turbulent at early times, compared to the thinner spirals typical at later times \citep[and references therein]{Glazebrook_2013,Tacconi_2020, Forster_Schreiber_2020}. Additionally, while mergers rates at $z\sim 1-2$ were higher than at $z \sim 0$ \citep[eg.][]{Man_2016,Mundy_2017,Mantha_2018,OLeary_2021}, discs still dominated \citep{Forster_Schreiber_2020}.

Initially, the dynamics of high redshift SF galaxies were classified through the visual analysis of resolved kinematic maps \citep[eg.][]{Forster_Schreiber_2006,Genzel_2006,Wright_2009,Epinat_2009}, where early samples showed roughly equal fractions of galaxies dominated by rotation, dispersion and disrupted kinematics \citep{Forster_Schreiber_2009}. As samples grew and deeper data were obtained, automatic methods such as Kinemetry \citep{Krajnovic_2006,Shapiro_2008,Bellocchi_2012} were explored to distinguish discs and mergers by quantifying the amount of non-axisymmetric structure in observed rotation fields, supporting the application of disc models to derive physical parameters \citep[eg.][]{Epinat_2010,Davies_2011}. Classification schemes were also developed for lower resolution data \citep[eg.][]{Epinat_2012,Green_2014,Wisnioski_2015,Stott_2016,Rizzo_2022}.
Applying these methods to large samples demonstrated that most SF galaxies, out to at least $z \sim 2$, have disc-like dynamics \citep{Forster_Schreiber_2020}, consistent with structural measurements derived from galaxy morphologies \citep[eg.][]{Wuyts_2011,van_der_Wel_2014b}.
As disc galaxies dominate and can be well described by dynamical models, kinematic studies at $z>1$ have focused their analysis on disc-like systems
\citep[eg.][]{Stott_2016,Di_Teodoro_2016,Forster_Schreiber_2018,Wisnioski_2019,Tiley_2021,Espejo_Salcedo_2022,Nestor_Shachar_2023,Parlanti_2023,Rizzo_2023}, leaving disrupted and merging galaxy kinematics largely unexplored \citep[though see][]{Epinat_2012,Lopez_Sanjuan_2013,Forster_Schreiber_2018,Genzel_2023,Tsukui_2024a}.

A number of factors make analysing galaxy kinematics at $z > 1$ difficult, predominantly the observational effect of beam smearing due to the combination of seeing limitations and spatial pixel scale. Beam smearing spatially blurs data according to the point spread function (PSF) and in discs decreases the magnitude of measured rotational velocities and increases velocity dispersions \citep{Davies_2011,Johnson_2018}.
Low signal-to-noise (S/N) and instrument spectral resolution also impact analysis, as kinematics can be poorly constrained and accurately measuring intrinsic velocity dispersions below the spectral resolution is difficult \citep[eg.][]{Wisnioski_2015}.

Significant effort over the last $20$ years has been dedicated to accounting for the impact of beam smearing and a number of approaches have been developed, broadly categorised as either 2D or 3D deconvolution methods.
2D methods are applied to observed kinematic maps (or 1D profiles) of rotation and velocity dispersion. Using correction factors is a simplistic, though highly efficient 2D method, where observed rotation and dispersion values are scaled by correction factors derived from PSF-convolved disc models dependent on numerous galaxy properties \citep[eg.][]{Burkert_2016,Johnson_2018}. 2D forward modelling approaches are more adaptable to individual galaxies, as 3D disc models are convolved with the PSF and collapsed to fit to observed 2D kinematics. Such models can be parameterised by specific flux and rotation profiles \citep[eg.][]{Stott_2016, Turner_2017} or flux and mass profiles using codes such as {\tt Dysmal} and {\tt DysmalPy}  \citep[and references therein]{Lee_2024}.
However, a disadvantage of 2D methods is that they rely on the pre-calculation of observed kinematic maps and can therefore struggle in low S/N regions as individual kinematic fits may be poorly constrained.

3D approaches aim to improve upon 2D methods by fitting directly to observed data cubes and a number of modelling codes have been developed, each with unique benefits and limitations \citetext{eg. {\tt GalPaK\textsuperscript{3D}}; \citealp{Bouche_2015}, {\tt KinMS}; \citealp{Davis_2013}, {\tt \textsuperscript{3D}BAROLO}; \citealp{Di_Teodoro_2015}, {\tt BLOBBY3D}; \citealp{Varidel_2019}, {\tt Dysmal} and {\tt DysmalPy}; \citealp[and references therein]{Price_2021,Lee_2024}}.
3D modelling methods use all available information provided by IFS and interferometric observations \citep{Di_Teodoro_2015,Forster_Schreiber_2020} and generally perform better than 2D methods in low S/N observations.
Individual codes also have specific benefits which allow for more flexible recovery of kinematics, for example {\tt BLOBBY3D} fits individual flux blobs to reproduce complex flux profiles and the tilted-ring nature of {\tt\textsuperscript{3D}BAROLO} allows it to fit varying rotation profiles in a less constrained way.
These two types of deconvolution methods have been successfully applied to both adaptive optics-assisted and seeing-limited observations to determine many galaxy properties across a range of redshifts \citep[and references therein]{Forster_Schreiber_2020}.

Despite the focus of most kinematic surveys at $z>1$ being disc-like systems, it is important to kinematically analyse the full range of galaxy types, which include disrupted and merging galaxies.
Whilst pioneering work has been done and mergers have been observed, detailed kinematic analysis of these complex systems has only been carried out for a small number of the best-quality galaxies at higher redshift \citep[eg.][]{Epinat_2012,Lopez_Sanjuan_2013,Forster_Schreiber_2018,Genzel_2023,Tsukui_2024a}.
These studies have identified kinematic signatures of accretion and potential interactions to inform constraints on gas inflow rates, merger fractions and the dynamic effect of close companions.
However, large galaxy samples with consistent selection criteria and modelling approach must be built to statistically analyse these findings. This currently necessitates the use of seeing-limited observations which are substantially impacted by beam smearing.
Though current modelling techniques can allow for either more flexible flux or rotation profiles, no technique exists to deconvolve high redshift kinematics with both non-parametric flux and rotation distributions. Additionally, detailed kinematic modelling techniques such as Kinemetry still assume some level of ordered rotation, so are not necessarily appropriate for highly disrupted systems which cannot be approximated as disc-like \citep[eg.][]{Law_2009}.
Even when aiming to study purely disc-like galaxies, it can be difficult to distinguish mergers from discs at high redshift in seeing-limited data \citep{Simons_2019}.

Therefore, to study dynamically complex systems or more clearly identify discs within large samples, a spatially non-parametric approach to kinematic modelling is needed, where beam smearing can be accounted for independent of assumptions on galaxy structure or rotation.
Whilst the increased model freedom could make the method less suitable for low S/N observations, it would allow for more detailed analysis of kinematically complex systems in large galaxy samples.}

In this paper, we introduce a code called {\tt ROHSA-SNAPD} that forward models deconvolved kinematic maps from 3D IFS emission-line data. {\tt ROHSA-SNAPD}, based on {\tt ROHSA} ``Regularized Optimization for Hyper-Spectral Analysis'' \citep{Marchal_2019}, utilises kinematic regularisation when fitting to data \revonechanged{a possible tool to model complex kinematics}. In this work, we build upon the core features of {\tt ROHSA} to include the effects of the observational PSF and instrument resolution in a spatially non-parametric way. In Section \ref{sec:RSNAPD} we discuss the fitting procedure used by the code to model deconvolved kinematic maps. In Section \ref{sec:evaluation_idealised_disc_galaxy} we apply {\tt ROHSA-SNAPD} to a mock observation of an idealised disc galaxy as a proof of concept and to demonstrate the effect of altering hyper-parameters in a simplistic, controlled case. We evaluate {\tt ROHSA-SNAPD} on three more complex mock observations with varying levels of kinematic disturbance generated from a cosmological simulation in Section \ref{sec:evaluation_tng_galaxies}.

We assume cosmological parameters from the \citet{PlanckCollaboration_2016} results: $H_0 = 67.74$ km s$^{-1}$ Mpc$^{-1}$, $\Omega_m = 0.3089$ and $\Omega_\Lambda = 0.6911$.

\section{The {\tt ROHSA-SNAPD} code}
\label{sec:RSNAPD}
We adapt the existing emission-line fitting code {\tt ROHSA}, presented by \citet{Marchal_2019} to create the Python-based code {\tt ROHSA-SNAPD}, "Spatially Non-parametric Approach to PSF Deconvolution using {\tt ROHSA}". Originally developed to perform phase separation of 21\,cm data, {\tt ROHSA} simultaneously fits multi-component Gaussian profiles to spectra and returns two-dimensional maps of the Gaussian parameters, amplitude, central position (rotational velocity) and standard deviation (velocity dispersion). A key feature of {\tt ROHSA} is that it enforces spatial coherence in the kinematic parameter maps during fitting. This ensures the returned model is physically meaningful, as it is expected that spatially close regions should have similar properties. We have used the core features of {\tt ROHSA} to fit emission-lines in optical/near-IR IFS data that can be modelled with a single Gaussian component and have included accounting for a number of observational effects in a spatially non-parametric way. 
We forward model beam smearing due to both seeing and the finite spatial sampling of observations (ie. pixel size), as well as broadening caused by instrument spectral resolution. The spatial coherence provided by {\tt ROHSA} is necessary for {\tt ROHSA-SNAPD} to ensure that the deconvolved parameter maps are physically meaningful without needing to assume an underlying rotation model.

\subsection{{\tt ROHSA-SNAPD} model}
\label{subsec:model}
\begin{figure*}
    \centering
    \includegraphics[width=\linewidth,trim={0cm 4cm 4cm 4cm},clip]{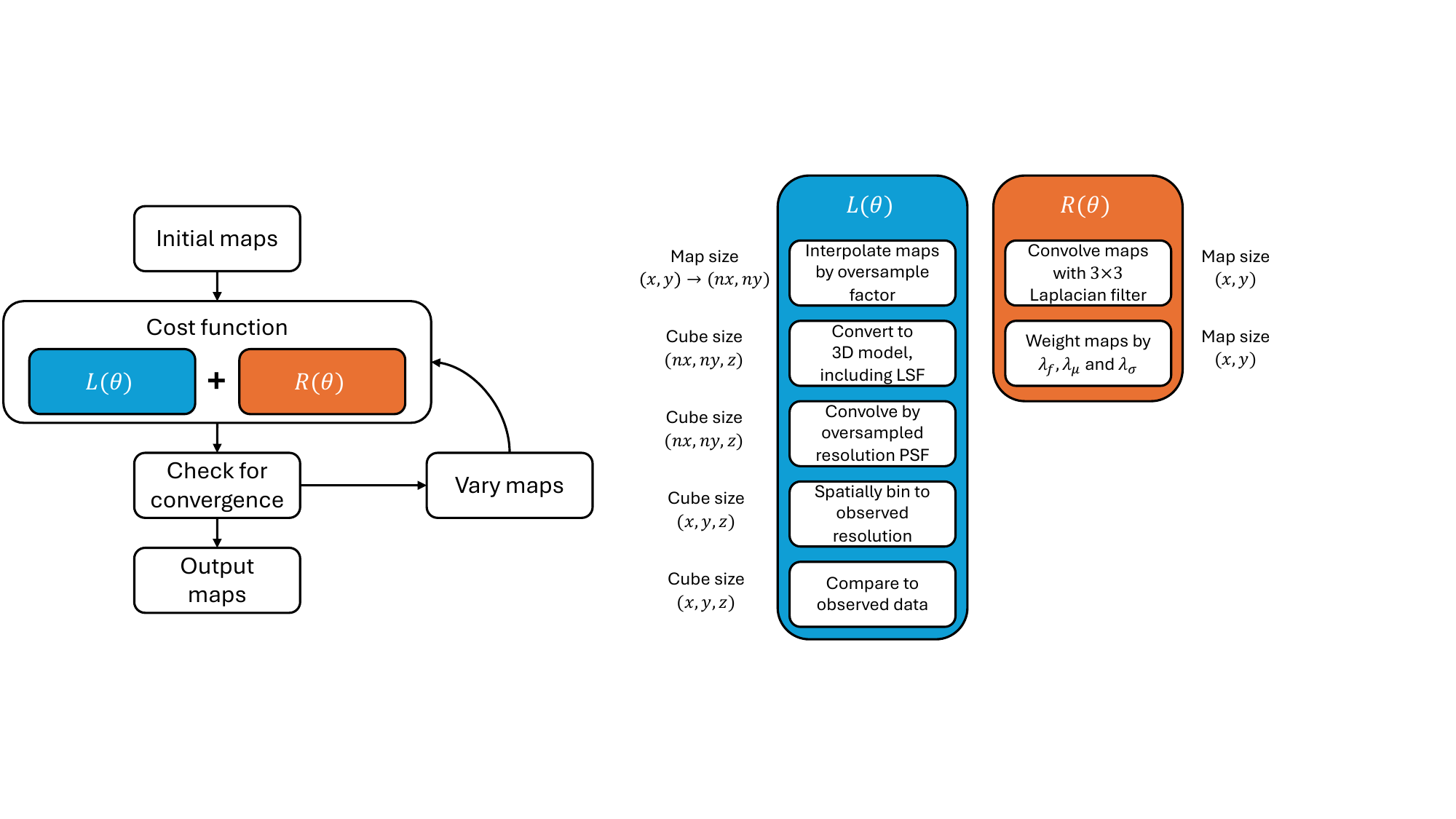}
    \caption{
    \revonechanged{A schematic demonstrating the fitting procedure of {\tt ROHSA-SNAPD}. Map and cube sizes are shown for observed data with spatial size $(x,y)$, $z$ spectral channels and an oversample factor of $n$.}
    }
    \label{fig:flowchart}
\end{figure*}

\revonechanged{Our method uses a forward modelling approach applied directly to 3D data cubes to obtain deconvolved kinematics.} \revonechanged{A general overview of the fitting process is shown in Figure \ref{fig:flowchart}.} We describe each line-of-sight by a single Gaussian function,
\begin{align}
    G(v_z,\thetab(\rb)) &= \dfrac{\bm{f}(\rb)}{\sqrt{2 \pi} \, \sigma_{\textnormal{tot}}(\rb,v_z)} \, \exp{\left(\frac{-\left( v_z - \bm{\mu}(\rb) \right)^2}{2 \, \sigma_{\textnormal{tot}}(\rb,v_z)^2}\right)} \, , \label{eqn:G_v_theta}
\end{align}
\noindent for,
\begin{align}
    \sigma_{\textnormal{tot}}(\rb,v_z) &= \sqrt{\bm{\sigma}(\rb)^2+\sigma_{\textnormal{inst}} (\rb,v_z)^2} \, , \label{eqn:add_sigma_inst}
\end{align}
where $v_z$ is the radial velocity at each velocity channel of the data cube and the 2D parameter maps $\thetab(\rb) \, = \, \left( \bm{f}(\rb),\bm{\mu}(\rb), \bm{\sigma}(\rb) \right)$ describe the Gaussian flux $\bm{f}$, central position $\bm{\mu}$ and standard deviation $\bm{\sigma}$ at spatial pixel coordinate $\rb$. We optimise for flux rather than amplitude to avoid degeneracies between amplitude and standard deviation when fitting. Instrument spectral resolution, $\sigma_{\textnormal{inst}}(\rb,v_z)$ is added in quadrature to $\bm{\sigma}(\rb)$ to account for spectral broadening. In general, $\sigma_{\textnormal{inst}}(\rb,v_z)$ is derived observationally and can vary both spatially and spectrally depending on the specific instrument and emission-line of interest.
The spectral parameters $v_z$, $\bm{\mu}(\rb)$, $\bm{\sigma}(\rb)$ and $\sigma_{\textnormal{inst}}(\rb,v_z)$ are expressed in terms of velocity channels of the observed data cube.

Whilst the dispersion map $\bm{\sigma}(\rb)$ varies for each line of sight by default, it can also be held spatially constant where a single dispersion value, $\sigma_0$, is optimised rather than an entire map. This can be beneficial, for example, in low-resolution observations to reduce the number of free parameters in the model. \revonechanged{For an observed data cube with $x$ spatial pixels, the model optimises $2x+1$ parameters if a spatially constant velocity dispersion is assumed, or $3x$ parameters otherwise.}

The Gaussian functions $G(v_z,\thetab(\rb))$ are spatially convolved with a user-defined 2D PSF kernel, $P$, at each velocity channel, creating a convolved model cube,
\begin{align}
    \tilde{F}(v_z, \thetab(\rb)) = P \otimes G(v_z,\thetab(\rb)) \, ,
\end{align}
which is then compared to the observed data through the cost function, discussed in Section \ref{subsec:cost_function}. By iteratively changing the underlying parameters $\thetab(\rb)$ such that the cost of the fit is minimised, {\tt ROHSA-SNAPD} can produce a convolved model cube that closely matches the observed data, assuming the PSF and instrument spectral resolution are well described.

\subsection{Cost function}
\label{subsec:cost_function}
A cost function is used to quantify a model's goodness of fit, which allows an optimisation algorithm to iteratively approach the desired solution by varying parameters such that the cost decreases.
In {\tt ROHSA-SNAPD}, the cost function is a combination of two main terms and builds upon the function presented by \citet{Marchal_2019},
\begin{align}
    C \left(\thetab \right) =& \, L \left(\thetab \right) + R \left(\thetab \right). \label{eqn:adapted ROHSA cost func}
\end{align}

The first term, $L\left(\thetab \right)$, is the standard likelihood, which describes how closely the convolved {\tt ROHSA-SNAPD} model matches observed data with respect to the noise level.
In addition to the PSF convolution and instrument resolution discussed in Section \ref{subsec:model}, observed data is also impacted by the finite spatial sampling of observations.
This effect must also be included in our forward model to ensure the deconvolved kinematics approximate the intrinsic kinematics as closely as possible. Therefore, we first bilinearly interpolate the $\thetab(\rb)$ kinematic maps onto a finer grid, with pixel area $n^{-2}$ times the original, where the oversampling factor $n$ is a user-defined integer. Following Section \ref{subsec:model}, we use the oversampled maps and a PSF kernel at the same resolution to create a high-spatial-resolution convolved model (with the same spectral resolution as the original cube). The cube is then binned to the original resolution to introduce spatial sampling effects (though see Appendix \ref{appendix:limitations_bilinear_interpolation} for a discussion on the limitations of this method).
$L \left(\thetab \right)$ is therefore the least-square difference between the binned model, $\tilde{F}_{\textnormal{ binned}}(v_z, \thetab(\rb))$, and the observed data, $F(v_z, \rb)$,
\begin{align}
    L \left(\thetab \right) =& \, \frac{1}{2} \, \sum_{v_z, \, \rb} \left[ \frac{ \tilde{F}_{\textnormal{ binned}}(v_z, \thetab(\rb)) - F(v_z, \rb)}{\bm{\Sigma}(v_z, \rb)} \right]^2 \, ,
\end{align}
where $\bm{\Sigma}(v_z, \rb)$ is the root-mean-square noise for each pixel of $F(v_z, \rb)$.

The second term, $R\left(\thetab \right)$, enforces spatial coherence in the deconvolved model by penalising pixels in the parameter maps that differ from their surroundings.
Spatial coherence hyper-parameters, $\lambda_{f}$, $\lambda_{\mu}$ and $\lambda_\sigma$, weight a convolution with a Laplacian kernel to set the strength of regularisation on each parameter map $\bm{f}(\rb),\bm{\mu}(\rb)$ and $\bm{\sigma}(\rb)$ respectively,
\begin{align}
    \begin{split}
    R \left(\thetab \right) = \, \frac{1}{2} &\Bigl[\lambda_{f} ||\bm{d \otimes f}(\rb)||^2_2 \\
    &+ \lambda_{\mu} ||\bm{d \otimes \mu}(\rb)||^2_2 \\
    &+ \lambda_\sigma ||\bm{d \otimes \sigma}(\rb)||^2_2 \Bigr] \, ,
    \end{split}
\end{align}
where $d$ is the Laplacian kernel,
\begin{align}
    d = \begin{bmatrix}
        0 & -1 & 0\\
        -1 & 4 & -1\\
        0 & -1 & 0
        \end{bmatrix} \, / \, 4 \label{eqn:laplacian kernel}
\end{align}

\revtwochanged{We use a $3\times3$ Laplacian kernel to measure pixel-to-pixel variations, which are then penalised to enforce kinematic smoothness.}

\subsection{Optimisation}
\label{subsec:optimisation}
To decrease the cost of a fit and therefore converge on a solution, {\tt ROHSA-SNAPD} relies on the Limited-memory Broyden–Fletcher–Goldfarb–Shanno with Bounds (L-BFGS-B) algorithm \citep{Byrd_1995,zhu_algorithm_1997} provided by the {\tt SciPy} package \citep{Virtanen_2020_SciPy}. The L-BFGS-B algorithm is specifically designed for non-linear optimisation problems, utilising a function's gradient and limited memory Hessian matrix approximations to iteratively minimise the function with respect to user-defined bounds on each variable. Bounded optimisation is necessary to ensure variable values are physical, such as positive flux and velocity dispersions \citep{Marchal_2019}.
The algorithm terminates when either the projected gradient or absolute change in the cost function is sufficiently small or if the maximum number of iterations has been reached.

The L-BFGS-B algorithm requires the gradient of the cost function to be given at each iteration. To calculate the gradient we utilise reverse mode automatic differentiation provided by the {\tt JAX} {\tt Python} package \citep{Frostig_2018,jax2018github}.
\revtwochanged{In general, automatic differentiation utilises the fact that the derivative of complex mathematical functions can be expressed, using the chain rule, as the multiplication of the derivative of a number of simple functions with known derivatives \citep{Maclaurin_2016,Baydin_2018}. Forward and reverse mode automatic differentiation refers to the order in which the matrix multiplication of the derivatives is computed. In {\tt ROHSA-SNAPD}, reverse mode automatic differentiation is more efficient as the number of inputs is large compared to the number of outputs \citep{Baydin_2018}.}
This approach allows {\tt ROHSA-SNAPD} to fit faster and scale more effectively compared to using purely numeric differentiation techniques such as finite difference methods \citep[eg.][]{Baydin_2018}.
\revonechanged{We provide examples of computation time and memory usage for varying input data sizes and parameter choices in Appendix \ref{appendix:run_info}.}

\section{Evaluation on an idealised disc galaxy}
\label{sec:evaluation_idealised_disc_galaxy}
To test the accuracy of {\tt ROHSA-SNAPD}, we fit synthetic data with known kinematic properties. To demonstrate the non-parametric deconvolution in a simplistic, controlled case, we first fit an idealised disc galaxy generated with the KINematic Molecular Simulation ({\tt KinMS}) code \citep{Davis_2013}.
\subsection{Creating a mock observation with {\tt KinMS}}
\label{subsec:creating_mock_observations_with_KinMS}
To create a physically motivated mock observation with {\tt KinMS}, we set the galaxy parameters using observed scaling relations. We assumed a galaxy redshift of $z=2$ and stellar mass of $3 \times 10^{10} \textnormal{ M}_\odot$ which are generally similar to $z \sim 2$ galaxies observed by KMOS surveys \citep[eg.][]{Wisnioski_2015,Wisnioski_2019,Harrison_2016,Curti_2020,Tiley_2021}. 
In this work we adopt KMOS-like data characteristics as the seeing-limited KMOS instrument has produced the largest number of resolved IFS observations at $z \gtrsim 1$ to date \citep{Forster_Schreiber_2020}. This is motivated by our aim, described in Section \ref{sec:Introduction}, to develop a code that fits spatially non-parametric models to large statistical samples at high-redshift, where kinematic perturbation is expected. We leave the application of {\tt ROHSA-SNAPD} to real KMOS and higher resolution data to future work.

The size-mass relation from \citet{van_der_Wel_2014} was used to determine the galaxy's \revonechanged{effective $(0\textnormal{"}.413)$ and exponential scale $(0\textnormal{"}.246)$ radii} and create exponential light and \revonechanged{exponential disc} rotation profiles, assuming a position angle of $270^\circ$ and inclination of $45^\circ$. 
\revonechanged{The exponential rotation profile had a maximum velocity of $153.8$ km s$^{-1}$ at $0\textnormal{"}.530$ from the centre.
We used a constant $20$ km s$^{-1}$ velocity dispersion for the whole galaxy.} These profiles were passed to {\tt KinMS} which generated a $4\textnormal{”} \times 4\textnormal{”} \times 800\textnormal{ km s}^{-1}$ high-resolution intrinsic model cube with $0.025$ arcsecond$^2$ spatial binning and $10$ km s$^{-1}$ spectral binning ($160\times160\times80$ pixels). Observational effects were applied by spectrally convolving the cube with a KMOS-like spectral resolution of $\sigma_{\textnormal{inst}}(\rb,v_z) = 30$ km s$^{-1}$ and spatially convolving with a normalised 2D Moffat PSF model sampled at $0.025$ arcsecond$^2$ resolution on an \revonechanged{$196\times196$} pixel grid.
The PSF parameters were obtained from the KMOS\textsuperscript{3D} survey, which has Moffat PSF FWHMs that generally range between $0.3$ and $0.9$ arcseconds with a mean of around $0\textnormal{"}.482$. \revonechanged{The chosen model had a major and minor FWHM of $0\textnormal{"}.527$ and $0\textnormal{"}.478$ respectively.} The convolved cube was then binned spatially by a factor of $8$ and spectrally by a factor of $4$ to a KMOS-like resolution of $0\textnormal{”}.2 \times 0\textnormal{”}.2 \times 40\textnormal{ km s}^{-1}$ ($20\times20\times20$ pixels).  Finally, white Gaussian noise was added to complete the mock data cube, with uniform RMS error across the whole cube.

When displaying observed kinematics in figures throughout this work, S/N maps are used to mask noisy pixels. To generate a S/N map of the idealised disc, we first used $200$ noise realisations of the observed data to find the median intensity of every pixel in the cube. We then cropped this cube around the emission-line and summed over each line-of-sight to create an observed flux map. We created a noise map by summing in quadrature the root mean square of the noise for each pixel in the cropped cube. The S/N map was created by dividing the observed flux and noise maps. \revonechanged{Any observed kinematics in regions with S/N $< 1$ are masked. A similar method was used to determine the total S/N of the idealised disc galaxy given in Table \ref{tab:galaxy_sample}, $\textnormal{S/N} = 39.43$. Instead of summing the median intensity cube along each line-of-sight, values within a $1\textnormal{"}.5$ aperture were summed to a single value and divided by the root mean square of the noise of each pixel within the aperture, summed in quadrature.}

\subsection{Applying {\tt ROHSA-SNAPD} to the idealised disc galaxy}
\label{subsec:idealised_disc_applying_RSNAPD}
We now detail the general steps and parameters chosen to fit the idealised disc mock observation.
Data cubes passed to {\tt ROHSA-SNAPD} are assumed to be continuum-subtracted, cropped spectrally around an emission-line and scaled such that integrated flux values generally lie between $1$ and $100$. The flux must be scaled to ensure the L-BFGS-B optimisation algorithm can vary the parameter maps effectively. As only star-forming gas was used to create the synthetic data no subtraction or cropping was needed, though the data and RMS noise values were scaled to ensure the integrated flux was appropriate. After fitting, we re-scaled the deconvolved flux map back to the original flux level of the data.

\revonechanged{In this work we adopt the simplest case optimising for a spatially uniform dispersion for each galaxy.}
Assuming a constant velocity dispersion \revonechanged{can be} motivated both physically and computationally.
Physically, a spatially uniform velocity dispersion is consistent with it being isotropic \citep{Genzel_2011}\revonechanged{, a reasonable assumption in disc galaxies observed at this resolution. S}ubstantial deviations from this assumption have not been observed in most seeing-limited or adaptive optics-supported observations of star-forming galaxies at $z\sim2$, nor have significant trends been observed between velocity dispersion and galaxy properties such as stellar mass, inclination or \revonechanged{star formation} rate density \citep[][and references therein]{Ubler_2019,Forster_Schreiber_2020}.
From an optimisation perspective, assuming a single value for dispersion reduces the number of parameters in the model, simplifying the fitting process. \revtwochanged{A single dispersion fit means that the deconvolved dispersion would approximate the S/N-weighted median intrinsic velocity dispersion of the field. This approach is appropriate in observations with minimal variations in velocity dispersion relative to the data quality, but may produce poor fits with substantial velocity residuals in fields which have clearly non-uniform dispersions, such as highly disrupted merger systems.
Therefore, during testing, each observation was also} fit assuming a varying velocity \revtwochanged{dispersion. Whilst} the fits for some observations were reasonably accurate, a spatially varying velocity dispersion was poorly constrained in other systems, as shown in Appendix \ref{appendix:varying_dispersion_profiles}. This is most likely due to the low spatial and spectral sampling of our mock observations, in addition to other factors such as PSF size and noise level.
\revonechanged{The data quality and resolution would need to justify moving to a spatially non-uniform dispersion. 
One approach to identify potential intrinsic variations would be to use a map of velocity gradient as a proxy for the impact of beam smearing on the observed dispersion \citep[eg.][]{Varidel_2016,Stott_2016,Oh_2022}. Significant features in the observed dispersion map that do not align with the velocity gradient map could suggest intrinsic dispersion variation. A poor fit to the observed spectra when assuming a constant dispersion could also indicate that there is additional underlying kinematic structure which cannot be successfully modelled by the convolved ROHSA-SNAPD fit when assuming a constant dispersion. In the simulated galaxies used in this work,
}
the increase in the reduced $\chi_r^2$ of the resulting fits was only minimal when assuming a constant dispersion profile as opposed to a varying dispersion, \revonechanged{suggesting that assuming a varying velocity dispersion profile is not supported in this galaxy sample.}
For example, the most kinematically complex system in our sample, the ongoing merger introduced in Section \ref{sec:evaluation_tng_galaxies}, had a reduced $\chi_r^2$ of $1.080$ when using a varying velocity dispersion and $1.085$ when keeping dispersion constant, calculated from the convolved {\tt ROHSA-SNAPD} models within an observed S/N of $1$. \revtwochanged{Further, there were no significant velocity dispersion residuals which would observationally support the use of a varying velocity dispersion.} This suggested that there was little benefit to assuming a varying velocity dispersion for our mock observations, though we will study the recovery of spatially varying velocity dispersion maps in more detail in future work.

\revtwochanged{In this work we assumed that the PSF and instrument spectral resolution were known perfectly, which allowed the deconvolution process to be demonstrated as accurately as possible, given the data quality. However, in reality, the PSF and spectral resolutions are not known exactly and can potentially vary both spatially and spectrally. This could introduce systematic uncertainties into the model, increasing the overall fitting error. Therefore, using accurate PSF and spectral resolution models are crucial when applying {\tt ROHSA-SNAPD} to observations.}

We used an oversampling factor of $n=4$ within the cost function to approximate the effects of finite spatial sampling on the observations. \revonechanged{This meant that for our mock data of size $(20 \times 20 \times 20)$, within the cost function an oversampled model cube with dimensions $(80 \times 80 \times 20)$ was convolved with a PSF kernel of size $(85 \times 85)$ sampled at $0\textnormal{"}.05$ resolution.} In general, a larger oversampling factor would better account for the finite spatial resolution, assuming linear interpolation is appropriate for the system. However, it would also increase the \revonechanged{computation} time, so the choice of $n$ is a trade-off between these effects. We found $n=4$ to be appropriate for our KMOS-like data, as discussed in Section \ref{subsec:impact_spatial_binning}.
This is also the same oversampling factor chosen by \citet{Turner_2017} when fitting KMOS observations, though using a high-resolution disc model rather than linear interpolation.

To ensure fitting stability, the L-BFGS-B optimisation algorithm must be initialised with the approximate kinematic properties of the data. In this work, we obtained an initial flux map by spectrally summing the observed data cube. We chose the initial rotational velocity to be the same value for each pixel in the map, set as the central velocity channel of the observed cube. This is a reasonable initial rotation as in this work the data is centred around the galaxy's systemic velocity. We set the dispersion initial condition to be $10$ km s$^{-1}$, converted to units of velocity channels. This is a physically motivated velocity dispersion lower limit due to thermal motions \citep{Shields_1990,Ubler_2019}.

The optimisation algorithm also requires bounds to be specified.
In this work we set bounds substantially wider than the observed kinematics of our data, flux = ($10^{-6}$,$10^{3}$), dispersion = ($10^{-6}$, $300$) km s$^{-1}$ and rotational velocity 400 km s$^{-1}$ either side of the systemic velocity. Flux and dispersion lower limits were set above $0$ to avoid non-physical or unconstrained solutions.

The {\tt ROHSA-SNAPD} hyper-parameters $\lambda_f$, $\lambda_\mu$ and $\lambda_\sigma$ are the most important user-defined variables as their values can significantly affect the output model. To demonstrate their effect, we test a number of different values of $\lambda_f$ and $\lambda_\mu$ in the following sections. Because we optimised a single dispersion for each galaxy, $\lambda_\sigma$ did not affect the fits.

\subsection{Results of {\tt ROHSA-SNAPD} fit to the idealised disc galaxy}
\label{subsec:idealised_disc_results}
Using the input parameters detailed in Section \ref{subsec:idealised_disc_applying_RSNAPD}, we applied {\tt ROHSA-SNAPD} to the idealised disc mock observation, fitting $200$ noise realisations of the galaxy, where every run terminated due to the change in the cost function being sufficiently small. For each realisation, we obtained deconvolved {\tt ROHSA-SNAPD} kinematic maps, $\thetab(\rb)$. These maps were also converted back to convolved data cubes, $\tilde{F}(v_z, \thetab(\rb))$, and fit with Gaussian profiles so `convolved kinematic maps' could be obtained \revonechanged{and compared to the observed kinematics using consistent methodology}.
To test the accuracy of the fits, we compared the deconvolved and convolved {\tt ROHSA-SNAPD} kinematics with the intrinsic and observed kinematics of the mock data respectively. High-resolution intrinsic kinematic maps were obtained by taking moments of the high-resolution data and observed kinematic maps were acquired through fitting Gaussian profiles to the noisy KMOS-resolution data cubes.
The systemic velocity of each rotational map was set from the data's central velocity channel.

The idealised disc model allowed the effect of regularisation to be explored.
The main impact of flux regularisation is to stop the flux map from fitting to noise. Therefore, when fitting the $200$ noise realisations of this system we first tested increasingly larger values of flux regularisation until the deconvolved flux was no longer noisy, around $\lambda_f = 0.001$. During this process, rotational velocity regularisation was held constant at a reasonable value, $\lambda_\mu = 20$.

\subsection{Impact of velocity regularisation}
\label{subsec:Impact_of_velocity_regularisation}
Rotational velocity regularisation, $\lambda_\mu$, has two main effects, to stop the rotation map fitting to noise and to help break the degeneracy between rotational velocity and velocity dispersion by constraining the gradient of the rotation map. For very low velocity regularisation values the rotation map appears speckled, as it varies significantly pixel-to-pixel. When this `speckled' map is converted into a data cube and convolved, the small variations in velocity are combined to create broad emission profiles that can fit the noisy spectra almost perfectly, regardless of the deconvolved velocity dispersion. As a result the deconvolved velocity dispersion often drops to zero, returning an unrealistic solution. When rotational velocity regularisation is raised such that the rotation map no longer fits to the noise, the deconvolved velocity dispersion increases and the regularisation's second effect becomes important. 

\begin{figure*}
    \centering
    \includegraphics[width=0.9\linewidth]{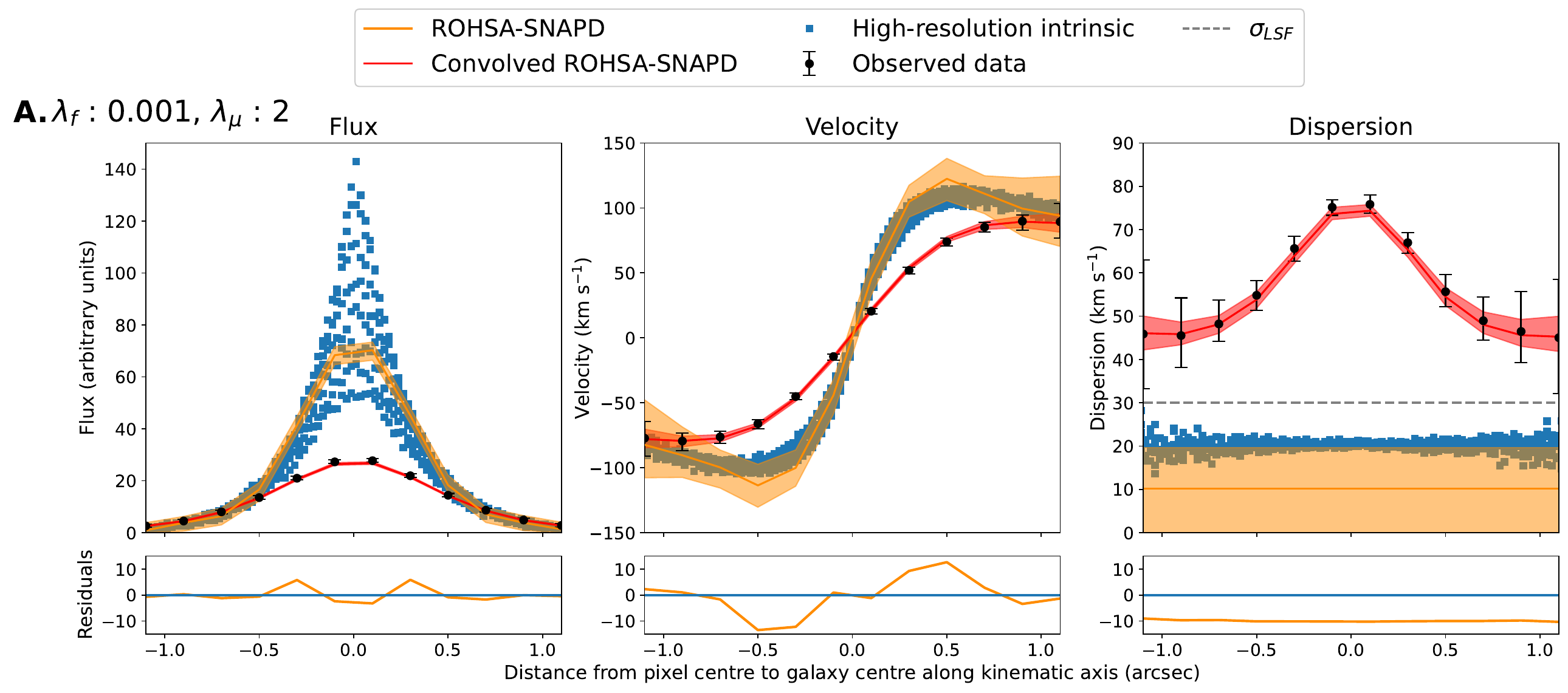} \hspace{5cm}
    
    \includegraphics[width=0.9\linewidth]{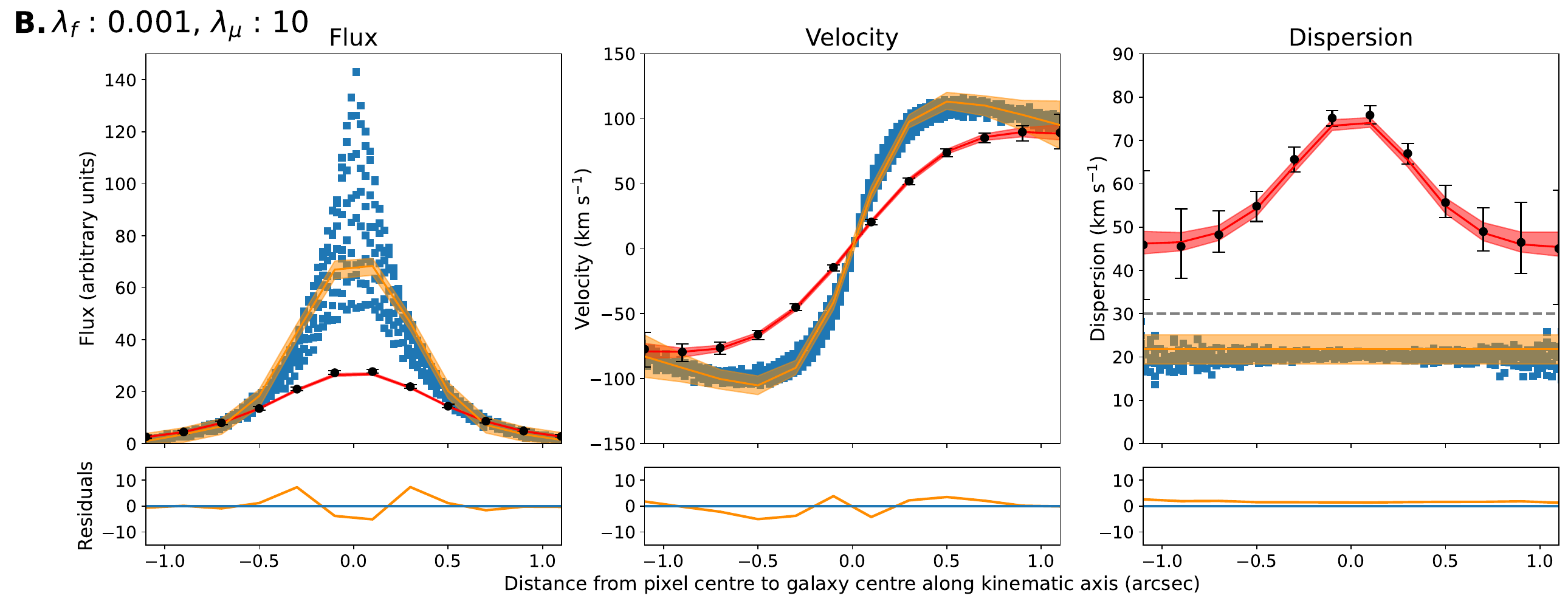}\hspace{5cm}
    
    \includegraphics[width=0.9\linewidth]{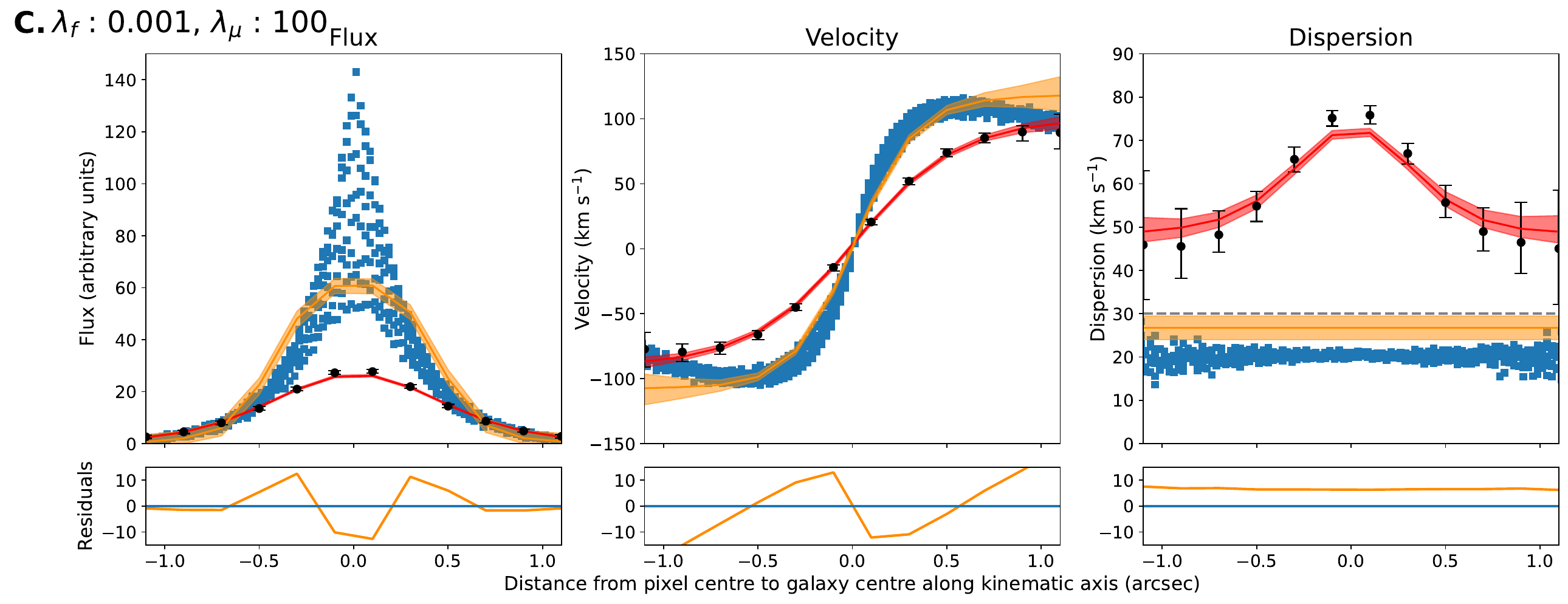}
    
    \caption{Each panel \textbf{A.}, \textbf{B.} and \textbf{C.} show the kinematic axis profiles and residuals for the same idealised disc galaxy, fit with different rotational velocity regularisation values, $\lambda_\mu = 2$, $10$ and $100$ respectively. In each panel, the three columns show the flux, rotational velocity and velocity dispersion profiles. The kinematic axis profiles are shown in the top row of each panel, where black points describe the observed data kinematics and associated errors and blue squares show the high-resolution intrinsic kinematics. The intrinsic flux has been scaled by a factor of $8^2$ to account for the smaller pixel scale. The orange and red regions show the deconvolved and convolved {\tt ROHSA-SNAPD} profiles. The solid lines show the median of the $200$ noise realisations and the shaded regions show the $16$th and $84$th percentiles. Observed and convolved {\tt ROHSA-SNAPD} kinematic profiles include the observational effects of beam smearing and instrument resolution, whereas the deconvolved {\tt ROHSA-SNAPD} profiles account for these effects and should closely match the intrinsic. The grey dashed line in the velocity dispersion panels shows the level of the instrument spectral resolution, $\sigma_{\textnormal{inst}}(\rb,v_z)$. The bottom row of each panel shows the percent error between the deconvolved {\tt ROHSA-SNAPD} and intrinsic profiles. For comparison between the differing pixel scales, the median of the intrinsic kinematic axis data is taken in the $0.2$ arcsecond bins to create observed resolution profiles.}
    \label{fig:KinMS_kinematic_axis_profiles}
\end{figure*}

To demonstrate the impact of regularisation on the gradient of the rotation map, we fit the disc galaxy with a range of rotational velocity regularisation values, holding flux regularisation constant. The three rows in Figure \ref{fig:KinMS_kinematic_axis_profiles}, \textbf{A.}, \textbf{B.} and \textbf{C.}, show the results of $\lambda_\mu = 2 \textnormal{, } 10 \textnormal{ and } 100$ respectively, using a mock long-slit along the kinematic axis of the galaxy.
The observed galaxy and convolved {\tt ROHSA-SNAPD} kinematic profiles, which both include observational effects, were obtained by averaging the data cubes spatially in PSF FWHM radius apertures and fitting Gaussian functions to the emission-line in averaged spectra along the kinematic axis (black data points). The intrinsic (blue data points) and deconvolved {\tt ROHSA-SNAPD} (orange model lines) data were not averaged and all high-resolution intrinsic kinematics within a $0\textnormal{"}.2$ wide slit were shown. The intrinsic flux profile was scaled by a factor of $8^2$ to account for the smaller pixel scale.
In each panel, the kinematic axis data is shown as a function of distance from the galaxy centre, and extends until the observed line-of-sight S/N drops below $1$.

When $\lambda_\mu = 2$ (Figure \ref{fig:KinMS_kinematic_axis_profiles} $\textbf{A.}$), rotational velocity regularisation is set too low and the gradient of the deconvolved rotational velocity is too steep, as reflected in the rotation residual profile which is low on the left side of the galaxy and too high on the right. When convolved with the PSF, this steeper gradient increases the line-width in the convolved solution and the optimiser must decrease the deconvolved velocity dispersion in order to match the observed dispersion profile. Therefore the median deconvolved {\tt ROHSA-SNAPD} velocity dispersion is lower than the intrinsic.
Conversely, when rotational velocity regularisation is too high (Figure \ref{fig:KinMS_kinematic_axis_profiles} $\textbf{C.}$), the deconvolved rotational velocity map is over-smoothed. This is seen when $\lambda_\mu = 100$, where the deconvolved rotation profile is too flat and the residuals are reversed. As a result, the deconvolved velocity dispersion must be increased above the intrinsic value to try to match the observed profile. 
If the velocity regularisation is set appropriately (Figure \ref{fig:KinMS_kinematic_axis_profiles} $\textbf{B.}$), which in this case occurs around $\lambda_\mu = 10$, the intrinsic rotational velocity and velocity dispersion are both recovered successfully and the profiles (blue squares) are both within error of the deconvolved {\tt ROHSA-SNAPD} fit (orange band).

Figure \ref{fig:KinMS_kinematic_axis_profiles} shows that regularisation is necessary to obtain meaningful results using {\tt ROHSA-SNAPD} but that the amount used can also substantially affect fit outcomes. It should be noted that there are often a large range of regularisation values that provide acceptable fits, though the most appropriate regularisation values will change depending on the specific galaxy sample and data quality. \revonechanged{We further demonstrate the impact of S/N and regularisation on the accuracy of deconvolved kinematics in Appendix \ref{appendix:SN_impact_regularisation}.}
\revtwochanged{There are numerous methods that can be used, at least as a starting point, to identify appropriate regularisation values for an observation. Generally, only the smallest amount of regularisation necessary should be used. As increasing regularisation necessarily increases the $\chi^2$ of the fit, a common method is to choose the smallest regularisation value which increases the $\chi^2$ of the fit above one standard deviation of the $\chi^2$ distribution, $N + \sqrt{2N}$, where $N$ is the degrees of freedom of the model \citep{Press_2007}. Another potential approach would be to use the L-curve method, which aims to find a balance between solutions driven by the data and driven by regularisation \citep{Hansen_1992}.}
\revonechanged{Determining regularisation parameters will be explored in more detail in an upcoming paper where we will apply {\tt ROHSA-SNAPD} to \revtwochanged{a} large, statistical sample of real observations.}

We now consider the most appropriate fit to the disc galaxy, $\lambda_\mu = 10$ (Figure \ref{fig:KinMS_kinematic_axis_profiles} \textbf{B.}), in more detail. As expected from the cost function minimisation, the convolved {\tt ROHSA-SNAPD} profiles (red) match almost perfectly with the observed mock data (black points). The deconvolved {\tt ROHSA-SNAPD} kinematics (orange) match well with the high-resolution profiles (blue points), although there are larger discrepancies due to the added complexity of the convolutions and oversampling.

A major source of uncertainty when trying to recover the intrinsic kinematics is accounting for spectral resolution, which is added in quadrature to the deconvolved {\tt ROHSA-SNAPD} velocity dispersion during fitting, as detailed in Equation \ref{eqn:add_sigma_inst}. Uncertainties increase the further the solution is below the spectral resolution of the instrument, because changes in the deconvolved {\tt ROHSA-SNAPD} velocity dispersion only have a small effect on the convolved model. This is why the deconvolved dispersion $16$th and $84$th percentiles are skewed to lower dispersions. Uncertainties also increase towards the edges of the galaxy, where flux is lower and the solution is less constrained. Although most of the intrinsic kinematics lie within the error, the most notable discrepancy is that the intrinsic flux profile peaks at a higher value and has a slightly different profile than the deconvolved {\tt ROHSA-SNAPD}. This is partially due to the intrinsic data having a factor of $8$ higher spatial sampling, but also because the PSF and the Laplacian kernel convolutions set a minimum spatial resolution that {\tt ROHSA-SNAPD} can deconvolve. As such, this imperfect recovery of small-scale structures is expected and is seen throughout the {\tt ROHSA-SNAPD} fits.

\revtwochanged{
Figure \ref{fig:KinMS_RSNAPD_fit_hist} demonstrates how {\tt ROHSA-SNAPD} iteratively converges to the solutions presented in Figure \ref{fig:KinMS_kinematic_axis_profiles}. The three panels \textbf{A.}, \textbf{B.} and \textbf{C.} show how the cost $C \left(\thetab \right)$ (left column) and constituent functions $L \left(\thetab \right)$ (middle column) and $R \left(\thetab \right)$ (right column) change with fit iteration for velocity regularisations $\lambda_\mu = 2$ (panel \textbf{A.}), $\lambda_\mu = 10$ (panel \textbf{B.}) and $\lambda_\mu = 100$ (panel \textbf{C.}). Each line describes a fit to one of the $200$ noise realisations of the idealised disc galaxy.
In general, for each regularisation value the standard likelihood term $L \left(\thetab \right)$ drives the total cost $C \left(\thetab \right)$ and smoothly decreases with iteration. The regularisation residual term $R \left(\thetab \right)$ is mostly influenced by the rotational velocity map residual (orange lines), due to the relative strengths of $\lambda_f$ and $\lambda_\mu$. As $\lambda_\mu$ is increased, the velocity residual, and therefore the $R \left(\thetab \right)$ term as a whole, has a greater impact on the total cost of the fit, resulting in smoother deconvolved rotational velocity maps.
The evolution of $R \left(\thetab \right)$ with iteration also reflects the fact that {\tt ROHSA-SNAPD} fits to high S/N regions of the data first. The two predominant peaks seen in the rotation map residuals occur when the code initially fits to the high S/N inner region (iterations $1-10$) and then the outer regions (iterations $50-100$).

\begin{figure*}
    \centering
    \includegraphics[width=\linewidth]{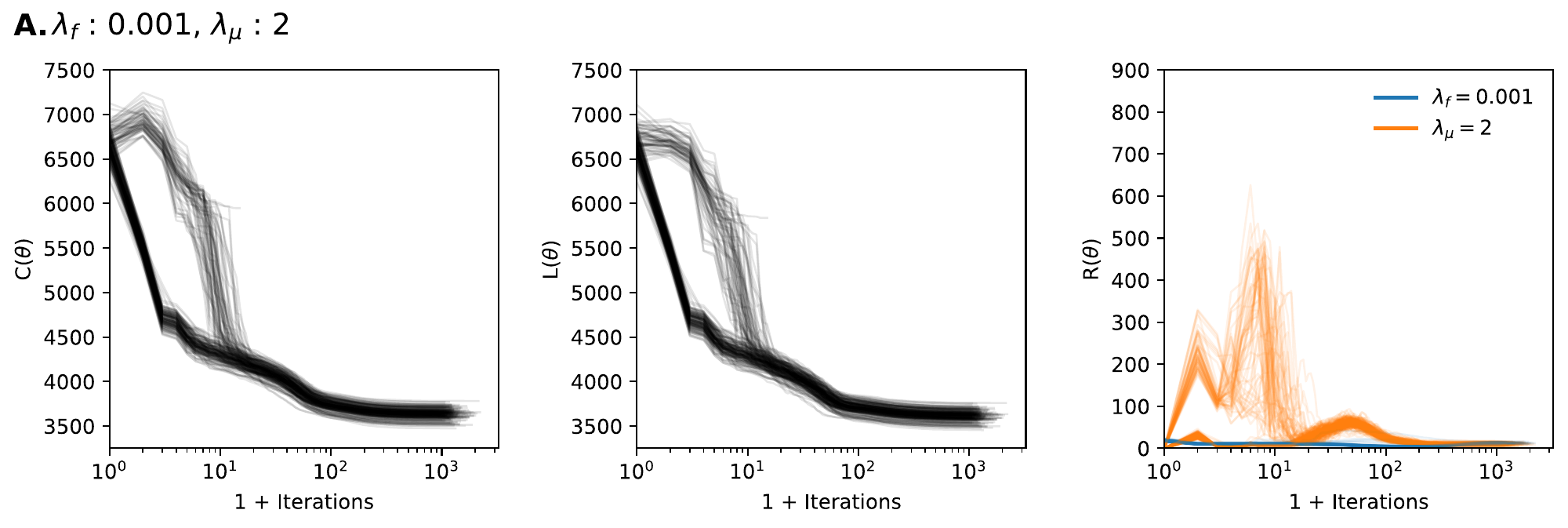}
    \includegraphics[width=\linewidth]{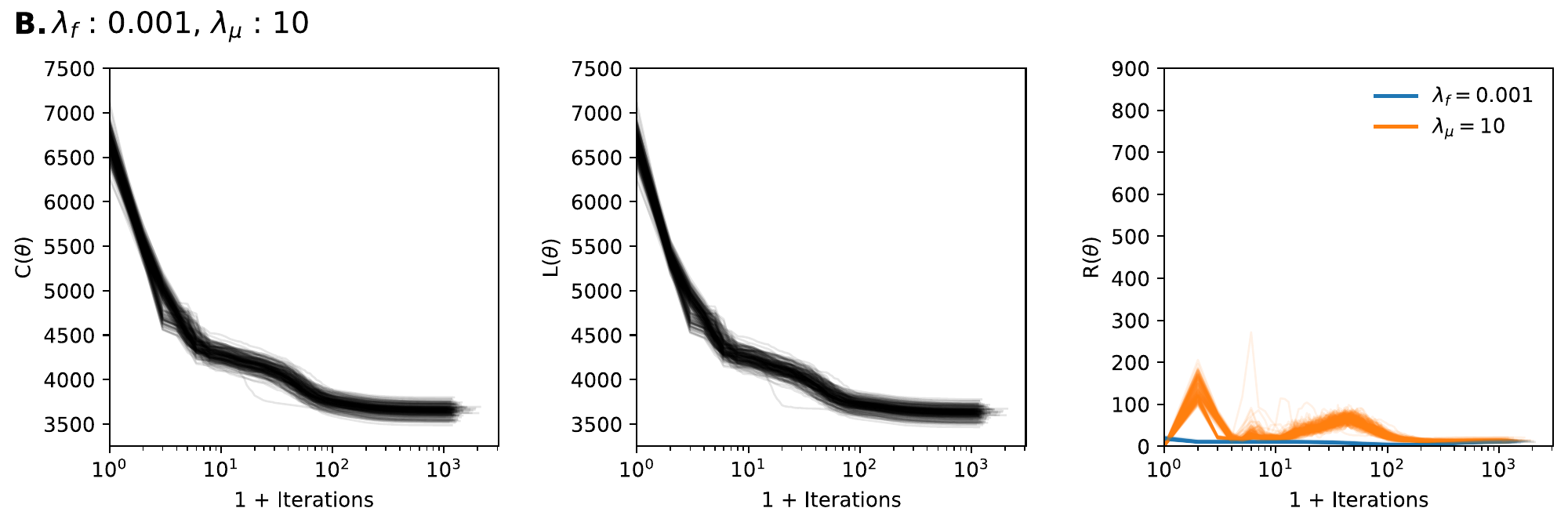}
    \includegraphics[width=\linewidth]{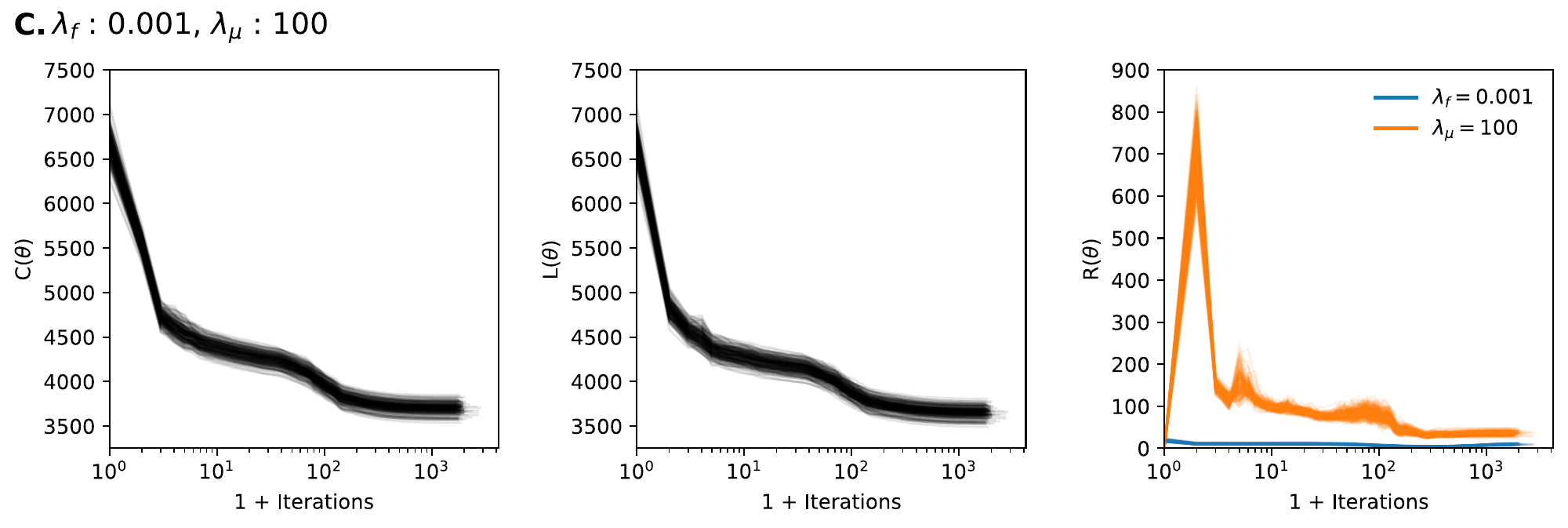}
    
    \caption{Panels \textbf{A.}, \textbf{B.} and \textbf{C.} show how the {\tt ROHSA-SNAPD} cost function varies when fitting the idealised disc galaxy with velocity regularisation values $\lambda_\mu = 2$, $10$ and $100$ respectively. The left column of each panel shows the change in total cost $C \left(\thetab \right)$ and the central column the change in standard likelihood $L \left(\thetab \right)$. The right column shows the change in regularisation residual $R \left(\thetab \right)$, split into contribution from the flux (blue lines) and rotation map residuals (orange lines). Each line in the figure is a fit to one of the $200$ noise realisations of the idealised disc galaxy.}
    \label{fig:KinMS_RSNAPD_fit_hist}
\end{figure*}
}

\subsection{Recovery of 2D kinematics}
\label{subsec:idealised_disc_recovery_of_2D_kinematics}
{\tt ROHSA-SNAPD} provides deconvolved 2D kinematic maps of galaxies as it is applied directly to data cubes.

\begin{figure*}
    \centering
    \includegraphics[width=0.49\linewidth,trim={3cm 3cm 2cm 0cm},clip]{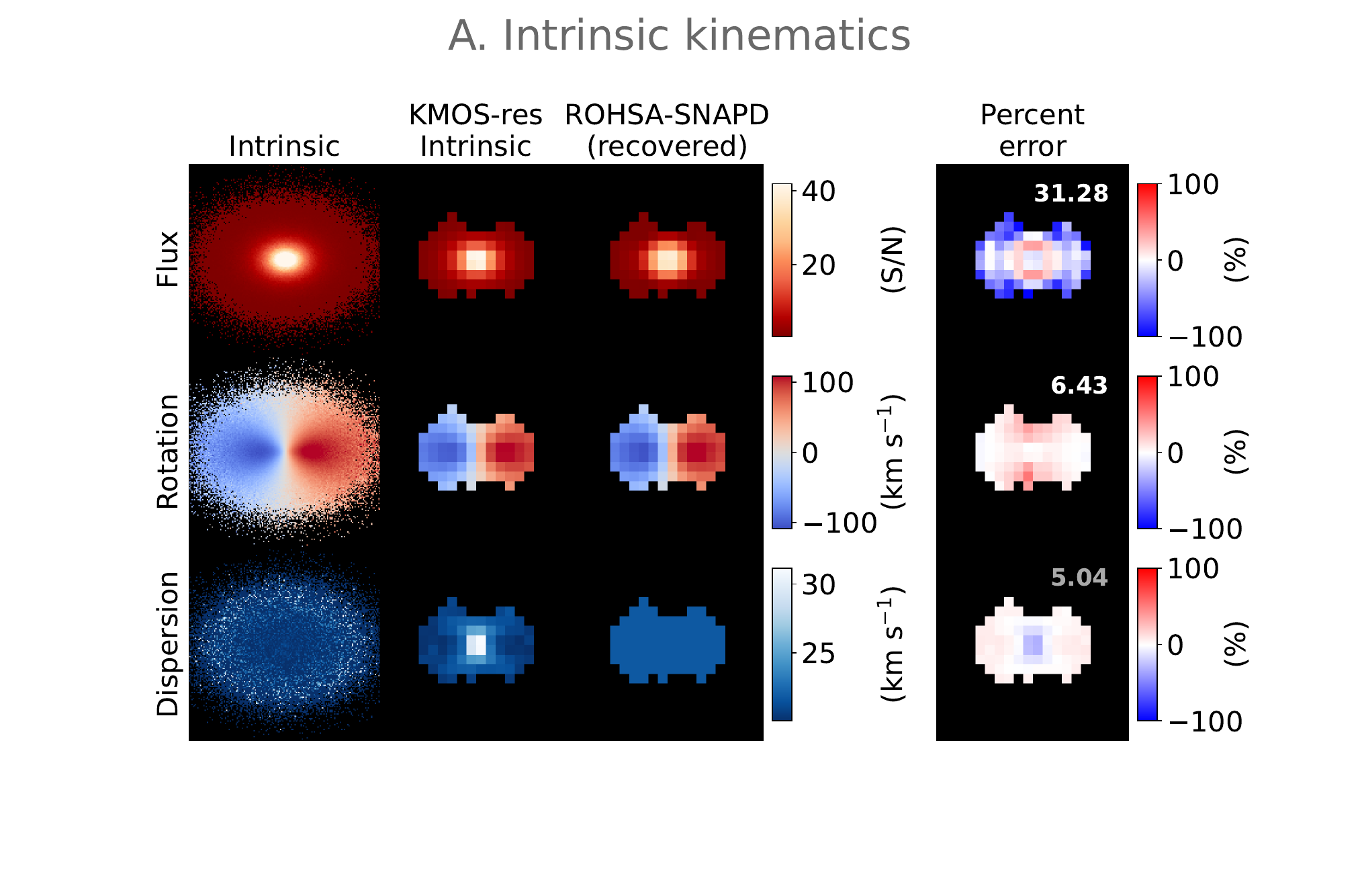}
    \includegraphics[width=0.49\linewidth,trim={3cm 3cm 2cm 0cm},clip]{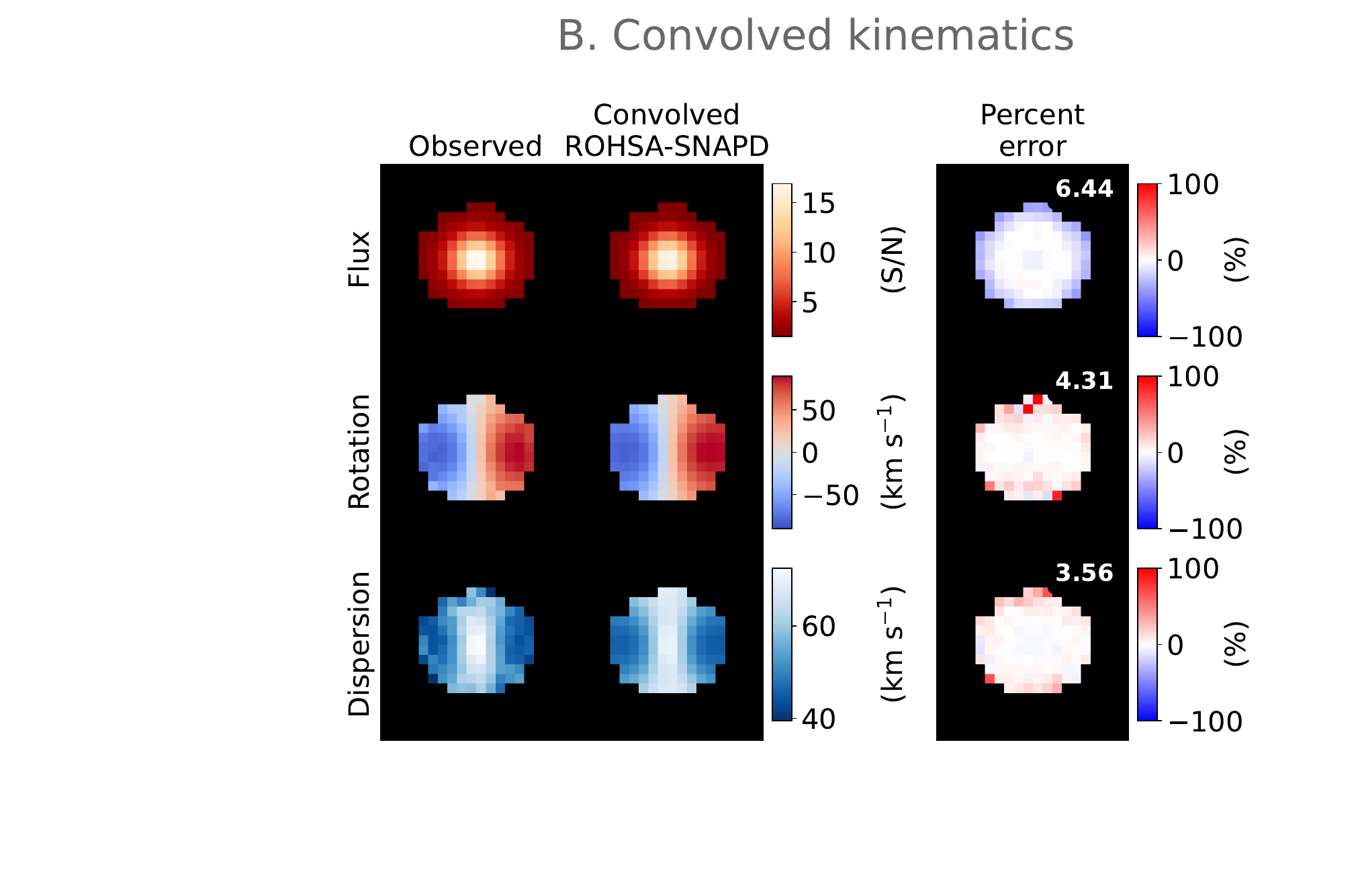}
    \caption{
    The results of the {\tt ROHSA-SNAPD} fits to the idealised disc galaxy using $\lambda_f = 0.001$ and $\lambda_\mu = 10$. The {\tt ROHSA-SNAPD} kinematics shown are the median deconvolved maps from the $200$ noise realisations.
    We compare the deconvolved {\tt ROHSA-SNAPD} kinematic maps with the intrinsic kinematics in \textbf{A.} and the convolved {\tt ROHSA-SNAPD} kinematic maps with the observed kinematics in \textbf{B.}. The rows of both panels show the flux, rotational velocity and velocity dispersion kinematic maps. \revonechanged{Every flux map has been scaled by the RMS noise level per spatial pixel (constant across the whole map) to present the flux in terms of S/N per spatial pixel.}
    As discussed in Section \ref{subsec:idealised_disc_recovery_of_2D_kinematics}, masks are created to show {\tt ROHSA-SNAPD} deconvolved flux (in the case of KMOS-resolution and deconvolved {\tt ROHSA-SNAPD} maps) and observed S/N > 1 (in the case of observed and convolved {\tt ROHSA-SNAPD} maps). All pixels with non-zero intrinsic flux are shown in the high-resolution intrinsic maps.
    The colourbar of each row is fixed between the $1$st and $99$th percentiles of the KMOS-resolution intrinsic and observed maps, removing any masked pixels. To ensure the rotation colourbars are symmetric, the bounds are set to $\pm$ the percentile with the largest magnitude.
    Percent error maps are shown in the final column of each row, defined here as ({\tt ROHSA-SNAPD} - data)/data$\times 100$ for each map. The median absolute percent error is given at the top left of each residual map, calculated the unmasked regions. The percent error comparing the KMOS-resolution intrinsic and deconvolved {\tt ROHSA-SNAPD} velocity dispersions is in grey to signify that the comparison between the two maps is impacted by additional observational effects introduced when creating the KMOS-resolution maps.
    }
    \label{fig:KinMS_kinematic_map_plots}
\end{figure*}

The deconvolved {\tt ROHSA-SNAPD} kinematic maps for the idealised disc, using $\lambda_f = 0.001$ and $\lambda_\mu = 10$, are shown in Figure \ref{fig:KinMS_kinematic_map_plots} \textbf{A.} alongside the intrinsic kinematics at high and KMOS-like spatial resolutions. The convolved {\tt ROHSA-SNAPD} kinematics \revonechanged{obtained by fitting Gaussian profiles to the convolved model cubes, as discussed in Section \ref{subsec:idealised_disc_results}}, are shown in Figure \ref{fig:KinMS_kinematic_map_plots} \textbf{B.}, compared with the mock observed kinematic maps. 
\revonechanged{To more clearly demonstrate the} \revtwochanged{code's ability} \revonechanged{to fit observed flux to low S/N, we scale every flux map the RMS noise level per spatial pixel (a constant value across the whole map) to present the flux in terms of S/N per spatial pixel.}

As was seen in the 1D kinematic axis profiles, the convolved and observed kinematic maps (Figure \ref{fig:KinMS_kinematic_map_plots} \textbf{B.}) match very closely, with the median absolute percent error below $7$ \revtwochanged{per cent} for all three maps.
The deconvolved and intrinsic maps also agree well (Figure \ref{fig:KinMS_kinematic_map_plots} \textbf{B.}), reflecting the kinematic axis profiles in Figure \ref{fig:KinMS_kinematic_axis_profiles}. The deconvolved {\tt ROHSA-SNAPD} flux generally matches the KMOS-resolution intrinsic flux, though the percent error increases at the galaxy centre because {\tt ROHSA-SNAPD} cannot reproduce the very small-scale structure. Additionally, the KMOS-resolution intrinsic flux is somewhat higher at the edges of the mask than the deconvolved {\tt ROHSA-SNAPD} flux, likely because the solution is less constrained in low S/N regions. The intrinsic rotation is very accurately recovered by {\tt ROHSA-SNAPD} with a median absolute percent error below $7$ \revtwochanged{per cent}.

Unlike the kinematic axis profiles, it is not straightforward to compare the high-resolution kinematics with those obtained with {\tt ROHSA-SNAPD} due to the difference in both spatial and spectral scales. 
For a reasonable comparison, we spatially bin the high-resolution data cubes and calculate moments to create `KMOS-resolution intrinsic maps' for comparison with the deconvolved {\tt ROHSA-SNAPD} model. The rebinned maps do not provide a one-to-one comparison, because by binning the high-resolution data to create the KMOS-resolution intrinsic maps, we are introducing observational effects due to finite spatial sampling, something which we are specifically accounting for in {\tt ROHSA-SNAPD} by oversampling the parameter maps. Despite this, the KMOS-resolution intrinsic maps still allow the general structure of the deconvolved and intrinsic kinematics to be compared, with some small discrepancies.

The kinematic maps in Figure \ref{fig:KinMS_kinematic_map_plots} are masked for visual clarity. The high-resolution intrinsic kinematic maps are masked to only show pixels with non-zero intrinsic flux. The KMOS-resolution intrinsic, deconvolved {\tt ROHSA-SNAPD} and percent error kinematic maps in Figure \ref{fig:KinMS_kinematic_map_plots} \textbf{A.} are masked to show only pixels with deconvolved {\tt ROHSA-SNAPD} flux above the flux lower bound.
The observed, convolved {\tt ROHSA-SNAPD} and percent error kinematic maps in Figure \ref{fig:KinMS_kinematic_map_plots} \textbf{B.} show all pixels with a S/N $> 1$, using a S/N map created from the observed data in the same way as in Section \ref{subsec:creating_mock_observations_with_KinMS}. These masks are not imposed during the {\tt ROHSA-SNAPD} fitting, only when displaying the maps.

\subsection{Impact of spatial sampling}
\label{subsec:impact_spatial_binning}
The clearest visual difference between the KMOS-resolution intrinsic and deconvolved {\tt ROHSA-SNAPD} kinematics in Figure \ref{fig:KinMS_kinematic_map_plots} is the velocity dispersion maps, particularly at the galaxy centre. The apparent discrepancy is due to observational effects introduced when spatially binning the high-resolution data to create the KMOS-resolution intrinsic kinematic maps and is most significant at the galaxy centre because a large range of rotational velocities overlap. 

\begin{figure}
    \centering
    \includegraphics[width=\linewidth]{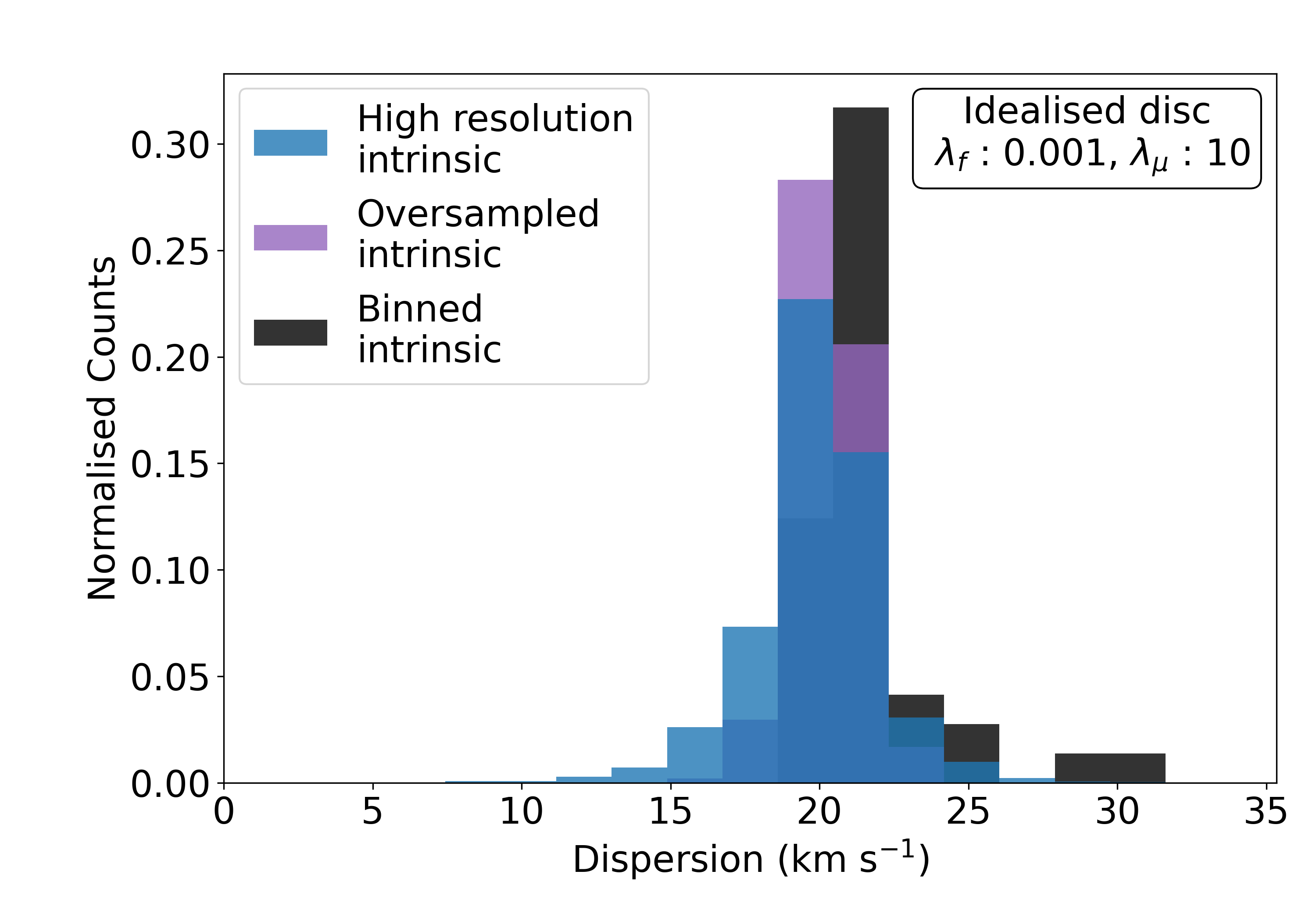}
    \includegraphics[width=\linewidth]{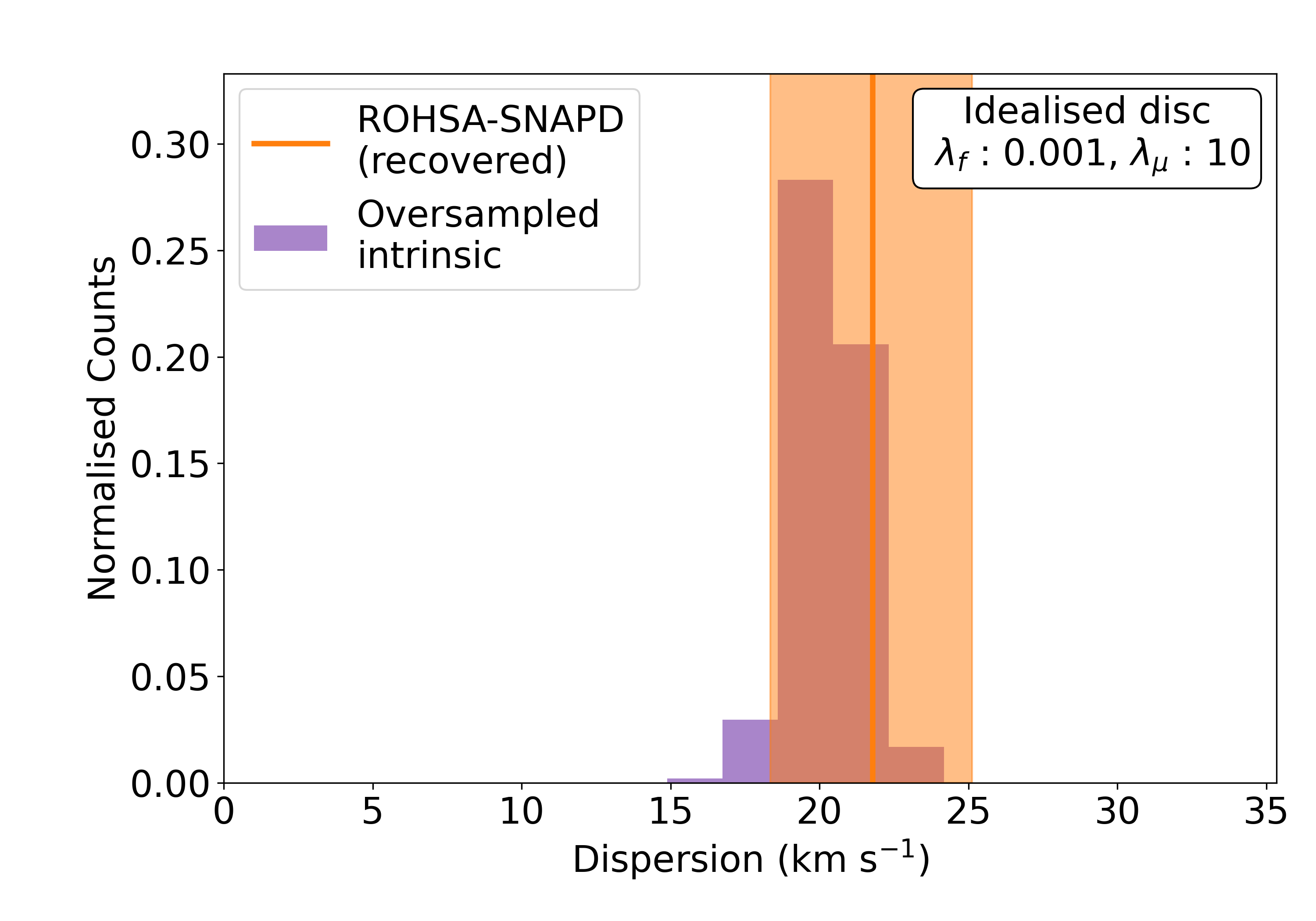}
    \caption{For the idealised disc, a comparison of the deconvolved {\tt ROHSA-SNAPD} velocity dispersion values (orange) with a histogram of the intrinsic velocity dispersion at high-resolution ($n=8$) (blue), at oversampled resolution ($n=4$) (purple) and at observed resolution ($n=1$) (black). For a clear comparison, the area under each distribution is normalised to $1$. From the $200$ noise realisations, the median deconvolved {\tt ROHSA-SNAPD} velocity dispersion is shown as a solid orange line, and the $16$th and $84$th percentiles are shaded. The axes of both panels span from $0$ to the maximum value of any resolution.}
    \label{fig:disp_hist}
\end{figure}

We demonstrate the impact of spatial sampling on the idealised disc in Figure \ref{fig:disp_hist}. In the top row of Figure \ref{fig:disp_hist} we compare the intrinsic dispersion distributions of the disc model when sampled at three different spatial scales: high-resolution ($n=8$ times higher spatial resolution than the observed data), oversampled resolution ($n=4$ times higher spatial resolution than the observed data) and observed resolution.
The oversampled resolution velocity dispersion map represents the most accurate dispersion that could be obtained using {\tt ROHSA-SNAPD} when accounting for spatial sampling effects using $n=4$ oversampling (though see Appendix \ref{appendix:limitations_bilinear_interpolation} for a discussion of why the interpolation method cannot completely recover the oversampled intrinsic kinematics).
The velocity dispersion maps were masked to only include pixels with deconvolved {\tt ROHSA-SNAPD} flux, scaling the mask extent relative to the map resolution. Therefore, the varying distributions in Figure \ref{fig:disp_hist} are purely due to spatial sampling effects.
In the bottom row of Figure \ref{fig:disp_hist} the deconvolved {\tt ROHSA-SNAPD} velocity dispersion is compared with the oversampled dispersion map values to validate the accuracy of the {\tt ROHSA-SNAPD} fit.

{\tt ROHSA-SNAPD} can effectively recover the intrinsic kinematics of the idealised disc, accounting for realistic observational effects. The velocity dispersion distributions of the three spatial resolutions in Figure \ref{fig:disp_hist} are very similar, though the KMOS-resolution histogram is skewed to higher values, reflecting the increased dispersion seen at the centre of the map in Figure \ref{fig:KinMS_kinematic_map_plots}. The oversampled dispersion values agree quite well with the high-resolution and are only minimally shifted, showing that the oversampled kinematic maps can accurately account for the effects of spatial binning in the idealised disc. The majority of the oversampled dispersion histogram is also within the error of the {\tt ROHSA-SNAPD} fit, showing that the deconvolved dispersion is accurate given the $n=4$ oversampling within the cost function.

\section{Evaluation on TNG100 galaxies}
\label{sec:evaluation_tng_galaxies}
We apply {\tt ROHSA-SNAPD} to a mock galaxy at three different merger stages created from the {\tt IllustrisTNG} cosmological simulation TNG100 (specifically TNG100-1) \citep{Marinacci_2018,Springel_2018,Pillepich_2018,Naiman_2018,Nelson_2018,Nelson_2019}. TNG100 galaxies have more complex kinematics than the idealised disc model, allowing us to demonstrate the main strength of {\tt ROHSA-SNAPD}, its ability to be applied to galaxies that deviate from purely disc-like kinematics.

\subsection{Creating mock observations from the TNG100 snapshots}
\label{subsec:creating_mock_observations_from_TNG}
We followed a generally similar process to that described in Section \ref{subsec:creating_mock_observations_with_KinMS} to create mock observations from the TNG100 simulation. We utilised merger catalogues \citep{Rodriguez-Gomez_2017,Eisert_2023} to identify a galaxy undergoing a major merging event at snapshot $33$ ($z=2.00$) of the simulation. By stepping forwards and backwards through time, we identified snapshots of the same galaxy in a pre- and post-merger stage. These three snapshots made up our TNG100 sample, as detailed in Table \ref{tab:galaxy_sample}, with a redshift range $1.8 < z < 2.1$, stellar mass range $2.7 < M_* \left[10^{10} \, \textnormal{M}_\odot \right] < 4.8$ and \revonechanged{star formation} rate (SFR) range $17.2 < \, $SFR$ \left[ \textnormal{M}_\odot \, \textnormal{yr}^{-1} \right] < 31.2$, all calculated within 2 stellar half-mass radii. \revonechanged{The identified TNG100 galaxy was chosen to have similar properties to $z \sim 2$ galaxies observed by KMOS surveys.}

\begin{table*}
    \centering
    \begin{tabular}{c c c c c c c c c c c c}
        \hline\hline \\ [-2ex]
        Name & Subhalo ID & Snapshot & $z$ & Lookback time & $\textnormal{M}_{*}$ & SFR & \multicolumn{2}{c}{$2 \, \times \, $R\textsubscript{half, stars}} & S/N & $\lambda_f$ & $\lambda_\mu$ \\ [0.5ex]
         &  &  &  & [Gyr] & $\left[10^{10} \, \textnormal{M}_{\odot} \right]$ & $\left[\textnormal{M}_{\odot} \, \textnormal{yr}^{-1} \right]$ & [kpc] & [arcsec] & & & \\
        \hline \\ [-3ex]
        Idealised disc & - & - & 2.00 & 10.51 & 3.00 & - & - & - & 39.43 & 0.001 & 10 \\
        Pre-merger & 31095 & 32 & 2.10 & 10.67 & 2.65 & 18.48 & 8.38 & 0.98 & 40.39 & 0.001 & 5 \\
        Ongoing merger & 34906 \& 34908 & 33 & 2.00 & 10.52  & 3.12 & 31.15 & 7.42 & 0.86 & 39.41 & 0.001 & 3 \\
        Post-merger & 40579 & 35 & 1.82 & 10.21 & 4.78 & 17.17 & 8.10 & 0.94 &  39.66 & 0.001 & 2 \\
        \hline \\ [-3ex]
    \end{tabular}
    \caption{Properties of the galaxy sample fit with {\tt ROHSA-SNAPD}. The first row details the parameters of the idealised disc created with {\tt KinMS} and the remaining rows are the TNG100 galaxy at different snapshots. The TNG100 masses and SFRs were calculated within twice the stellar half-mass radii. \revonechanged{The S/N of each galaxy was calculated within a $1.5$ arcsecond aperture, following the method described in Section \ref{subsec:creating_mock_observations_with_KinMS}}. The final two columns show the regularisation values used to successfully fit each observation.}
    \label{tab:galaxy_sample}
\end{table*}

\begin{figure*}
    \centering
    \includegraphics[trim={0cm 0cm 0cm 0cm},clip,width=\linewidth]{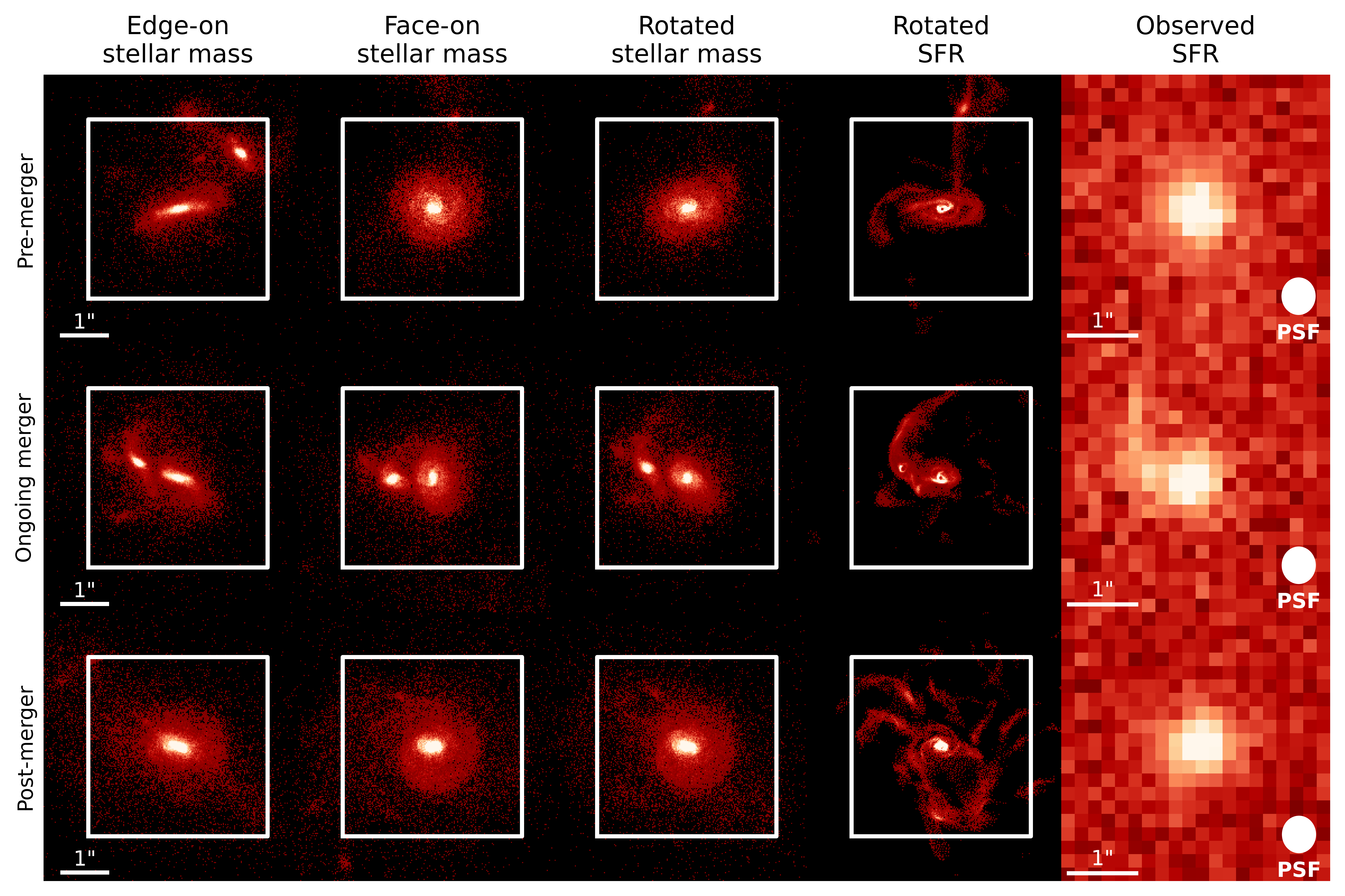}
    \caption{Flux and stellar mass maps of our TNG100 sample, comparing the high-resolution, noise-less simulation data (left panels) with the KMOS-like mock data which have been convolved with the PSF and instrument spectral resolution, binned spatially and spectrally and have had noise added (right panels). All intrinsic data with flux $> 0$ is shown. The intrinsic data is shown in $6 \times 6$ arcsecond apertures, with a white box denoting the KMOS-like $4 \times 4$ arcsecond field-of-view of the observed data. The FWHM ellipse of the PSF applied to the observed data is also shown in the bottom left of the mock data panels.}
    \label{fig:intrinsic_vs_observed_fluxes}
\end{figure*}

To create KMOS-like mock observations from the simulation data, we first aligned the galaxy face-on in each snapshot using the moment of inertia tensor, calculated from star-forming gas cells within twice the stellar half-mass radius\footnote{Following \url{https://www.tng-project.org/data/forum/topic/223/subhalo-face-on-vector-values/}}, before rotating the galaxy by 45 degrees.
\revtwochanged{In general the code is agnostic to a specific emission line, assuming it can be approximated by a Gaussian distribution. In this work we model H$\alpha$-like emission lines using gas cell instantaneous \revonechanged{star formation} rates to trace the ionised gas kinematics.}
\revtwochanged{We followed a similar methodology to \citet{Harborne_2020} to create 3D data cubes. The emission lines from each gas cell were assumed to be Gaussian}, with flux set by the \revonechanged{star formation} rate, central velocity set by the line-of-sight velocity and velocity dispersion equal to $\sqrt{\sigma_{\textnormal{inst}}(\rb,v_z)^2 + \sigma_{\textnormal{thermal}}}$, where $\sigma_{\textnormal{inst}}(\rb,v_z)$ is the instrument spectral resolution and $\sigma_{\textnormal{thermal}}$ is a $10$ km s$^{-1}$ thermal broadening term. Thermal broadening was added manually because it is not included in the simulation \citep{Pillepich_2019}.
The individual gas cell spectra were then spatially binned to create high-resolution data cubes, again at a $0\textnormal{”}.025 \times 0\textnormal{”}.025 \times 10\textnormal{ km s}^{-1}$ resolution. This also corresponds to a spatial binning of $\sim 0.64$ comoving kpc$^2$ at $z=2$, similar to the $0.5$ comoving kpc$^2$ resolution used by \citet{Pillepich_2019}.
The high-resolution cubes were then convolved by the same normalised Moffat PSF model as the idealised disc \revonechanged{(major and minor FWHM of $0\textnormal{"}.527$ and $0\textnormal{"}.478$ respectively)} and binned to KMOS-like resolution, $0\textnormal{”}.2 \times 0\textnormal{”}.2 \times 40\textnormal{ km s}^{-1}$ before adding white Gaussian noise. Flux and stellar mass maps of the galaxies are shown in the bottom three rows of Figure \ref{fig:intrinsic_vs_observed_fluxes}.

\subsection{Applying {\tt ROHSA-SNAPD} to the TNG100 mock observations}
\label{subsec:TNG100_applying_RSNAPD}
To apply {\tt ROHSA-SNAPD} to the TNG100 mock observations, we used the same bounds and oversample factor as when fitting the idealised disc and generated initial conditions specific to each observation, following the method described in Section \ref{subsec:idealised_disc_applying_RSNAPD}.
We fit each snapshot with appropriate regularisation values, as detailed in the final two columns of Table \ref{tab:galaxy_sample}. \revonechanged{We demonstrate the effect of using the same regularisation value for the full sample in \revtwochanged{in both Section \ref{subsec:fitting_all_TNG100_with_lam_10} and} Appendix \ref{appendix:all_same_regularisation_values}, where every snapshot is fit using $\lambda_\mu = 10$, the most appropriate regularisation value determined from the idealised disc.}
We compare the deconvolved {\tt ROHSA-SNAPD} kinematics with high- and KMOS-resolution intrinsic kinematic maps, obtained directly from the \revtwochanged{gas cells} using moment equations given by \citet{Harborne_2020}.

\subsection{Pre-merger}
\label{subsec:pre_merger_results}
\begin{figure*}
    \centering
    \includegraphics[width=0.49\linewidth,trim={3cm 3cm 2cm 0cm},clip]{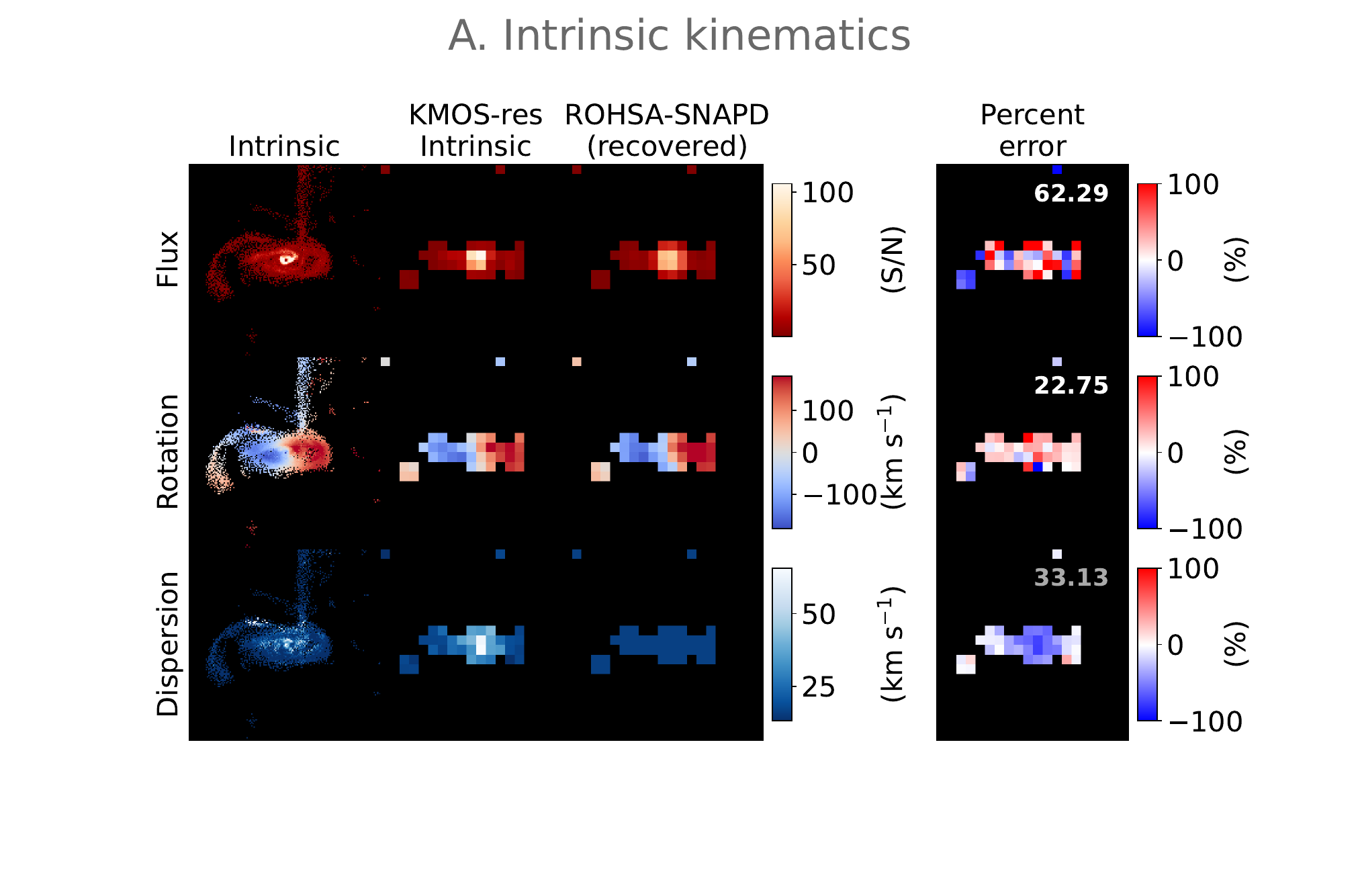}
    \includegraphics[width=0.49\linewidth,trim={3cm 3cm 2cm 0cm},clip]{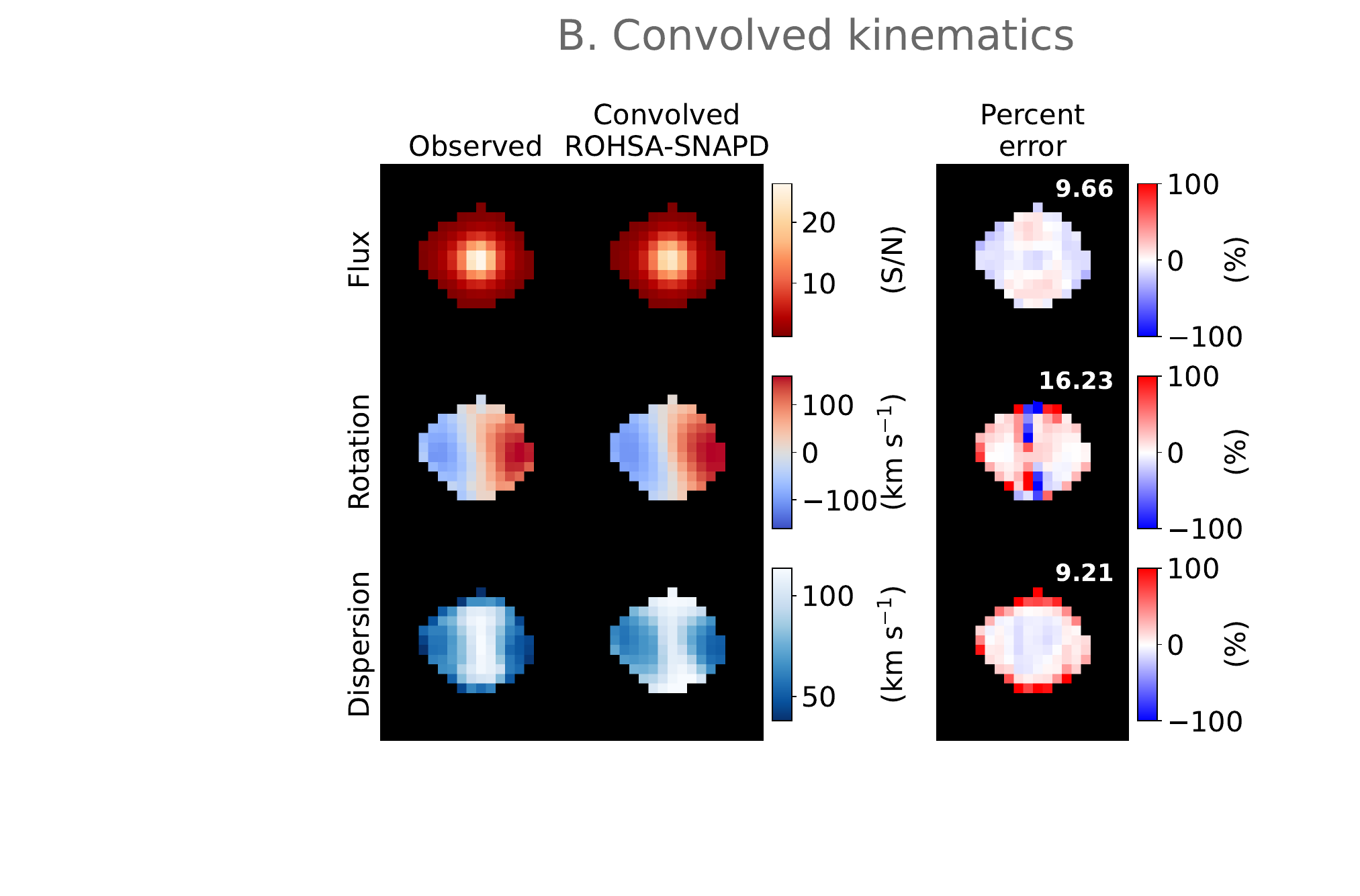}
    \caption{The results of the {\tt ROHSA-SNAPD} fits to the pre-merger snapshot using $\lambda_f = 0.001$ and $\lambda_\mu = 5$. The {\tt ROHSA-SNAPD} maps shown are the median fits to the $200$ noise realisations. This figure follows the same layout as Figure \ref{fig:KinMS_kinematic_map_plots}.}
    \label{fig:pre_merger_kinematic_maps}
\end{figure*}

In the pre-merger snapshot shown in Figure \ref{fig:pre_merger_kinematic_maps}, the galaxy is a mostly disc-like system, with a flattened flux distribution, a slightly warped rotation curve and a reasonably constant velocity dispersion profile. The companion galaxy can also be seen in each orientation of the snapshot. When observational effects are applied, the galaxy's flux is very symmetric as the PSF convolution heavily influences its distribution. The rotation warp is mostly smoothed over so that the galaxy appears to have ordered rotation and a characteristic central increase in velocity dispersion.
We fit this snapshot with {\tt ROHSA-SNAPD} using $\lambda_f=0.001$ and $\lambda_\mu=5$. The convolved {\tt ROHSA-SNAPD} model fits the observed galaxy successfully, matching the characteristic disc-like kinematics with a median absolute error of less than $17$ \revtwochanged{per cent} for all maps. The largest errors occur at the low S/N galaxy edges because the solution is less well constrained and near the centre of the rotation map where values are close to zero, so small deviations cause large percent errors.

\begin{figure*}
    \centering
    \includegraphics[width=0.33\linewidth]{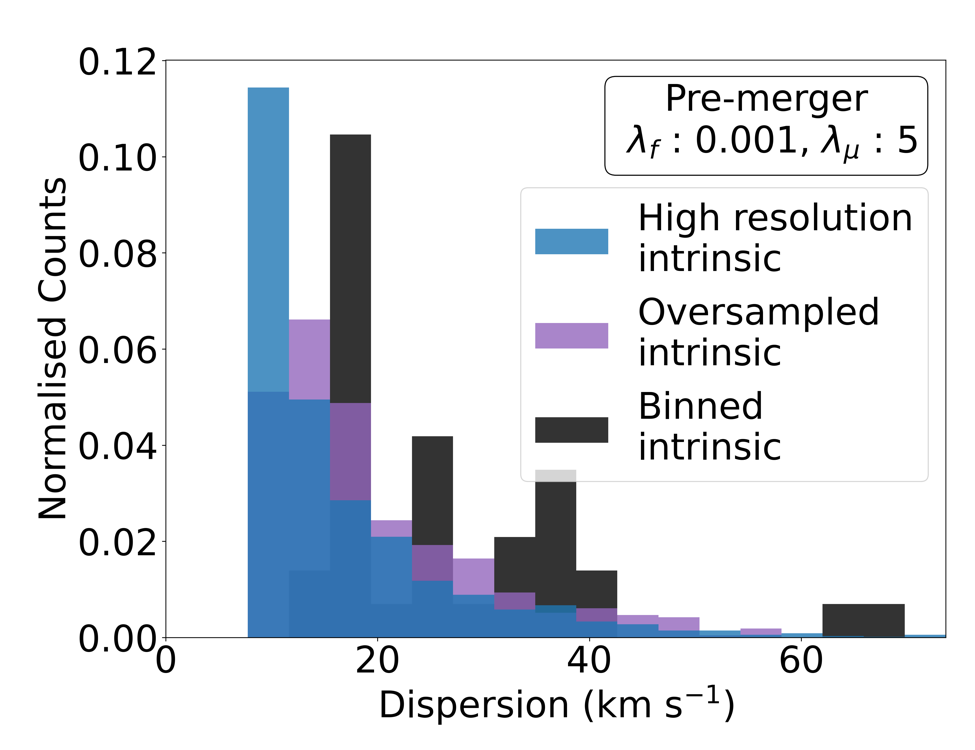}
    \includegraphics[width=0.33\linewidth]{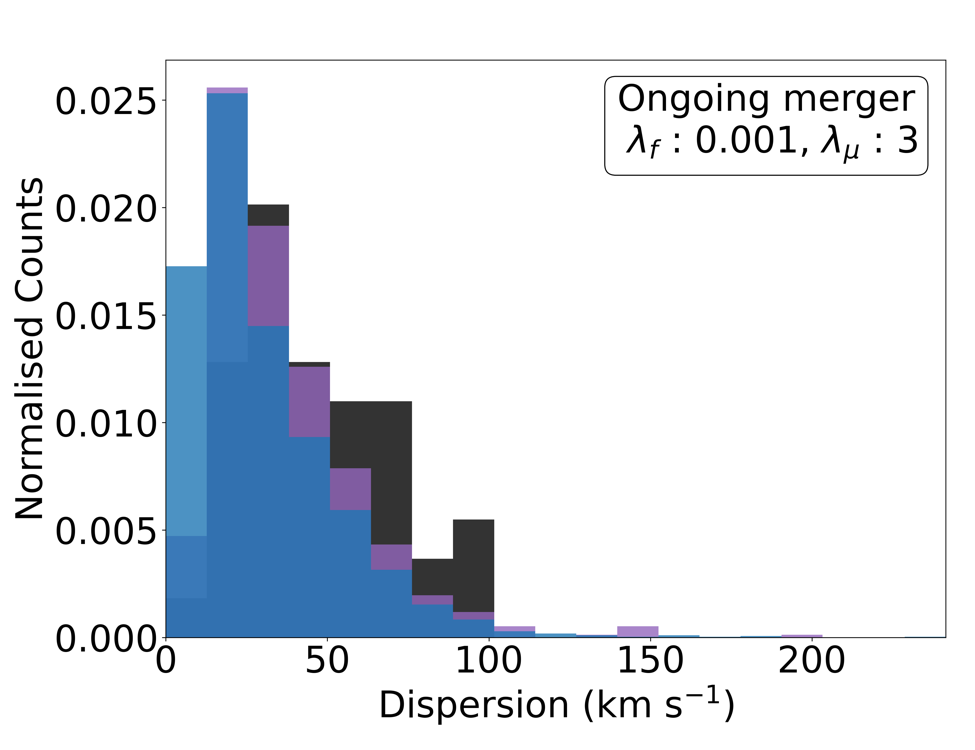}
    \includegraphics[width=0.33\linewidth]{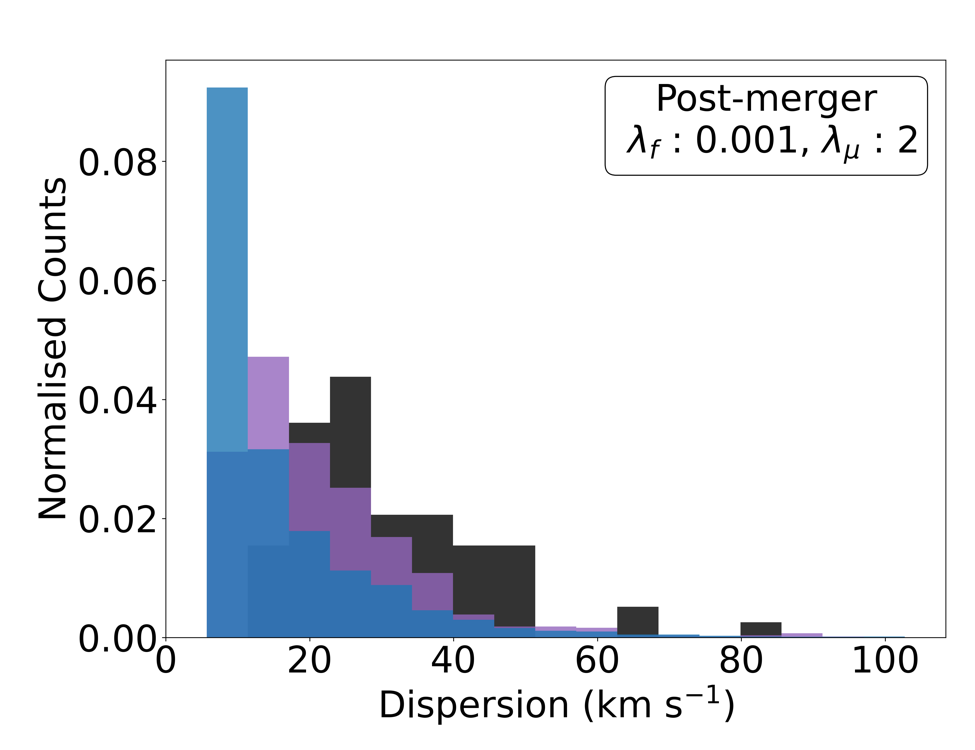}
    \includegraphics[width=0.33\linewidth]{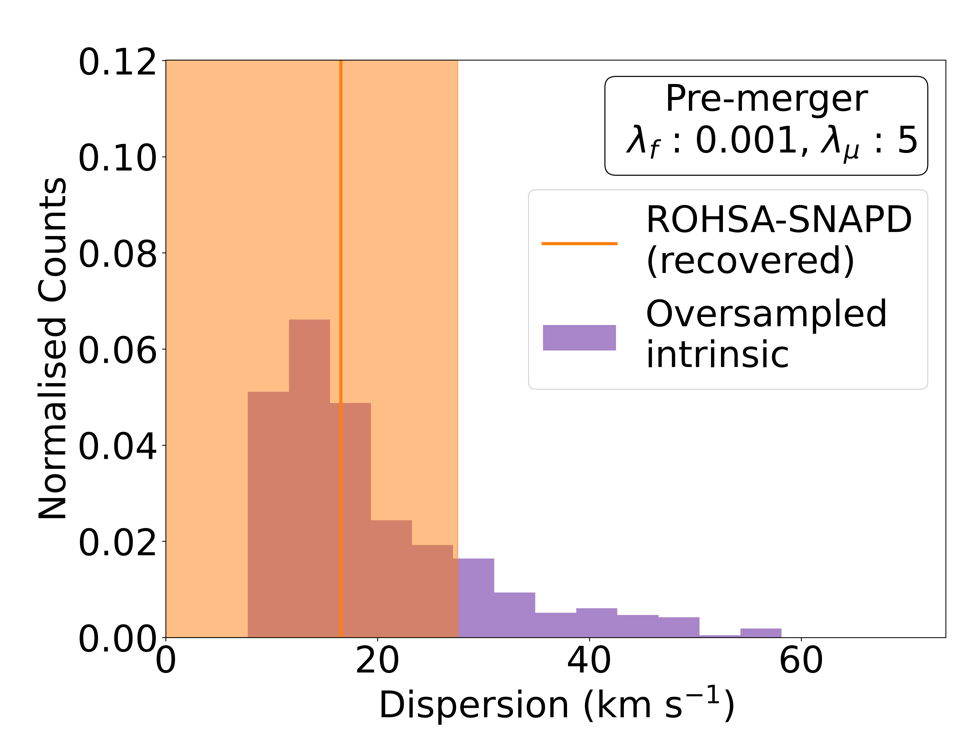}
    \includegraphics[width=0.33\linewidth]{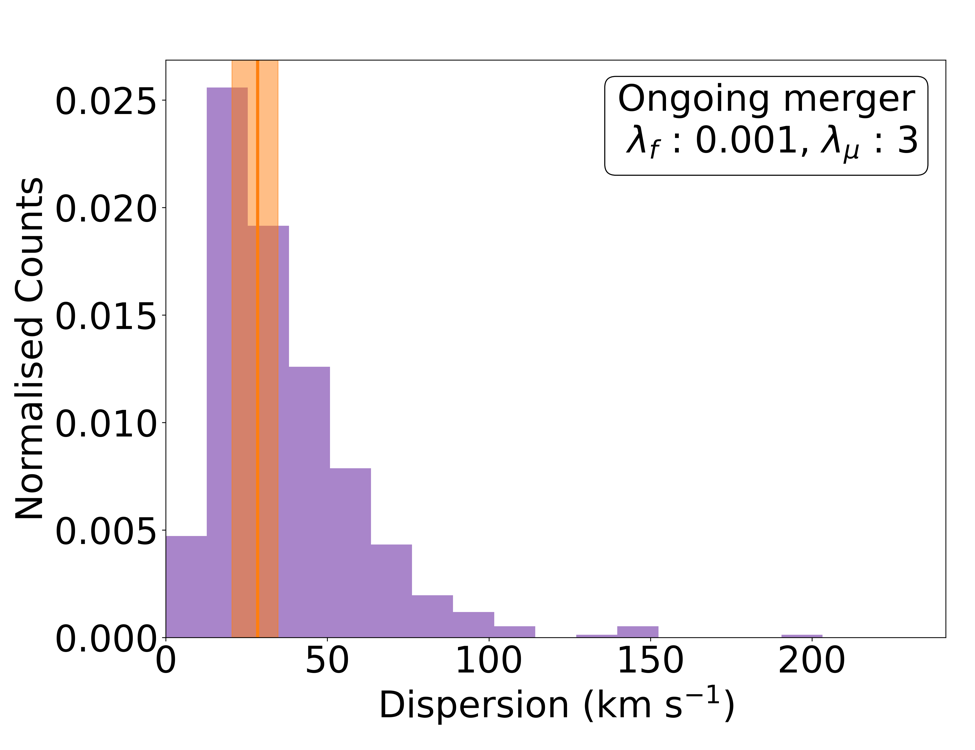}
    \includegraphics[width=0.33\linewidth]{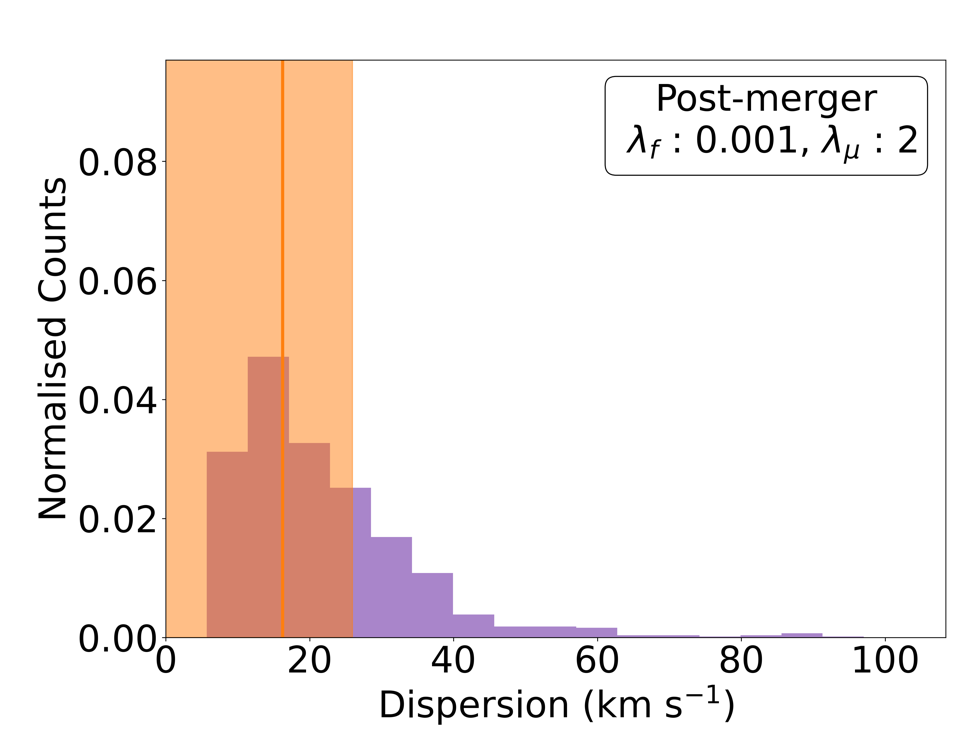}
    \caption{For each TNG100 snapshot, a comparison of the deconvolved {\tt ROHSA-SNAPD} velocity dispersion values (orange) with a histogram of the idealised disc's intrinsic velocity dispersion at high-resolution ($n=8$) (blue), at oversampled resolution ($n=4$) (purple) and at observed resolution ($n=1$) (black). For a clear comparison, the area under each distribution is normalised to $1$. From the $200$ noise realisations, the median deconvolved {\tt ROHSA-SNAPD} velocity dispersion is shown as a solid orange line, and the $16$th and $84$th percentiles are shaded. The \revonechanged{x-axis} of each panel \revonechanged{spans} from $0$ to the maximum value of any resolution.}
    \label{fig:disp_hist_TNG}
\end{figure*}

Despite the significant impact of the beam, {\tt ROHSA-SNAPD} still recovers the intrinsic kinematics of the galaxy.
The deconvolved flux map agrees well with the KMOS-resolution intrinsic flux,
apart from some small-scale pixel-to-pixel variations that {\tt ROHSA-SNAPD} cannot reproduce because the kernel convolutions set a minimum spatial scale that can be deconvolved.
The deconvolved rotation values also match well with the KMOS-resolution map, although
the warped rotation curve is only somewhat resolved due to the resolution of the data.
The increased error at the very centre of the rotation map is partially because low values can give rise to large percent errors, but also because there are sub-pixel variations in the rotation gradient of the intrinsic rotation map which cannot be accounted for by {\tt ROHSA-SNAPD}, discussed further in Appendix \ref{appendix:limitations_bilinear_interpolation}.
As mentioned in Section \ref{subsec:idealised_disc_recovery_of_2D_kinematics}, the KMOS-resolution intrinsic and deconvolved {\tt ROHSA-SNAPD} velocity dispersion maps cannot be directly compared due to spatially sampling effects introduced when binning the high-resolution intrinsic maps. Therefore, we compare the effects of spatial binning in Figure
\ref{fig:disp_hist_TNG}, which demonstrates the need for oversampling within the cost function as the KMOS-resolution intrinsic dispersions are substantially larger than the high-resolution and oversampled resolution values.
Figure \ref{fig:disp_hist_TNG} also shows that the deconvolved {\tt ROHSA-SNAPD} velocity dispersion matches well with the peak of the oversampled resolution intrinsic distribution, though there is a large spread of values that cannot be recovered completely with a single velocity dispersion.

\subsection{Ongoing merger}
\label{subsec:ongoing_merger_results}
\begin{figure*}
    \centering
    \includegraphics[width=0.49\linewidth,trim={3cm 3cm 2cm 0cm},clip]{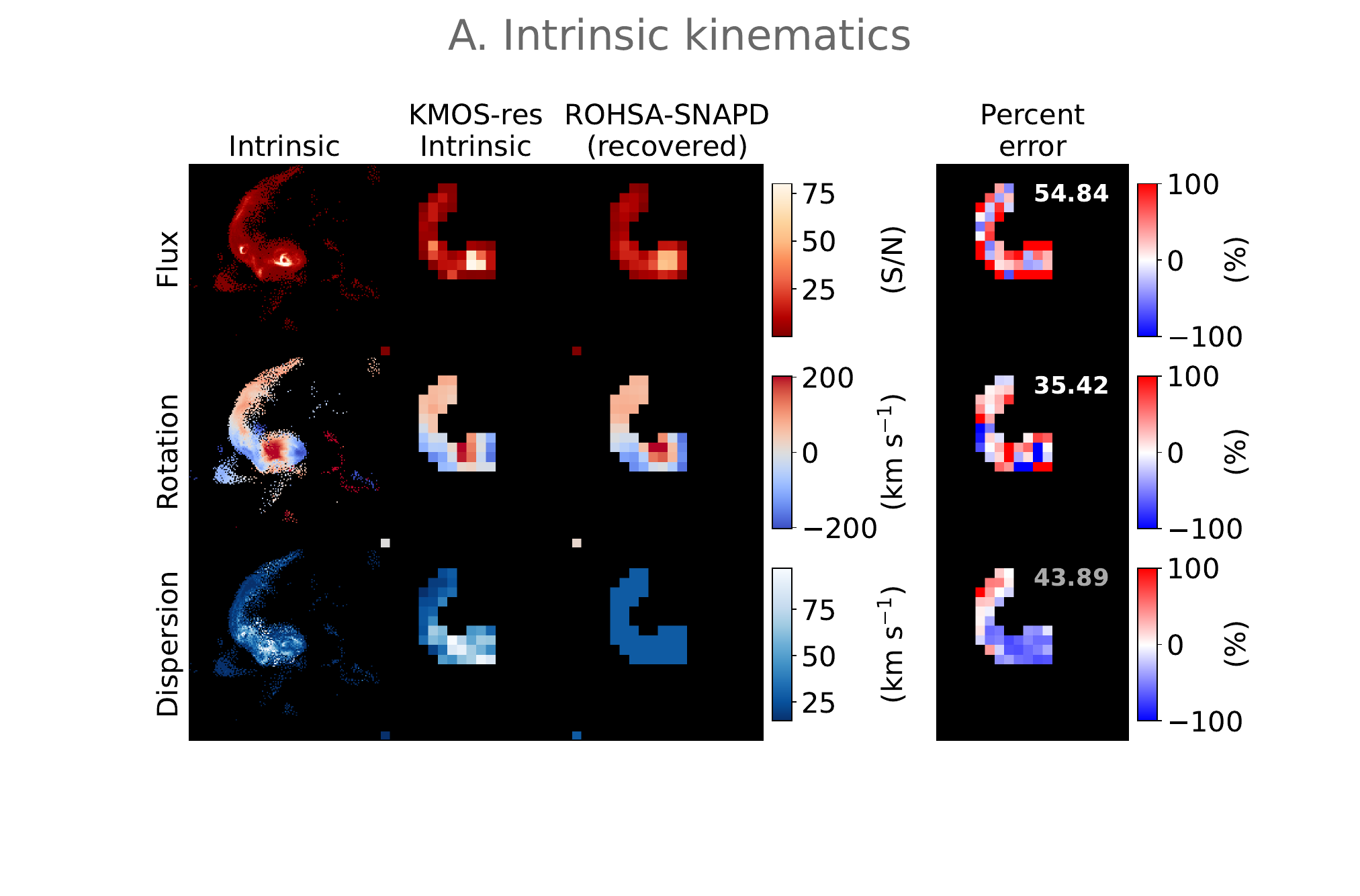}
    \includegraphics[width=0.49\linewidth,trim={3cm 3cm 2cm 0cm},clip]{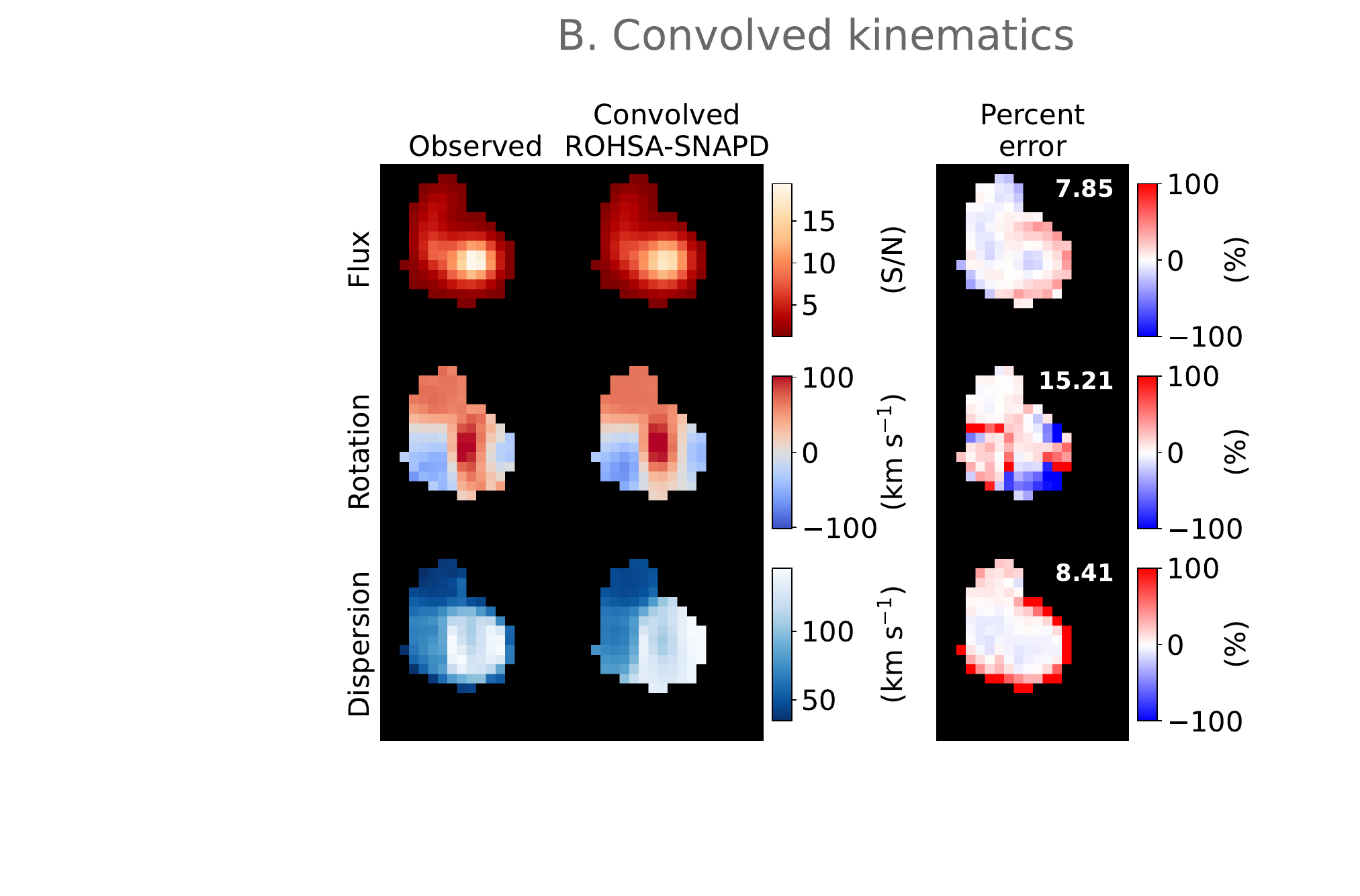}
    \caption{The results of the {\tt ROHSA-SNAPD} fits to the ongoing merger snapshot using $\lambda_f = 0.001$ and $\lambda_\mu = 3$. The {\tt ROHSA-SNAPD} maps shown are the median fits to the $200$ noise realisations. This figure follows the same layout as Figure \ref{fig:KinMS_kinematic_map_plots}.}
    \label{fig:ongoing_merger_kinematic_maps}
\end{figure*}

The ongoing merger snapshot shown in Figure \ref{fig:ongoing_merger_kinematic_maps} is the most kinematically complex system analysed in this work and is an ideal use case for the {\tt ROHSA-SNAPD} code. The companion galaxy in the top left of the field, Subhalo $34908$, is merging with the main galaxy at the centre of the field, Subhalo $34906$, with a stellar mass ratio of $1/2.37$ and gas mass ratio of $1/8.70$ using all particles of the subhalos, following the methodology of \citet{Rodriguez-Gomez_2015}. This creates highly non-disc-like features in both the stellar mass and \revonechanged{star formation} maps, most notably the long tail of stripped gas.
The intrinsic rotation and dispersion maps also show the effects of merging, where the mostly ordered rotation of the central galaxy is combined with the velocity of the infalling galaxy, and the colliding gas creates a disrupted, non-uniform dispersion map. Merging signatures are also seen in the observed kinematics, particularly where the asymmetric intrinsic rotation causes complex features in the convolved velocity dispersion map. We fit this system with {\tt ROHSA-SNAPD} using $\lambda_f=0.001$ and $\lambda_\mu=3$. The convolved {\tt ROHSA-SNAPD} fits are again very similar to the observed galaxy, with the median absolute percent error below $16$ \revtwochanged{per cent}. Most significantly, the convolved {\tt ROHSA-SNAPD} velocity dispersion map matches the observed accurately, demonstrating that beam smearing can produce highly complex observed dispersion profiles, even when assuming a constant underlying velocity dispersion.

The deconvolved {\tt ROHSA-SNAPD} kinematics match the intrinsic kinematics well, particularly the flux variations and rotation structure that distinguish the central and infalling galaxies much more clearly than in the observed maps. Because the system contains two different galaxies and complex kinematics, a constant dispersion is not a perfect fit, though {\tt ROHSA-SNAPD} does recover the peak of the oversampled resolution dispersion distribution well, as shown in Figure \ref{fig:disp_hist_TNG}. When fitting this system with a non-constant velocity dispersion there was little increase in the accuracy of the fit, suggesting that, given the quality of the data, a constant velocity dispersion is appropriate for this system.

\subsection{Post-merger}
\label{subsec:post_merger_results}
\begin{figure*}
    \centering
    \includegraphics[width=0.49\linewidth,trim={3cm 3cm 2cm 0cm},clip]{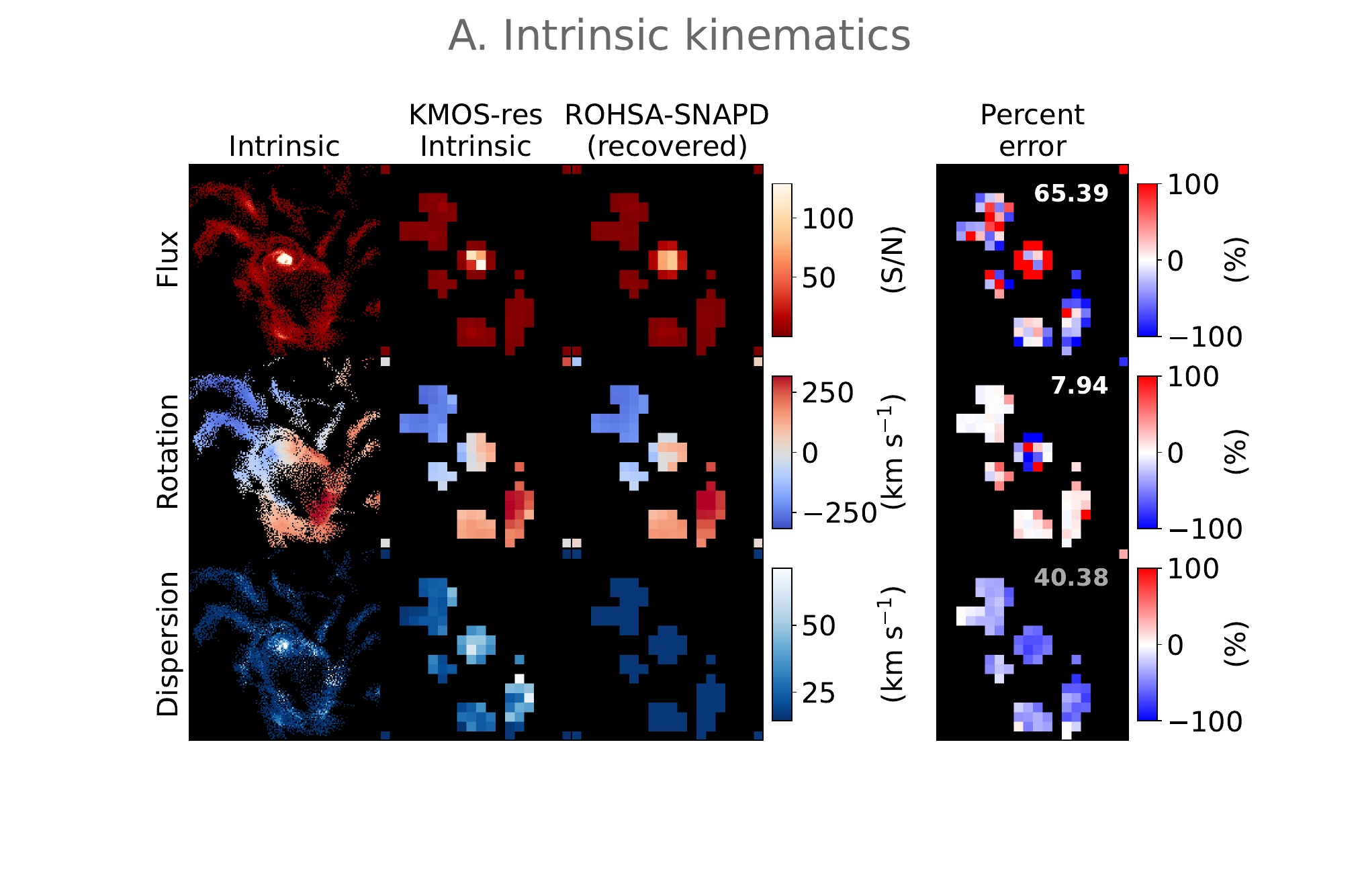}
    \includegraphics[width=0.49\linewidth,trim={3cm 3cm 2cm 0cm},clip]{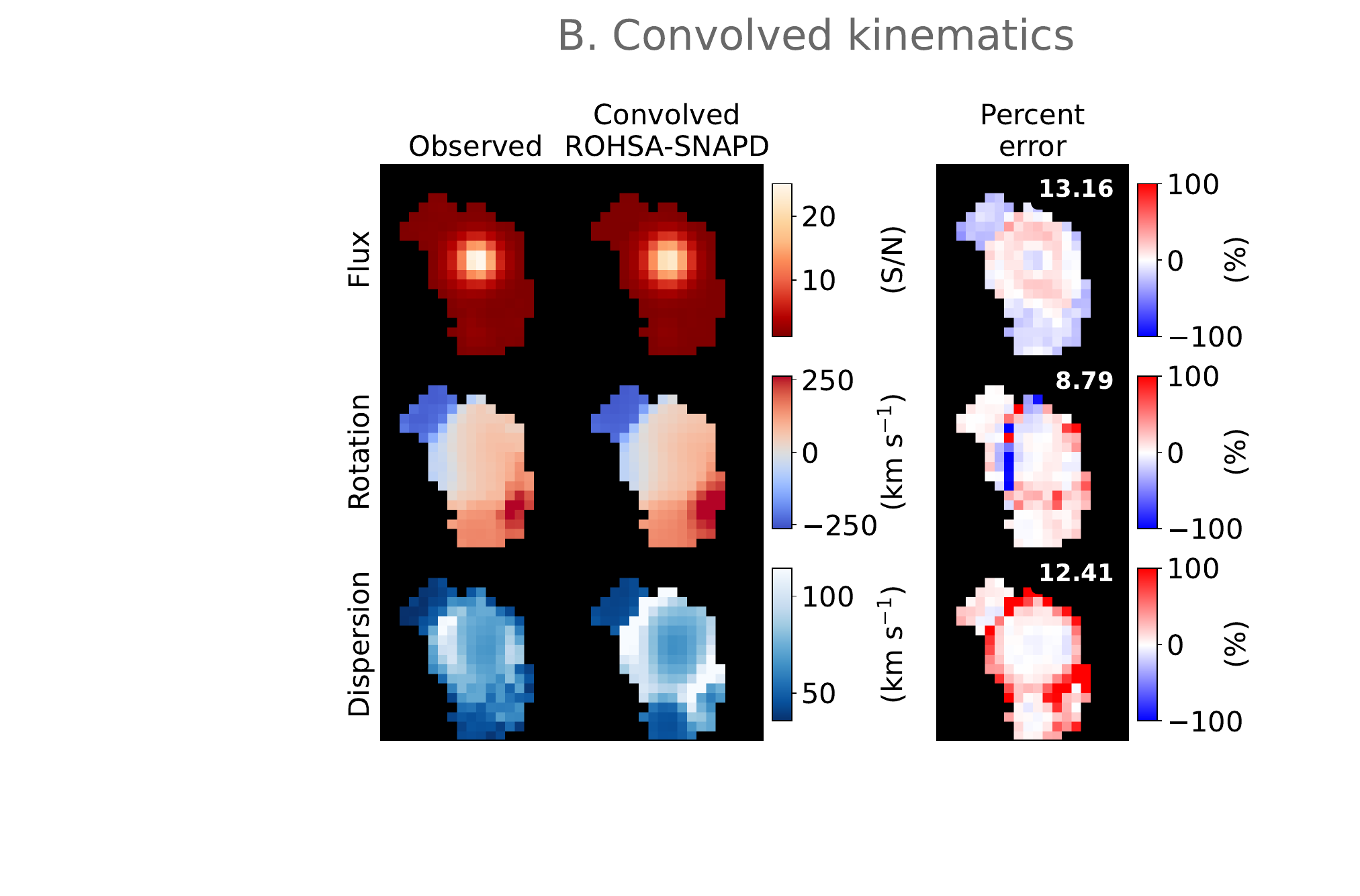}
    \caption{The results of the {\tt ROHSA-SNAPD} fits to the post-merger snapshot using $\lambda_f = 0.001$ and $\lambda_\mu = 2$. The {\tt ROHSA-SNAPD} maps shown are the median fits to the $200$ noise realisations. This figure follows the same layout as Figure \ref{fig:KinMS_kinematic_map_plots}.}
    \label{fig:post_merger_kinematic_maps}
\end{figure*}

The post-merger snapshot in Figure \ref{fig:post_merger_kinematic_maps} shows the galaxy has a compact central flux peak with extended faint emission from outlying gas disrupted by the merging event. The stellar mass maps in Figure \ref{fig:intrinsic_vs_observed_fluxes} are also indicative of a post-merger galaxy, showing shell-like structure to the bottom right of the central galaxy and stellar streams to the upper left. Whilst the rotation and velocity dispersion in the galaxy centre is reasonably uniform, the extended kinematics are more complex. The observed galaxy appears less disrupted because beam smearing smooths over the features and obscures most of the extended structure, although the velocity dispersion map is clearly non-disc-like. This snapshot was fit using using $\lambda_f=0.001$ and $\lambda_\mu=2$. The convolved {\tt ROHSA-SNAPD} maps match the observed kinematics well, with a median absolute error below $14$ \revtwochanged{per cent} for all maps. The complex observed velocity dispersion is again fit effectively, though the percent error does increase near the edges of the galaxy.

The deconvolved {\tt ROHSA-SNAPD} flux recovers most of the intrinsic extended structure, though some is missed close to the galaxy centre, likely because it is difficult to discern low level intrinsic flux from the bright beam-smeared central region. The {\tt ROHSA-SNAPD} rotation map also matches the intrinsic quite well, even in the extended, low flux regions.
The disrupted nature of the post-merger galaxy means there is a significant spread in intrinsic velocity dispersion values, though the deconvolved {\tt ROHSA-SNAPD} value shown in Figure \ref{fig:disp_hist_TNG} does capture the peak of the oversampled resolution distribution.

\revtwochanged{
\subsection{Fitting all systems with the same regularisation values}
\label{subsec:fitting_all_TNG100_with_lam_10}
In the previous sections, we showed the results of fitting the pre-, ongoing and post-merger galaxies with {\tt ROHSA-SNAPD} using $\lambda_\mu=5$, $3$ and $2$ respectively. To further demonstrate the impact of regularisation on the recovered velocity dispersion values, we show the results of using $\lambda_\mu=10$ for all galaxies, informed from the most appropriate idealised disc regularisation. In general, the $\lambda_\mu=10$ convolved models are very similar to the models in Sections \ref{subsec:pre_merger_results}, \ref{subsec:ongoing_merger_results} and \ref{subsec:post_merger_results}, as shown in Appendix \ref{appendix:all_same_regularisation_values}. However, Figure \ref{fig:disp_hist_TNG100_lam_10} shows that the deconvolved velocity dispersion values are significantly higher when using $\lambda_\mu=10$ and less accurately capture the intrinsic velocity dispersions, particularly for the post-merger galaxy.
These results highlight both the importance of choosing appropriate regularisation values, but also the difficulty of accurately recovering deconvolved velocity dispersions given the data quality, as despite significantly increased deconvolved dispersions, there are only minimal changes in the convolved models.

\begin{figure*}
    \centering
    \includegraphics[width=0.33\linewidth]{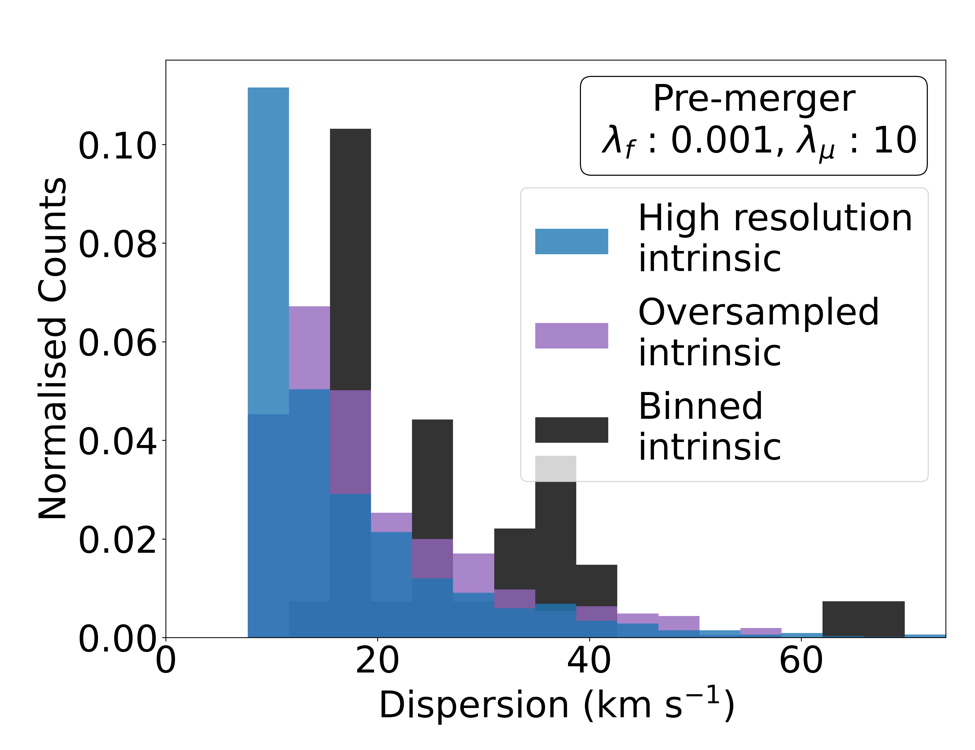}
    \includegraphics[width=0.33\linewidth]{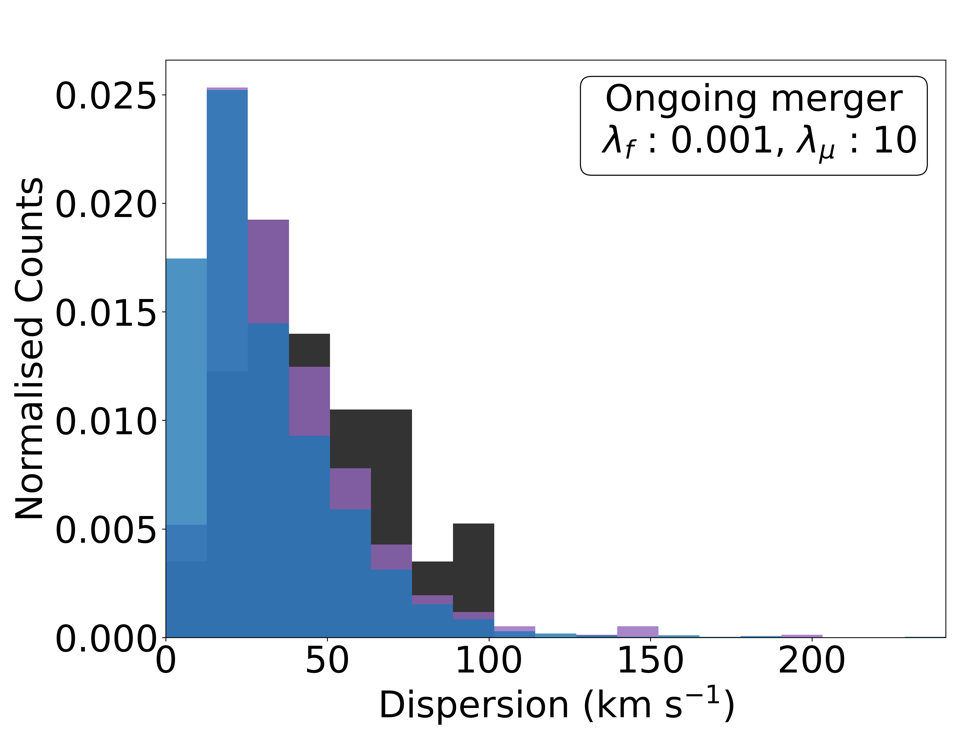}
    \includegraphics[width=0.33\linewidth]{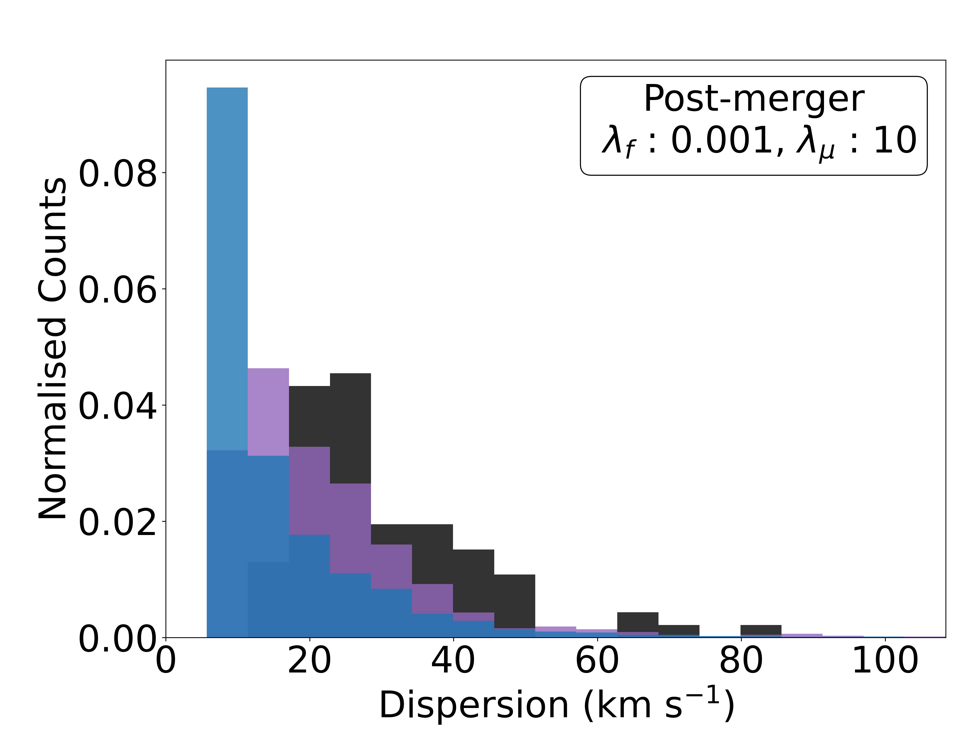}
    \includegraphics[width=0.33\linewidth]{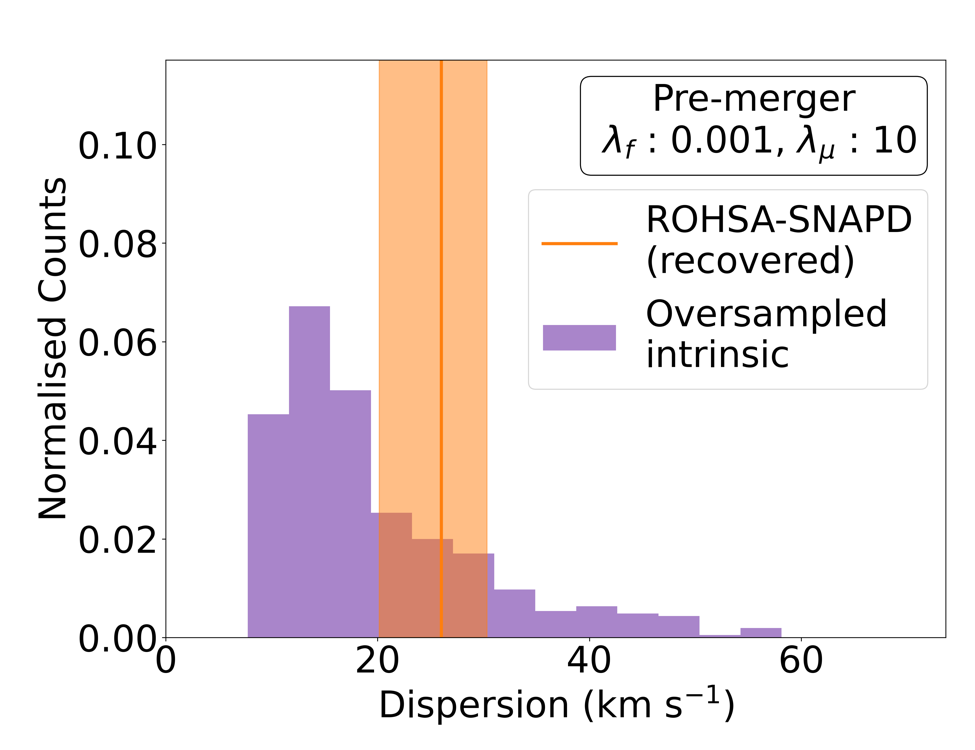}
    \includegraphics[width=0.33\linewidth]{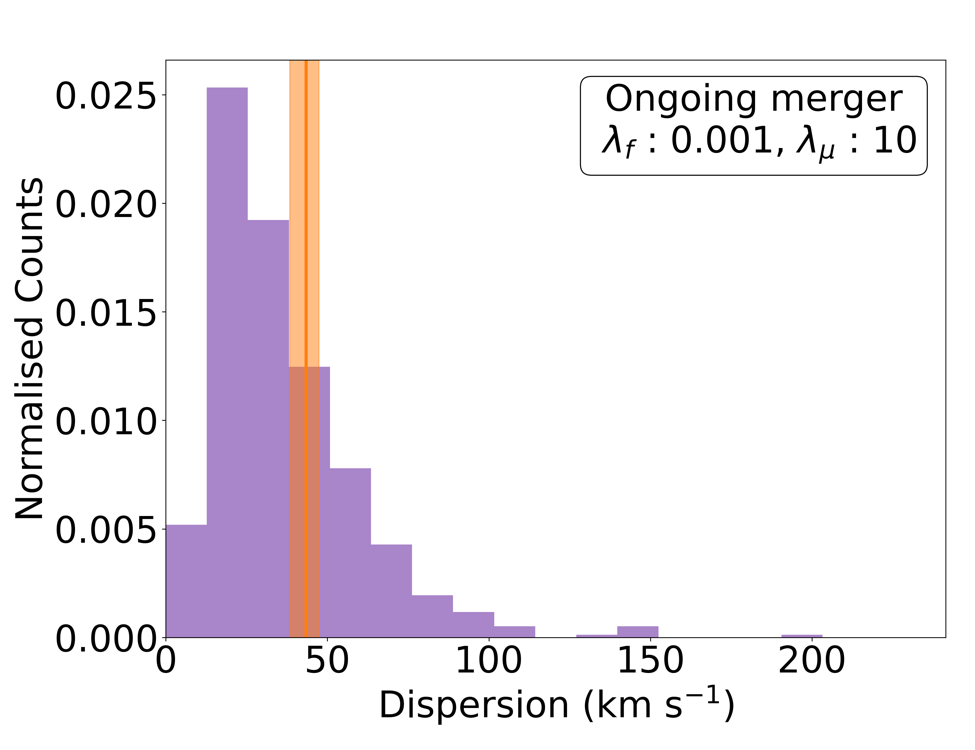}
    \includegraphics[width=0.33\linewidth]{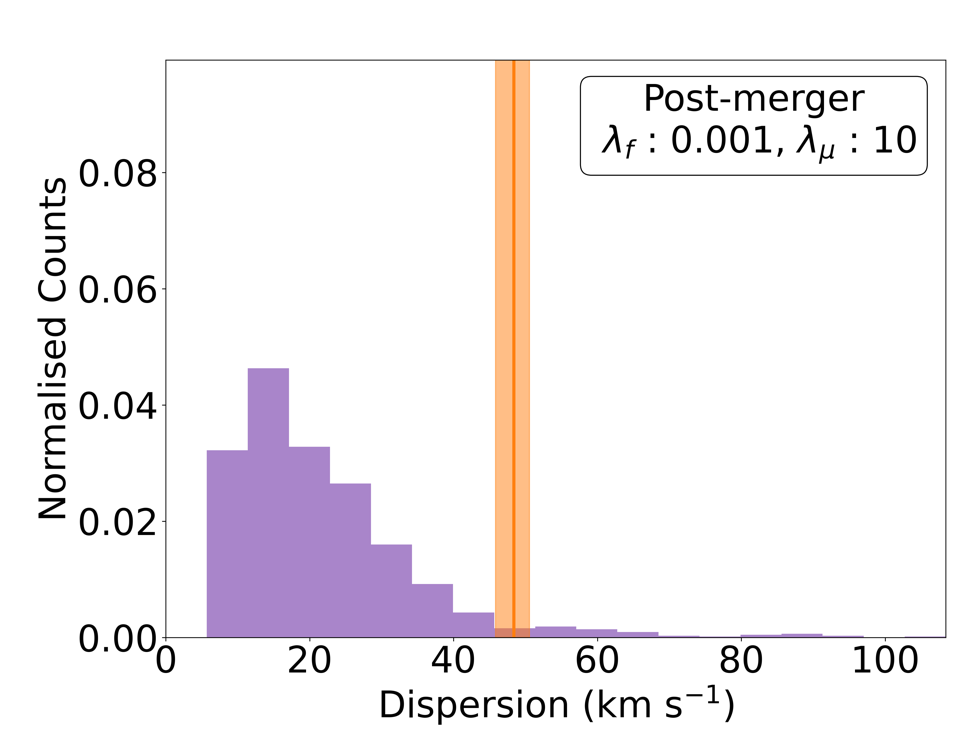}
    \caption{ This figure follows the same layout as Figure \ref{fig:KinMS_kinematic_map_plots} and compares the deconvolved {\tt ROHSA-SNAPD} velocity dispersion values (orange) with a histogram of the intrinsic velocity dispersion of the respective galaxy at high-resolution ($n=8$) (blue), at oversampled resolution ($n=4$) (purple) and at observed resolution ($n=1$) (black). For a clear comparison, the area under each distribution is normalised to $1$. From the $200$ noise realisations, the median deconvolved {\tt ROHSA-SNAPD} velocity dispersion is shown as a solid orange line, and the $16$th and $84$th percentiles are shaded. The \revonechanged{x-axis} of each panel \revonechanged{spans} from $0$ to the maximum value of any resolution.}
    \label{fig:disp_hist_TNG100_lam_10}
\end{figure*}

}

\section{Conclusions}
We have introduced a Python-based emission-line fitting code {\tt ROHSA-SNAPD}, "Spatially Non-parametric Approach to PSF Deconvolution using ROHSA", which can obtain deconvolved kinematics from 3D IFS data whilst accounting for the effects of beam smearing and instrument resolution.
\revtwochanged{A number of 3D kinematic modelling codes exist which are optimised for disc-like galaxies and can fit to clumpy flux maps or varying rotation profiles. However, {\tt ROHSA-SNAPD} has been specifically designed to be applied to kinematically complex systems such as disrupted and merging galaxies. The key feature of {\tt ROHSA-SNAPD} is that it enforces spatial correlation when forward modelling to obtain deconvolved kinematic maps in a spatially non-parametric way, without needing to assume an underlying galaxy rotation model.}
We highlight the following results:

\begin{itemize}[leftmargin=*]
    \setlength\itemsep{1em}

    \item We have evaluated {\tt ROHSA-SNAPD} on four $1.82 \leq \textnormal{z} \leq 2.10$ KMOS-like mock observations. First, an idealised disc galaxy to demonstrate the effect of regularisation and then three systems from a cosmological simulation with varying levels of kinematic disturbance, from pre-merger to post-merger state.
    
    \item In general, the convolved {\tt ROHSA-SNAPD} models matched the observed systems very closely ($<14$ \revtwochanged{per cent} error for flux and dispersion, $<17$ \revtwochanged{per cent} error for velocity) and demonstrated that observational effects can cause complex velocity dispersion profiles, even when assuming a constant underlying dispersion map.
    
    \item The deconvolved {\tt ROHSA-SNAPD} kinematic maps generally agreed well with the often complex intrinsic kinematics of the mock observations. Some small-scale structures were unable to be recovered, most likely due to the kernel convolutions setting a minimum spatial scale that the code could deconvolve.
\end{itemize}
\revonechanged{{\tt ROHSA-SNAPD} is publicly released on Github \githubcodelink.} In future works, we will apply {\tt ROHSA-SNAPD} to large statistical galaxy samples. Methods to determine an appropriate level of regularisation will also be explored, as well as the recovery of spatially varying velocity dispersions.

\section*{Acknowledgements}
We thank the anonymous referees for their constructive comments that have improved this work.
We also appreciate enlightening conversations with François Orieux.

\revtwochanged{\textit{Conflict of interest and funding:}
The authors declare no conflict of interest.}
IK acknowledges that this research was supported by an Australian Government Research Training Program (RTP) Scholarship.
JTM acknowledges support from the Australian Research Council (ARC) Discovery Project DP210101945.
Part of this research was conducted by the Australian Research Council Centre of Excellence for All Sky Astrophysics in 3 Dimensions (ASTRO 3D), through project number CE170100013.

\textit{Software used in this work:} JAX \citep{Frostig_2018,jax2018github}, Scipy \citep{2020SciPy-NMeth}, Numpy \citep{harris2020array_numpy}, Astropy \citep{astropy:2013,astropy:2018,astropy:2022}, Matplotlib \citep{Hunter:2007_matplotlib}, Uncertainties: a Python package for calculations with uncertainties, Eric O. LEBIGOT.

\section*{Data Availability}

\revonechanged{The {\tt ROHSA-SNAPD} code introduced in this work is publicly released on Github \githubcodelink.}
The {\tt IllustrisTNG} cosmological simulation data used within this work is available from the {\tt IllustrisTNG} website \url{https://www.tng-project.org/}. Idealised disc models were generated using the {\tt KINematic Molecular Simulation} code \citep{Davis_2013}. Mock observations derived from the simulated data are available on request.



\bibliographystyle{mnras}
\bibliography{papers} 




\appendix

\revonechanged{\section{Computation time and memory usage}
\label{appendix:run_info}
There are a number of factors which impact the computation time of {\tt ROHSA-SNAPD}, predominantly the spatial and spectral pixel size of the input data, the oversample factor $n$ used within the cost function, the choice of regularisation and the S/N of the data.
To demonstrate the effect of changing these factors, we created a series of mock observations of the idealised disc using the same underlying high resolution intrinsic model as in Section \ref{sec:evaluation_idealised_disc_galaxy}.
{\tt ROHSA-SNAPD} was applied to each observation using $16$ CPU cores \footnote{The specific processors used were two Intel E5-2680v4 CPUs.} and the same general parameters as in Section \ref{subsec:idealised_disc_applying_RSNAPD}. In all cases we optimised a single dispersion value. Apart from Appendix \ref{appendixsubsec:decreasing_data_SN_and_regularisation} in which fits were run to convergence, all other fits were terminated after $1000$ iterations to compare changes in computation time. After each run, we saved the wall time (the total elapsed time of the run), the total CPU usage (the number of CPU cores used in the run), the total CPU time (the summed wall time of each individual CPU core) and the maximum memory used in the run.

In Appendices \ref{appendixsubsec:increasing_data_spatial_res}, \ref{appendixsubsec:increasing_data_spec_res}, \ref{appendixsubsec:increasing_oversamp_fact} and \ref{appendixsubsec:decreasing_data_SN_and_regularisation} we show the median results from fitting $20$ noise realisations of each mock observation.

\subsection{Data spatial sampling}
\label{appendixsubsec:increasing_data_spatial_res}
If the number of spatial pixels in a data cube increase, so do the size of the resulting kinematic maps and therefore the number of parameters which must be optimised in the model.
To demonstrate the change in computation time with increased spatial size, we created simulated data at $3$ spatial pixel resolutions, $0\textnormal{"}.2$, $0\textnormal{"}.1$ and $0\textnormal{"}.05$. The noise added to the mock data was decreased to ensure the S/N within $1\textnormal{"}.5$ was kept constant for each resolution. The normalised Moffat PSF model FWHM was also decreased relative to the spatial pixel size to retain approximately Nyquist sampling of the PSF. To ensure consistency between resolutions, the oversampled PSF kernel size was kept at $(85 \times 85)$, the same as in Section \ref{sec:evaluation_idealised_disc_galaxy}. Further, the oversample factor was decreased with increasing spatial resolution, such that the oversampled model resolution was the same for the whole sample. This meant that the changes in computation time were only due to the increased number of parameters being optimised (the  $R\left(\thetab \right)$ convolution within the cost function is also impacted by spatial resolution, though it has a minimal impact on the total run time as it only involves $2$ convolutions per cost function evaluation).
The resulting CPU time, wall time and maximum memory usage are shown in Figure \ref{fig:spat_samp_tests}. In general, the total CPU time and CPU usage increased with the number of model parameters, particularly for the $0\textnormal{"}.05$ pixel size case with around $10 \, 000$ parameters. The total wall time increased less significantly, due to the parallelisation of the code. As the number of model parameters increased by a factor of $16.0$, the total CPU time, CPU usage and wall time increased by factors of $3.2$, $2.9$ and $1.1$ respectively. The maximum memory usage was only minimally impacted.

\begin{figure*}
    \centering
    \includegraphics[width=0.33\linewidth]{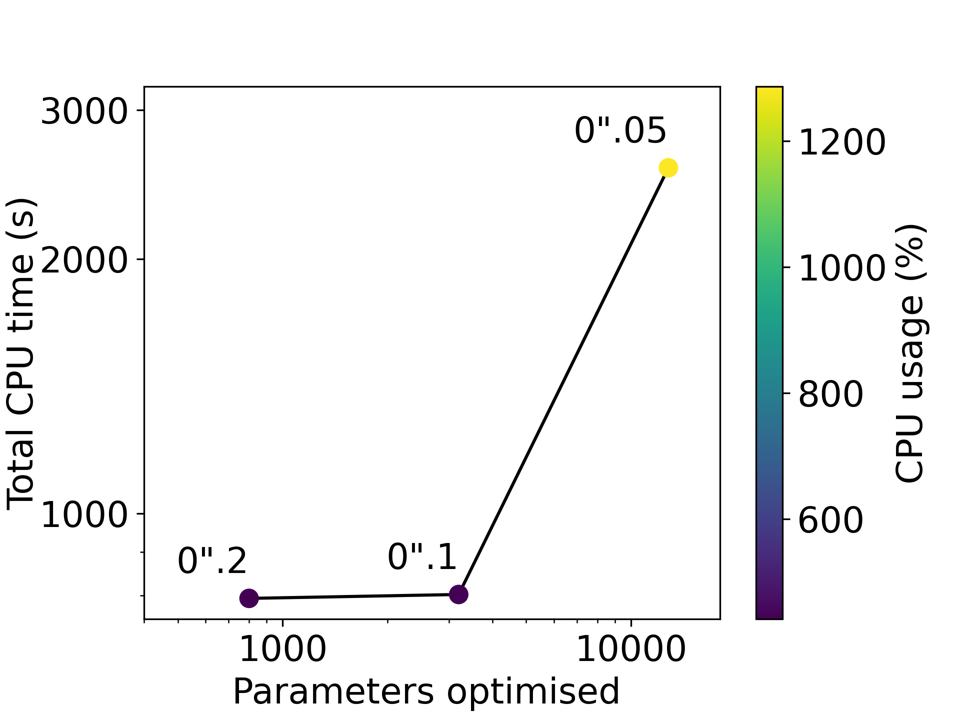}
    \includegraphics[width=0.33\linewidth]{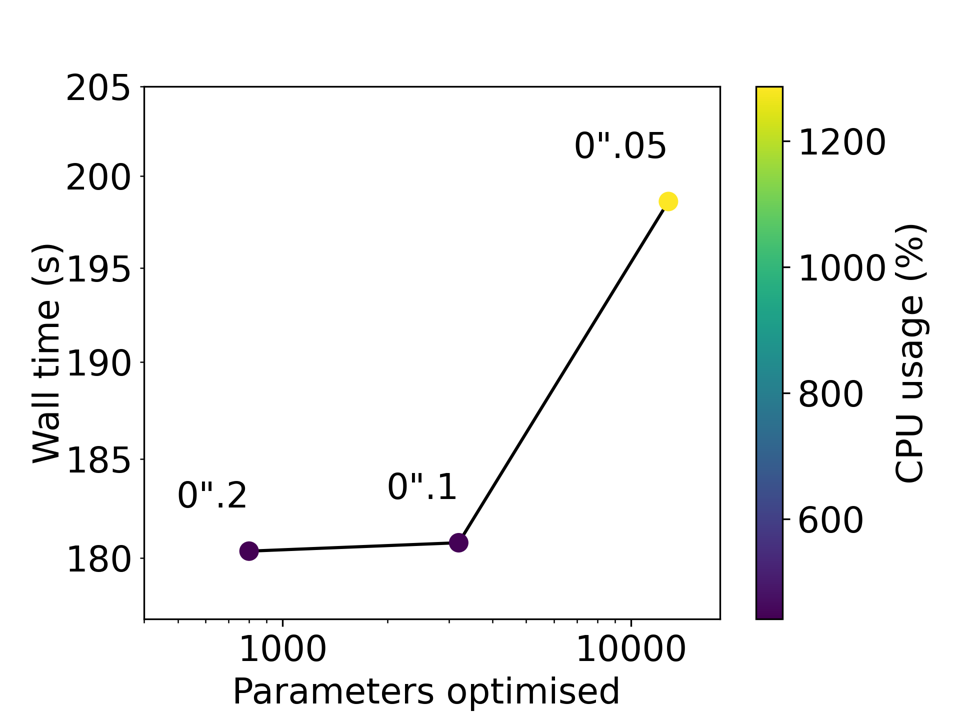}
    \includegraphics[width=0.33\linewidth]{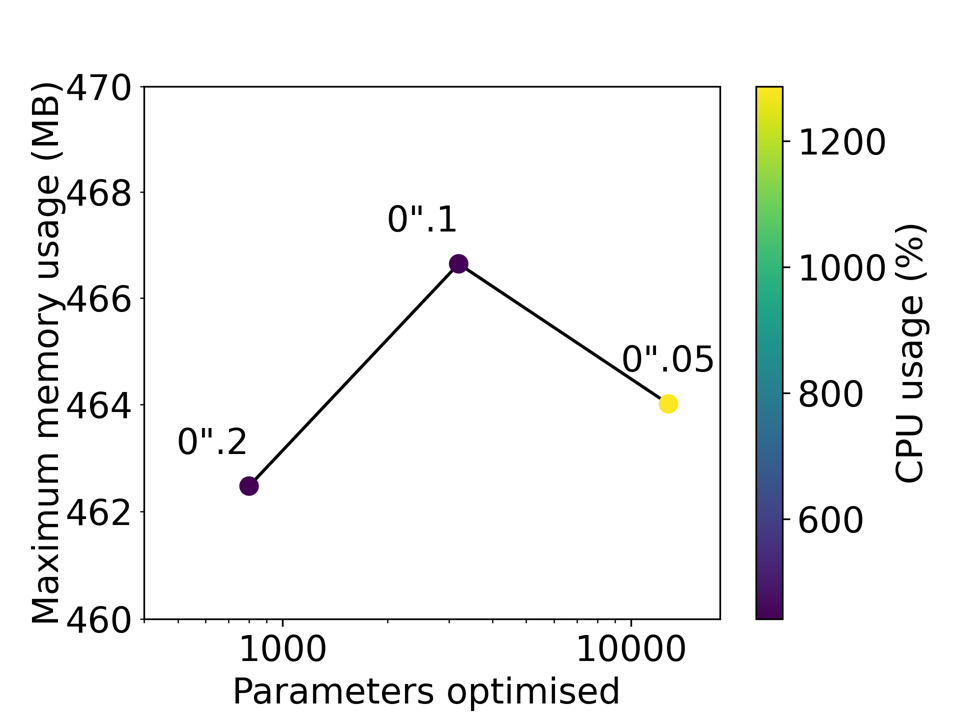}
    \caption{From left to right, the three panels describe the total CPU time, wall time and maximum memory usage when increasing the spatial sampling of the idealised disc galaxy introduced in Section \ref{sec:evaluation_idealised_disc_galaxy}. The spatial pixel sampling is included above each marker.}
    \label{fig:spat_samp_tests}
\end{figure*}

\subsection{Data spectral sampling}
\label{appendixsubsec:increasing_data_spec_res}
Within the cost function PSF convolutions are computed for each spectral channel of the data, therefore, if the number of spectral channels are increased, the number of convolutions required to calculate $L \left(\thetab \right)$ also increase.
To demonstrate the effect of spectral size on the computation time, we created simulated data cubes with $3$ different spectral samplings, $40\textnormal{ km s}^{-1}$, $20\textnormal{ km s}^{-1}$ and $10\textnormal{ km s}^{-1}$. Again, the noise added to the mock data was decreased to ensure the S/N within $1\textnormal{"}.5$ was kept constant for each resolution. As the spatial resolution was constant for all data cubes, the PSF and oversample factors were not changed.
The resulting computation time and maximum memory usage are shown in Figure \ref{fig:spec_samp_tests}. We note that for the simulated data with spectral sampling of $10$ km s$^{-1}$ and $20$ km s$^{-1}$, the fits did not take $1000$ iterations to converge (they took a median of $740.5$ and $615.5$ respectively). Therefore, the total CPU and wall times for these fits were scaled up to show the times for $1000$ iterations. As the number of spectral channels increased by a factor of $4$, the total CPU time, CPU usage and wall time increased by factors of $2.8$, $1.7$ and $1.6$ respectively. The maximum memory usage increased by a factor of $1.5$.

\begin{figure*}
    \centering
    \includegraphics[width=0.33\linewidth]{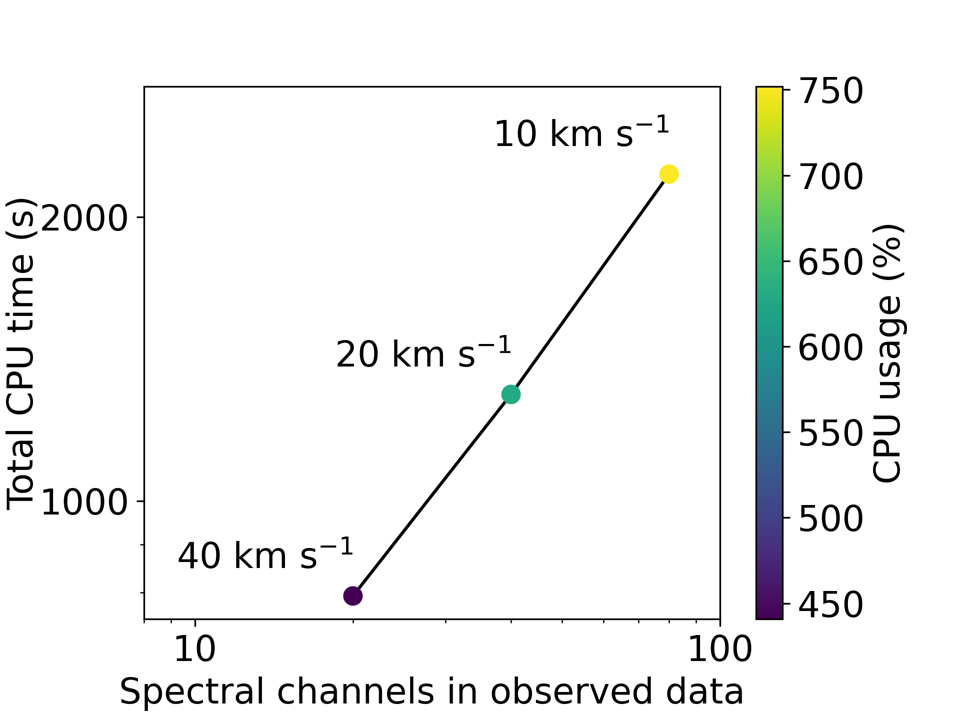}
    \includegraphics[width=0.33\linewidth]{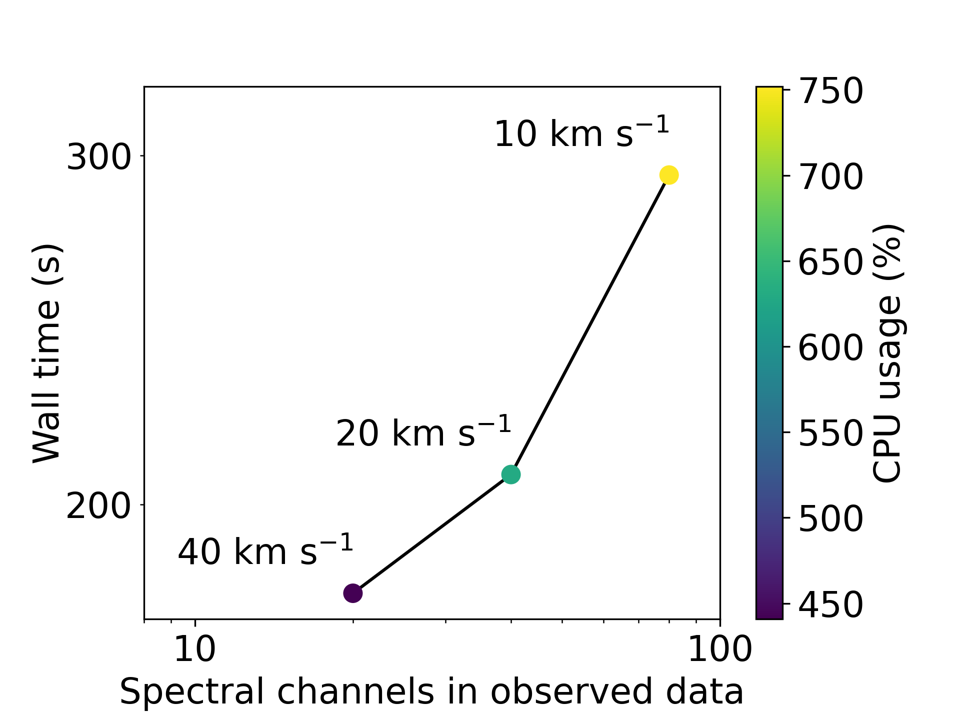}
    \includegraphics[width=0.33\linewidth]{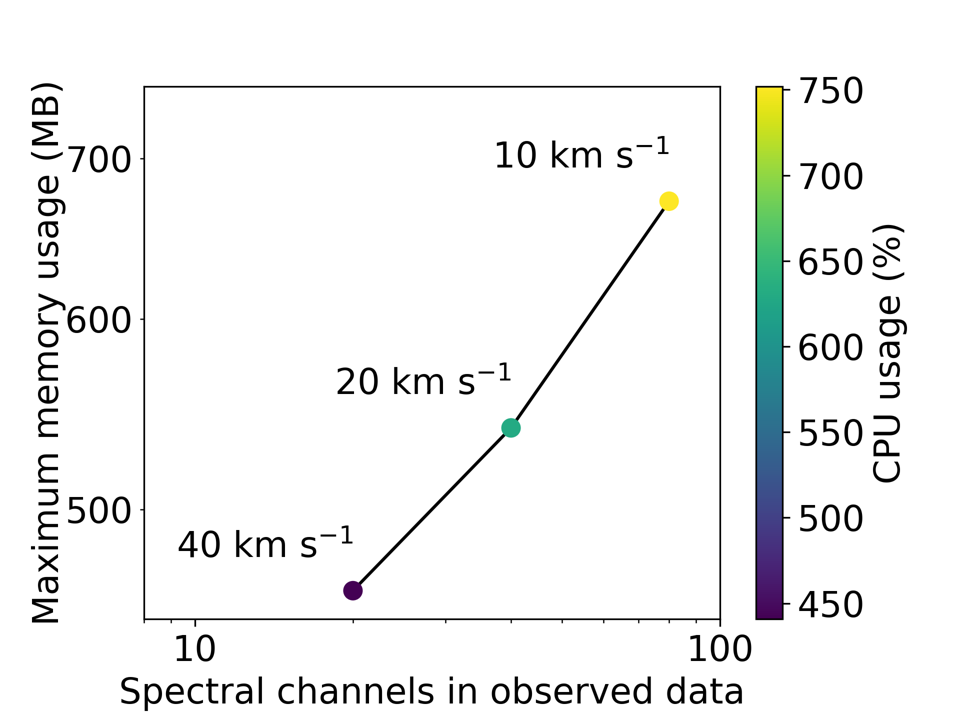}
    \caption{From left to right, the three panels describe the total CPU time, wall time and maximum memory usage when increasing the spectral sampling of the idealised disc galaxy introduced in Section \ref{sec:evaluation_idealised_disc_galaxy}. The spectral sampling is included above each marker.}
    \label{fig:spec_samp_tests}
\end{figure*}

\subsection{Oversample factor}
\label{appendixsubsec:increasing_oversamp_fact}
Larger oversample factors increase both the number of spatial pixels in the oversampled model and the PSF image resolution, therefore increasing the complexity of the convolutions required to calculate $L \left(\thetab \right)$. To demonstrate the impact on computation time, we fit an observed data cube using oversample factors of $n = 1$, $2$, $3$ and $4$. $n = 1$ means no oversampling is applied.
We show the computation time and maximum memory usage when applying {\tt ROHSA-SNAPD} assuming these oversample factors in Figure \ref{fig:oversample_tests}. As the number of spatial pixels in the oversampled model increased by a factor of $64$, the total CPU time, CPU usage and wall time increased by factors of $20.0$, $6.6$ and $3.0$ respectively. The maximum memory usage increased by a factor of $1.8$.

\begin{figure*}
    \centering
    \includegraphics[width=0.33\linewidth]{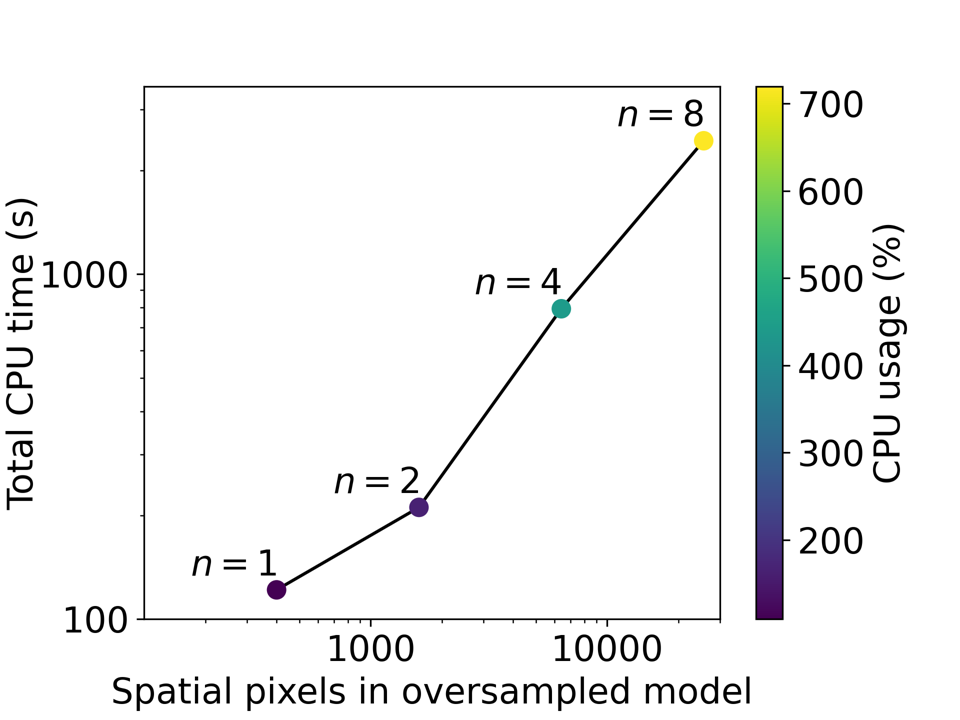}
    \includegraphics[width=0.33\linewidth]{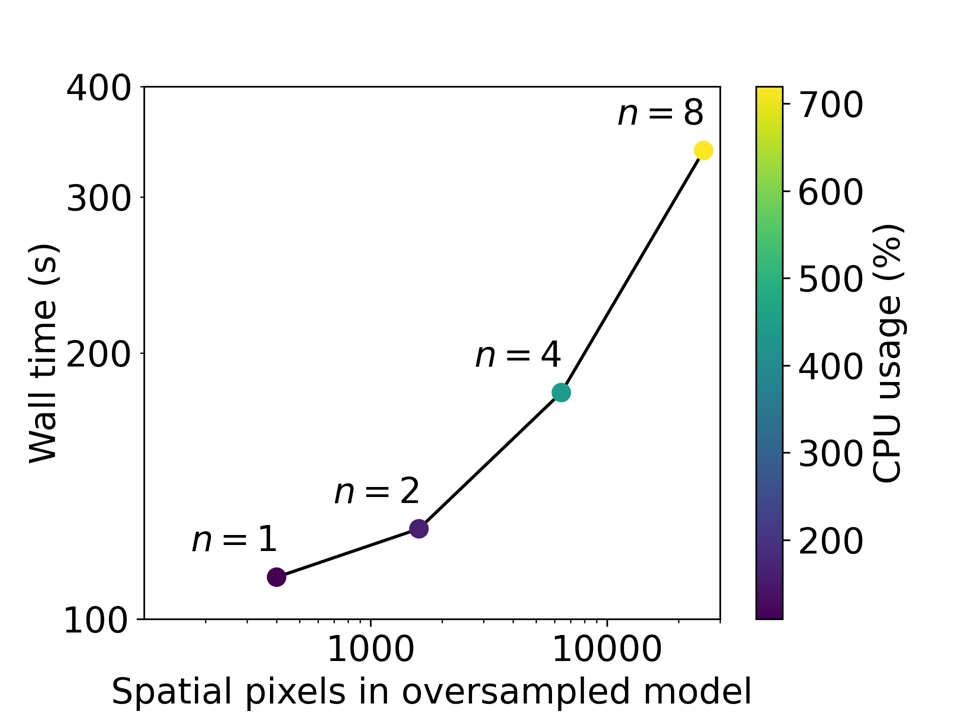}
    \includegraphics[width=0.33\linewidth]{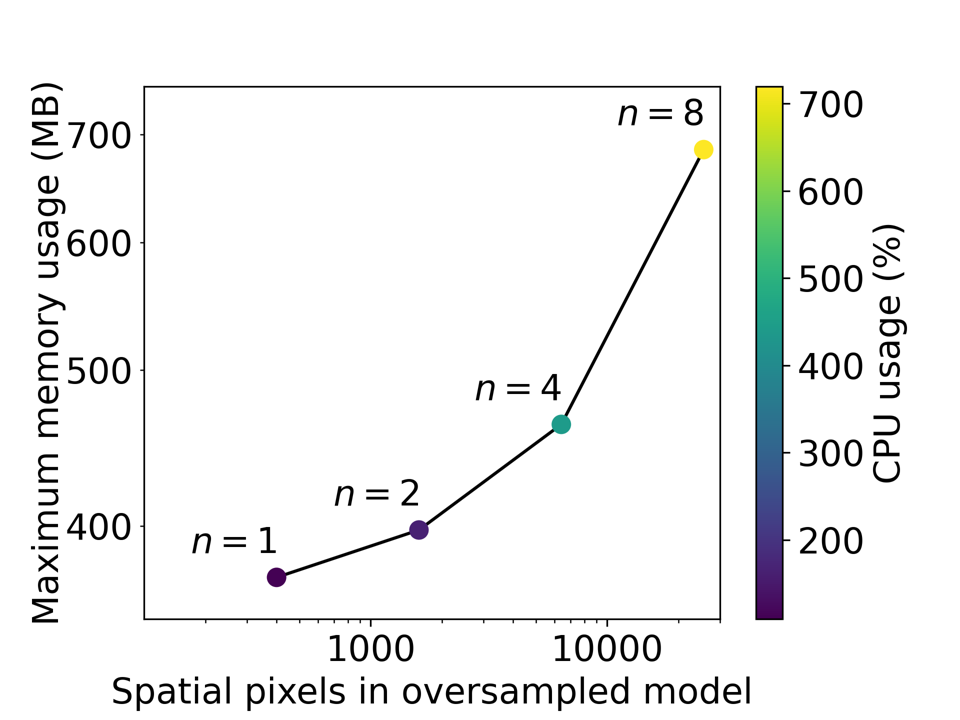}
    \caption{From left to right, the three panels describe the total CPU time, wall time and maximum memory usage when increasing the oversample factor used to fit the idealised disc galaxy introduced in Section \ref{sec:evaluation_idealised_disc_galaxy}. The oversample factor $n$ is included above each marker.}
    \label{fig:oversample_tests}
\end{figure*}

\subsection{Data S/N and regularisation hyper-parameters}
\label{appendixsubsec:decreasing_data_SN_and_regularisation}
Data S/N and the choice of regularisation hyper-parameters put constraints on the output model, therefore changing the number of iterations required to fit to data to the required precision. To demonstrate this effect, we created simulated data at four S/N levels (S/N $= 10, 20, 30$ and $40$ within a $1\textnormal{"}.5$ aperture) and fit them using a range of flux and velocity regularisation values. The high resolution intrinsic data for each galaxy was the same as in Section \ref{sec:evaluation_idealised_disc_galaxy}, only the RMS noise level differed.
Comparing the results in Figure \ref{fig:SN_regularisation_change_plot}, flux regularisation $\lambda_f$, shown by the coloured lines, had the largest impact on the number of iterations required to converge. This is because, as $\lambda_f$ was increased, the model became more constrained and therefore converged in less iterations. S/N, shown as solid through to dotted lines, also had an impact, where higher S/N observations generally took more iterations to converge. This is likely because the solution required further refining to fit within the observed error.

\begin{figure}
    \centering
    \includegraphics[width=\linewidth]{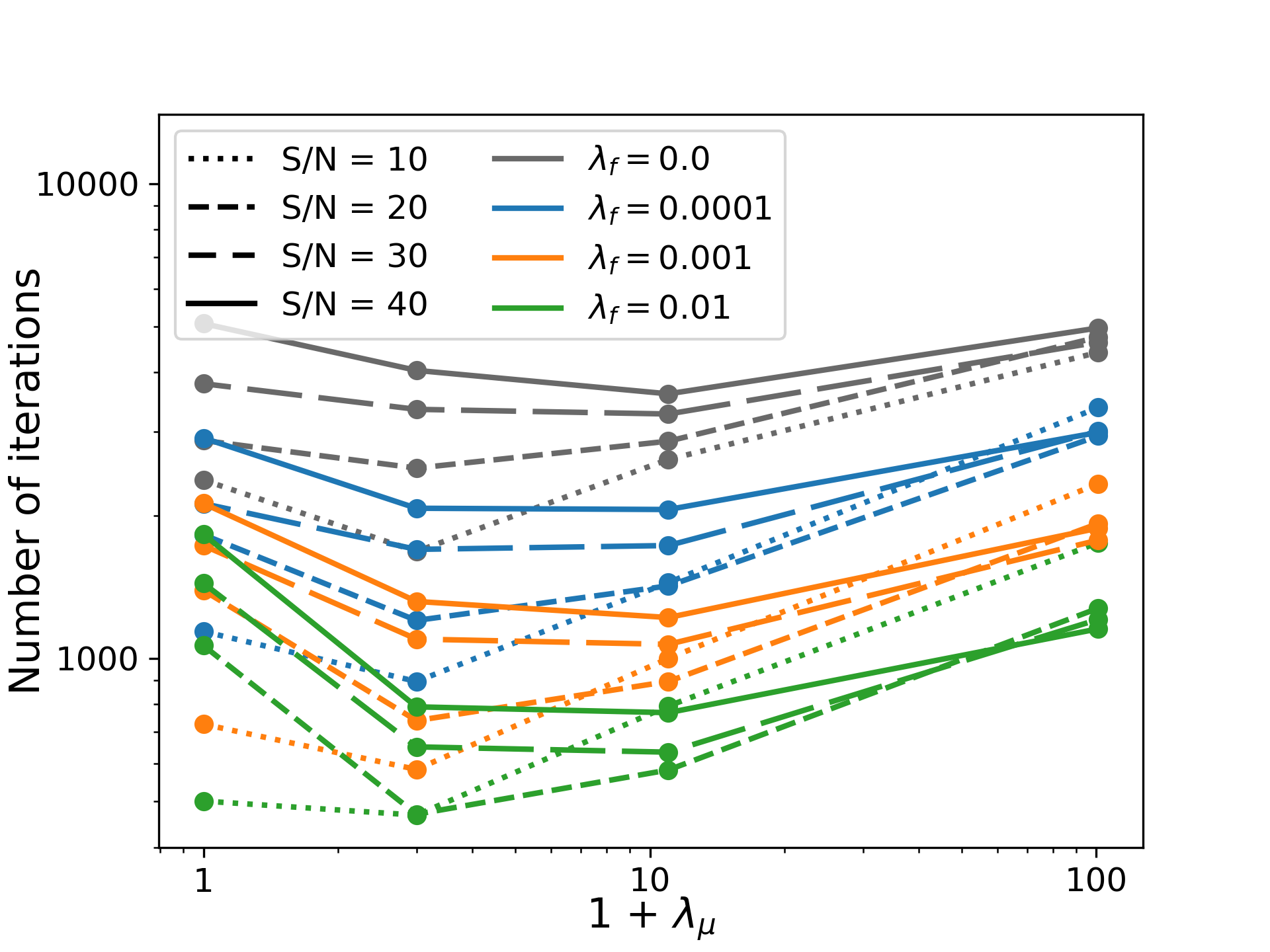}
    \caption{The number of iterations required for {\tt ROHSA-SNAPD} to converge for a range of S/N and regularisation values. The S/N level is shown by line type (solid through to dotted) and each flux regularisation value $\lambda_f$ has a different line colour.}
    \label{fig:SN_regularisation_change_plot}
\end{figure}

}

\section{Fitting all TNG100 galaxies with the same regularisation values}
\label{appendix:all_same_regularisation_values}
In Section \ref{subsec:fitting_all_TNG100_with_lam_10}, we demonstrated the impact of using a velocity regularisation of $\lambda_\mu=10$ for all galaxies and showed the impact on deconvolved velocity dispersion. Here we include additional details about the $\lambda_\mu=10$ fits.

\begin{figure*}
    \centering
    \includegraphics[width=0.49\linewidth,trim={3cm 3cm 2cm 0cm},clip]{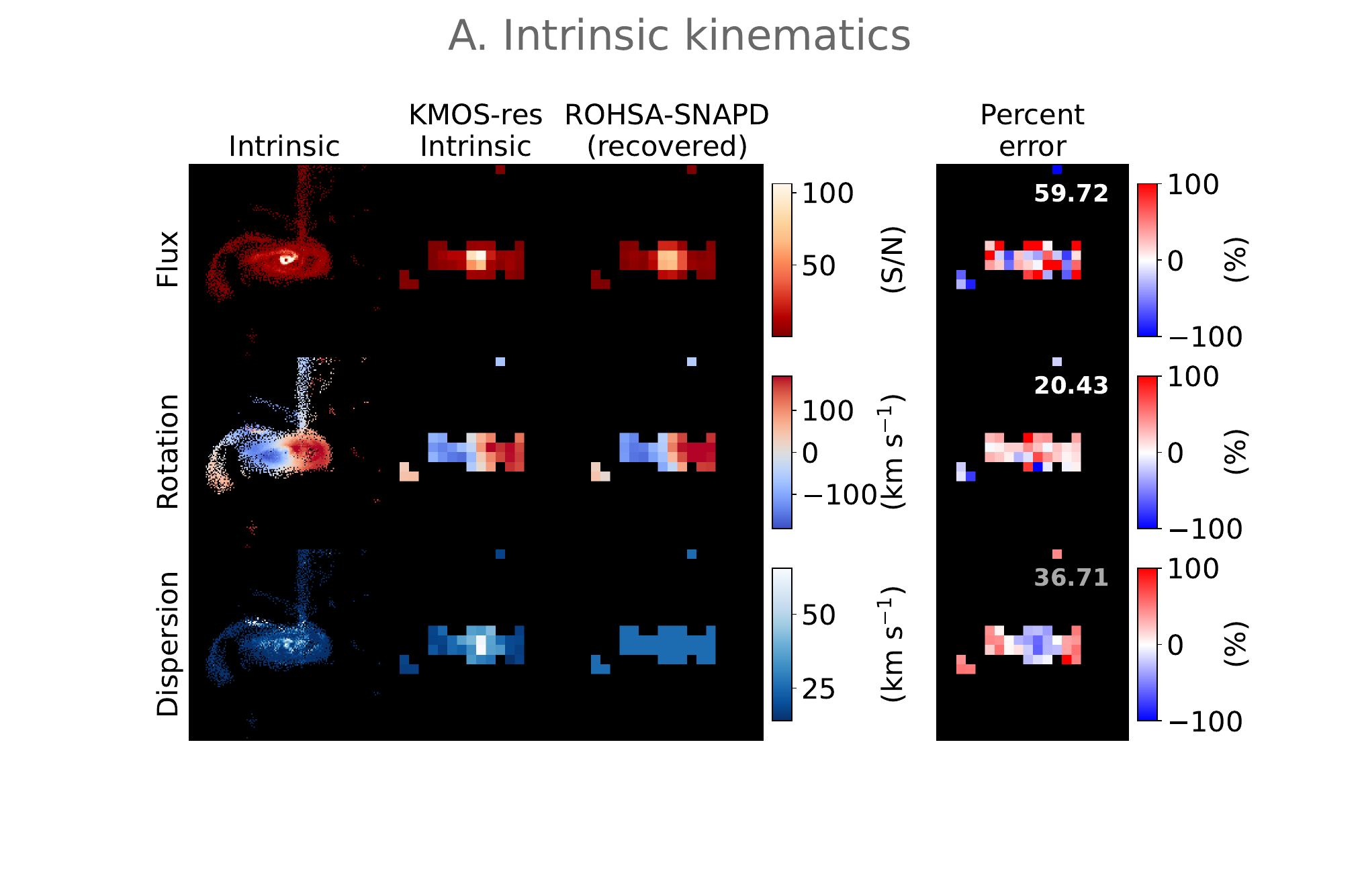}
    \includegraphics[width=0.49\linewidth,trim={3cm 3cm 2cm 0cm},clip]{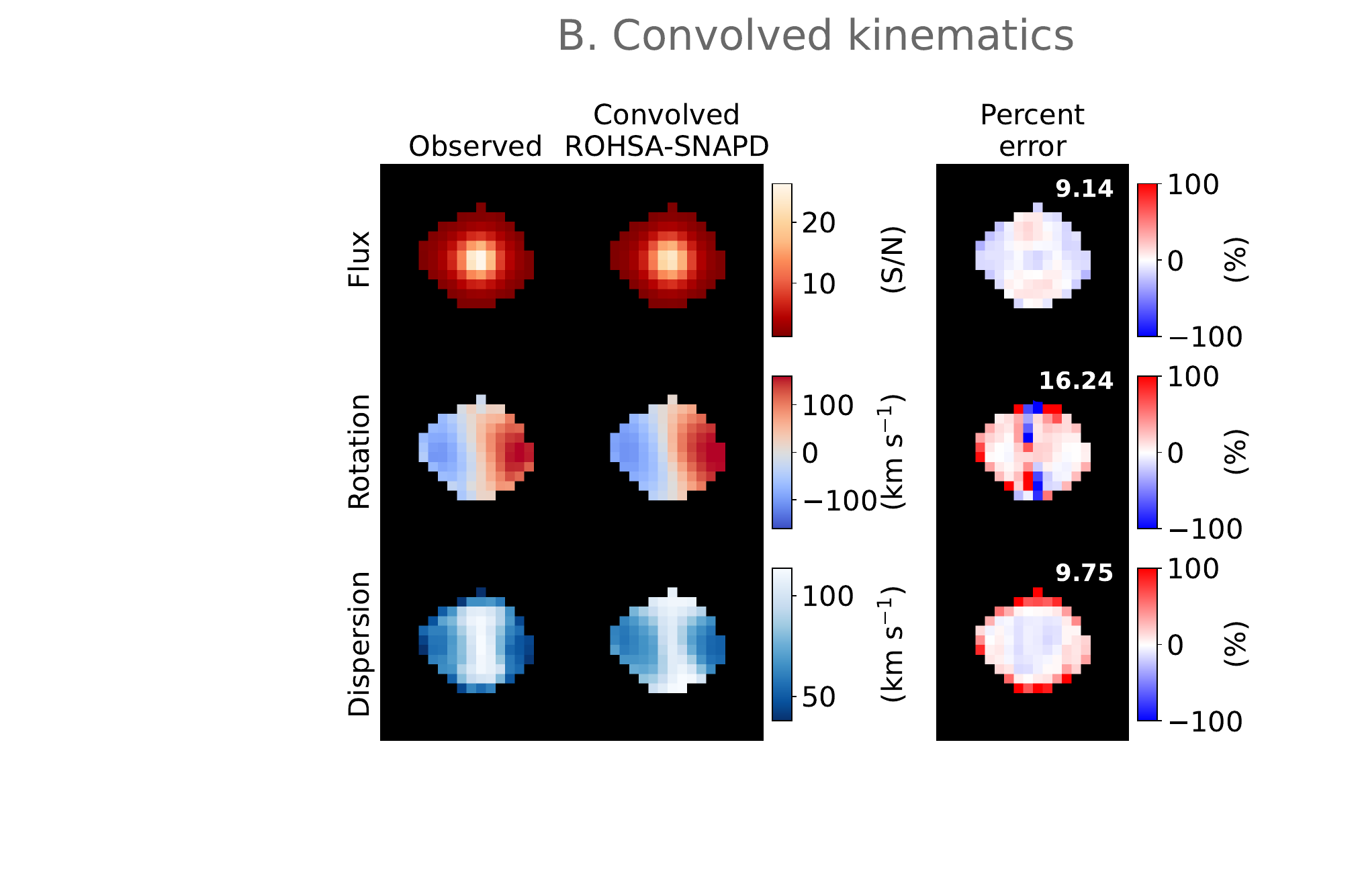}
    \caption{The results of the {\tt ROHSA-SNAPD} fits to the pre-merger snapshot using $\lambda_f = 0.001$ and $\lambda_\mu = 10$. The {\tt ROHSA-SNAPD} maps shown are the median fits to the $200$ noise realisations. This figure follows the same layout as Figure \ref{fig:KinMS_kinematic_map_plots}.}
    \label{fig:pre_merger_kinematic_maps_lambda_mu_10_appendix}
\end{figure*}

\begin{figure*}
    \centering
    \includegraphics[width=0.49\linewidth,trim={3cm 3cm 2cm 0cm},clip]{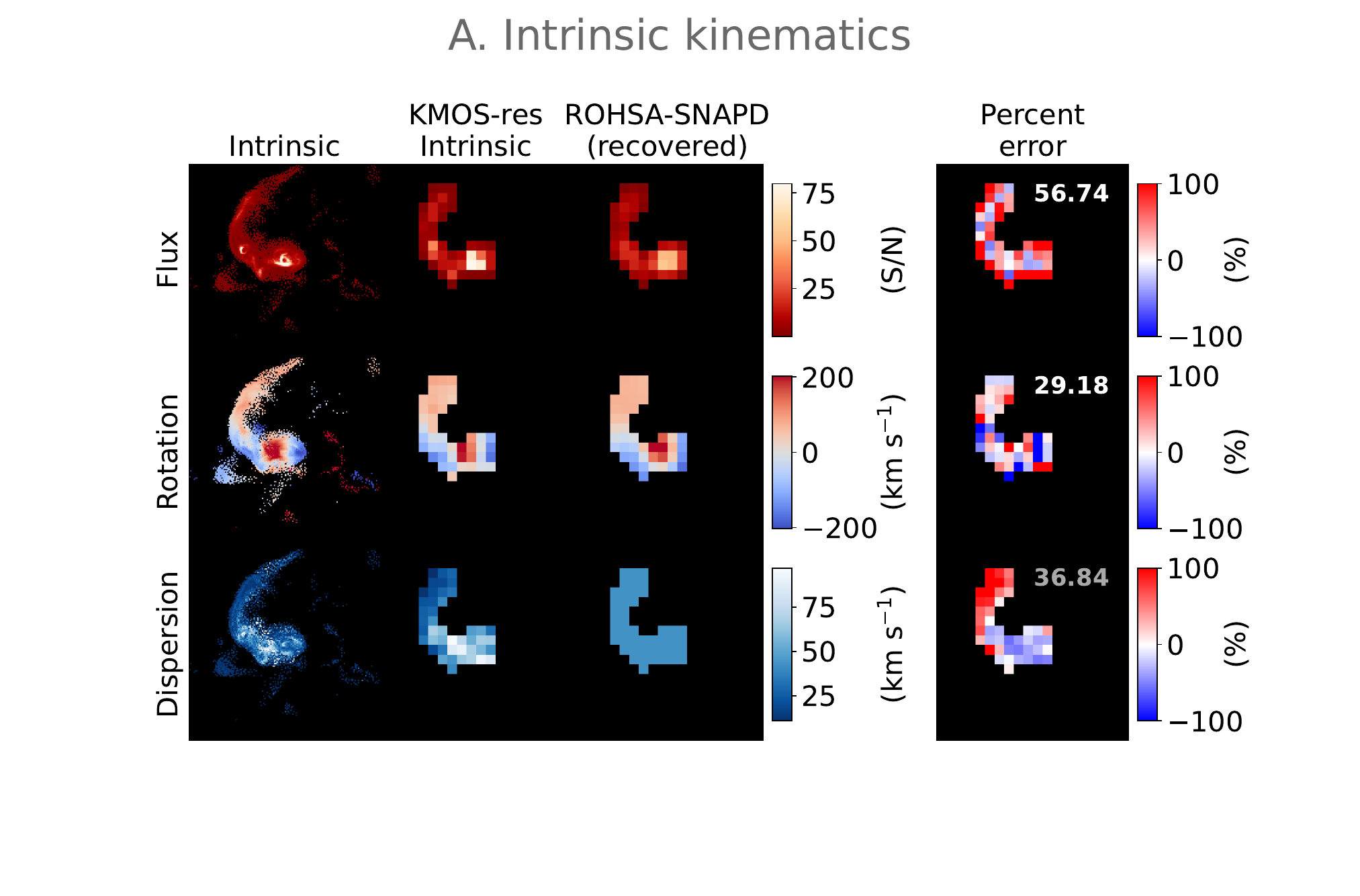}
    \includegraphics[width=0.49\linewidth,trim={3cm 3cm 2cm 0cm},clip]{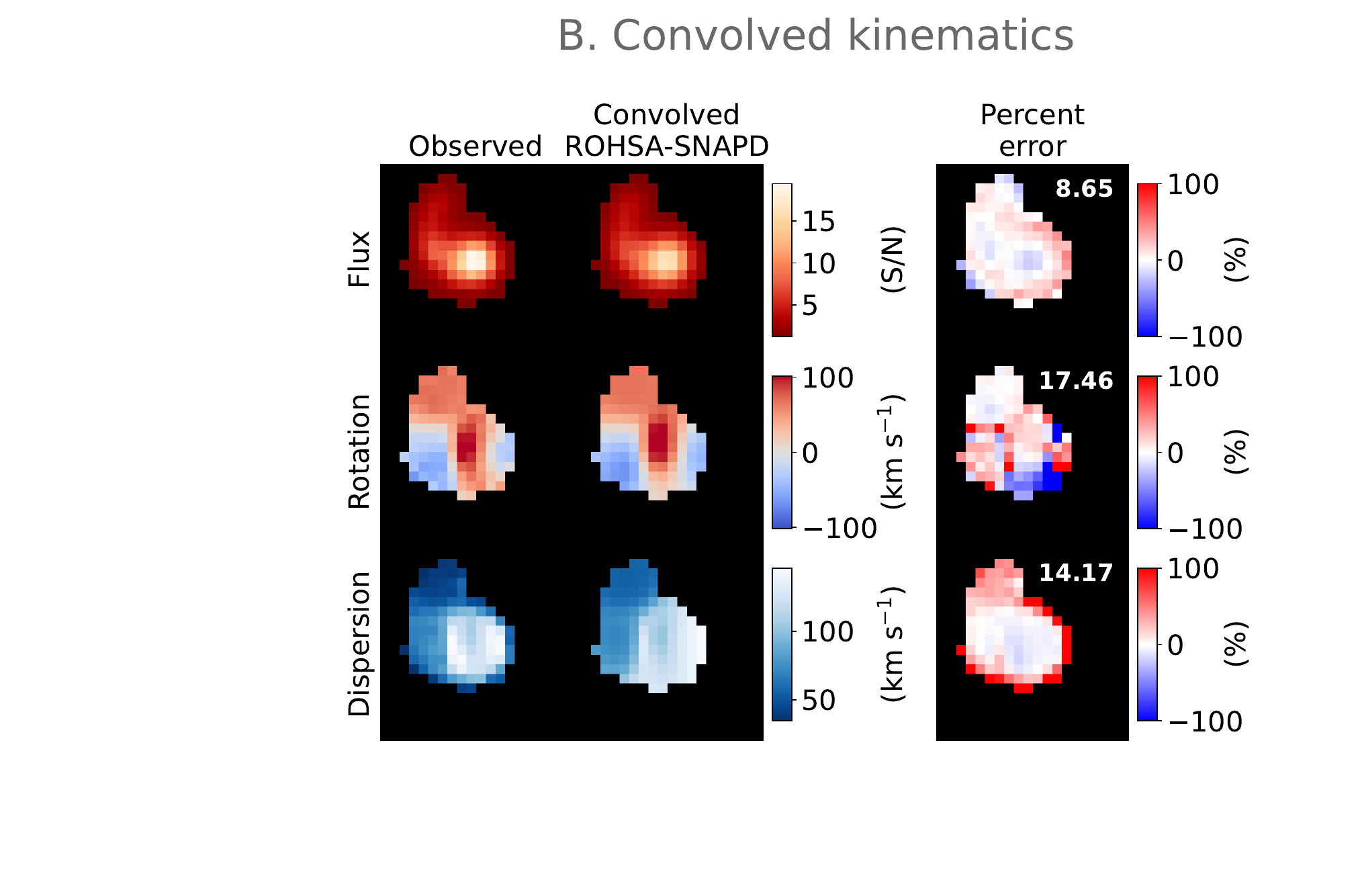}
    \caption{The results of the {\tt ROHSA-SNAPD} fits to the ongoing merger snapshot using $\lambda_f = 0.001$ and $\lambda_\mu = 10$. The {\tt ROHSA-SNAPD} maps shown are the median fits to the $200$ noise realisations. This figure follows the same layout as Figure \ref{fig:KinMS_kinematic_map_plots}.}
    \label{fig:ongoing_merger_kinematic_maps_lambda_mu_10_appendix}
\end{figure*}

\begin{figure*}
    \centering
    \includegraphics[width=0.49\linewidth,trim={3cm 3cm 2cm 0cm},clip]{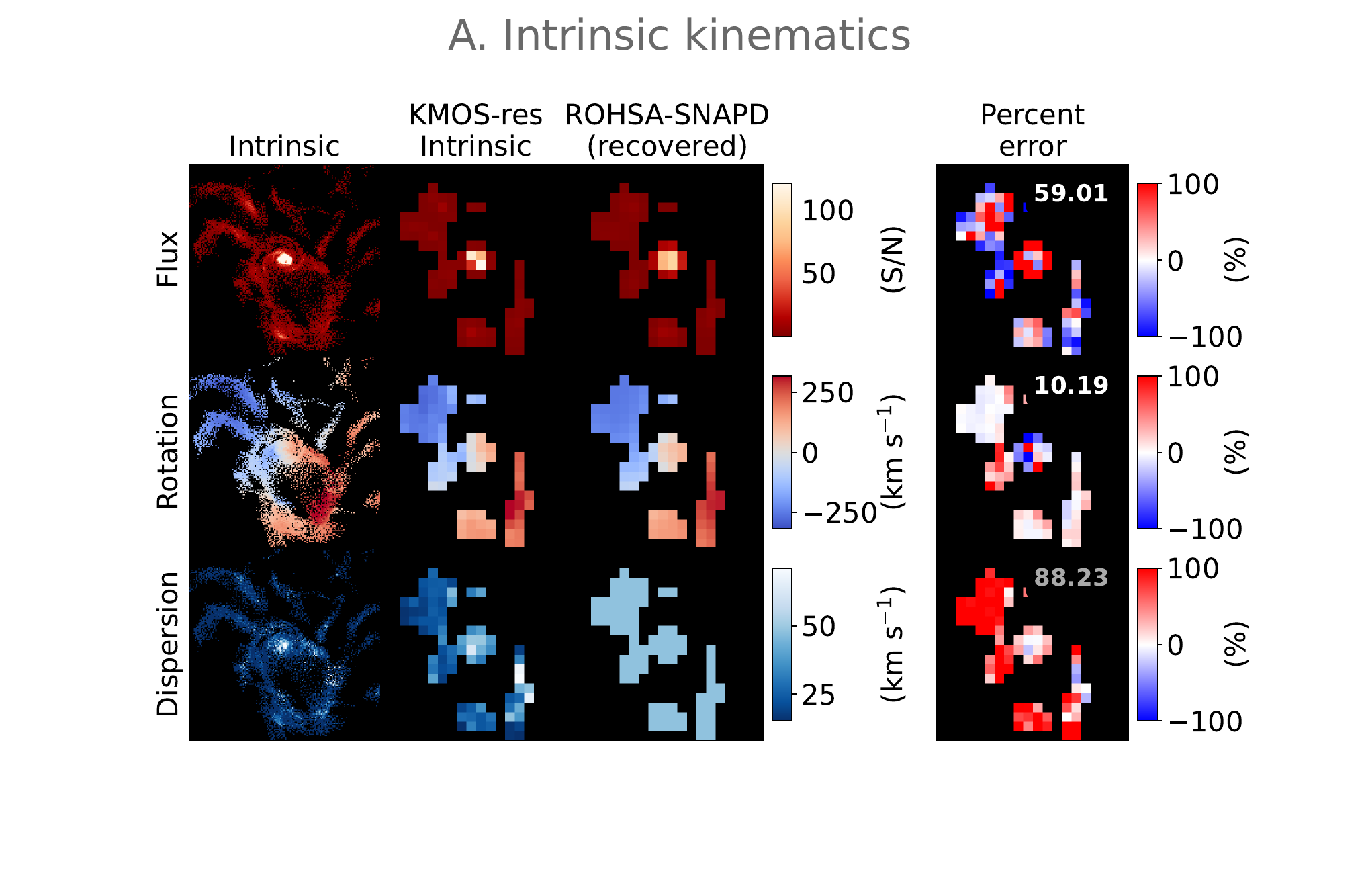}
    \includegraphics[width=0.49\linewidth,trim={3cm 3cm 2cm 0cm},clip]{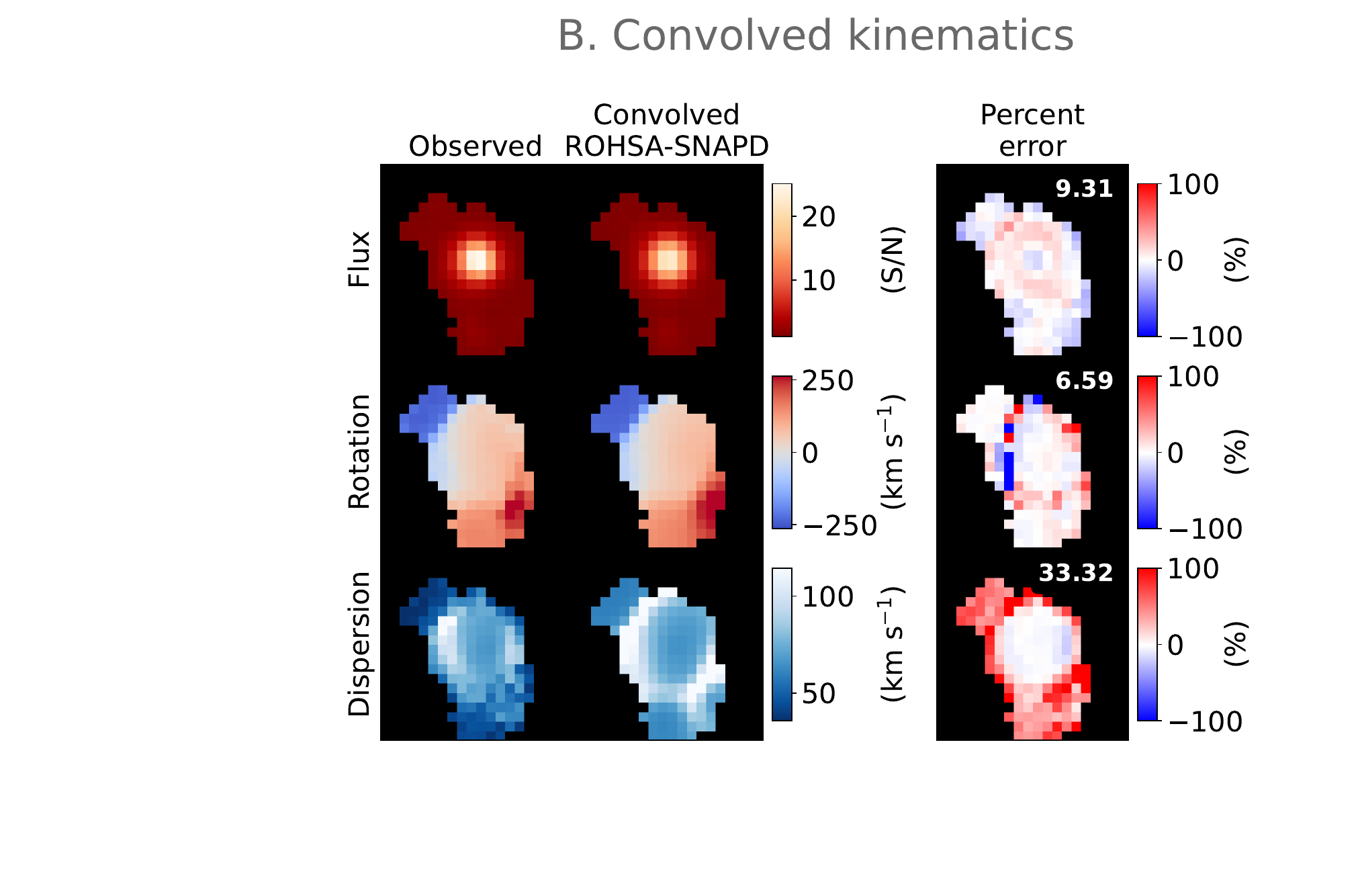}
    \caption{The results of the {\tt ROHSA-SNAPD} fits to the post-merger snapshot using $\lambda_f = 0.001$ and $\lambda_\mu = 10$. The {\tt ROHSA-SNAPD} maps shown are the median fits to the $200$ noise realisations. This figure follows the same layout as Figure \ref{fig:KinMS_kinematic_map_plots}.}
    \label{fig:post_merger_kinematic_maps_lambda_mu_10_appendix}
\end{figure*}

Figures \ref{fig:pre_merger_kinematic_maps_lambda_mu_10_appendix}, \ref{fig:ongoing_merger_kinematic_maps_lambda_mu_10_appendix} and \ref{fig:post_merger_kinematic_maps_lambda_mu_10_appendix} show that \revtwochanged{whilst the deconvolved velocity dispersions can differ substantially,} the deconvolved kinematic structure is very similar between each galaxy's most appropriate regularisation value and $\lambda_\mu=10$.

\section{Recovery of varying velocity dispersion profiles with {\tt ROHSA-SNAPD}}
\label{appendix:varying_dispersion_profiles}
\begin{figure*}
    \centering
    \includegraphics[width=0.49\linewidth,trim={3cm 3cm 2cm 0cm},clip]{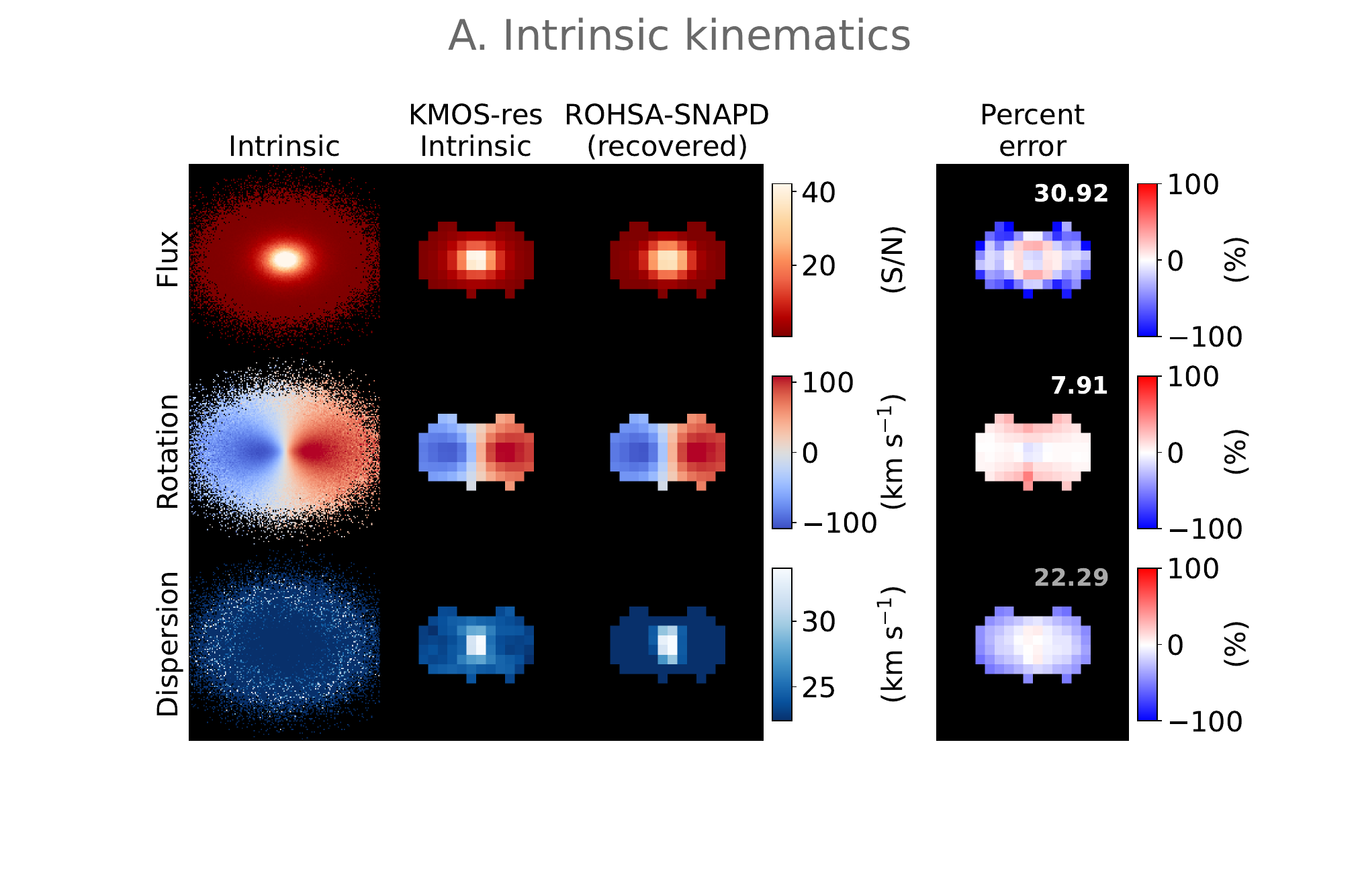}
    \includegraphics[width=0.49\linewidth,trim={3cm 3cm 2cm 0cm},clip]{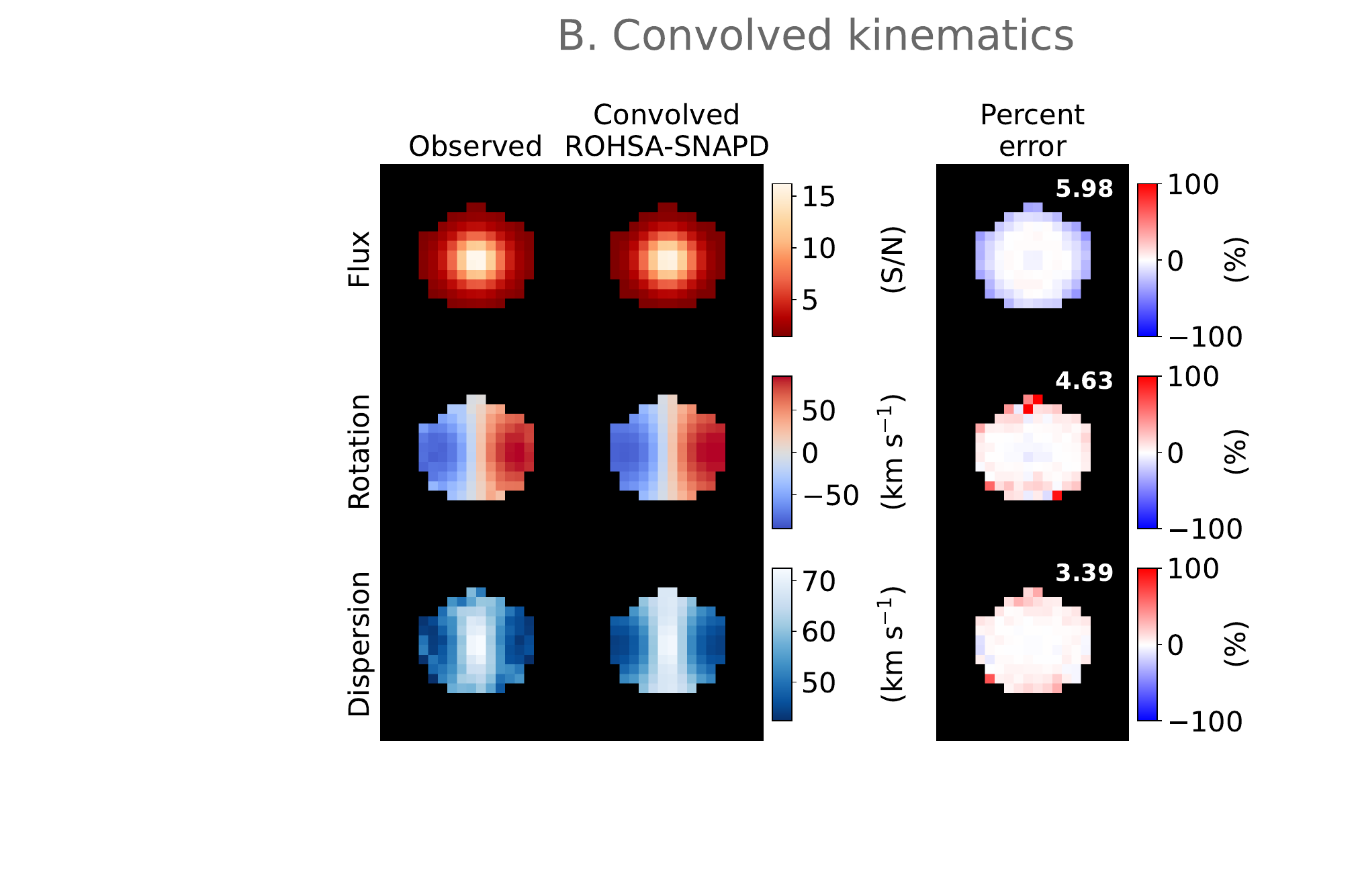}
    \caption{The results of the {\tt ROHSA-SNAPD} fits to the idealised disc galaxy using $\lambda_f = 0.001$, $\lambda_\mu = 10$ and $\lambda_\sigma = 10$, fitting with a varying velocity dispersion profile. The {\tt ROHSA-SNAPD} maps shown are the median fits to the $200$ noise realisations. This figure follows the same layout as Figure \ref{fig:KinMS_kinematic_map_plots}.}
    \label{fig:idealised_disc_varying_dispersion}
\end{figure*}

\begin{figure*}
    \centering
    \includegraphics[width=0.49\linewidth,trim={3cm 3cm 2cm 0cm},clip]{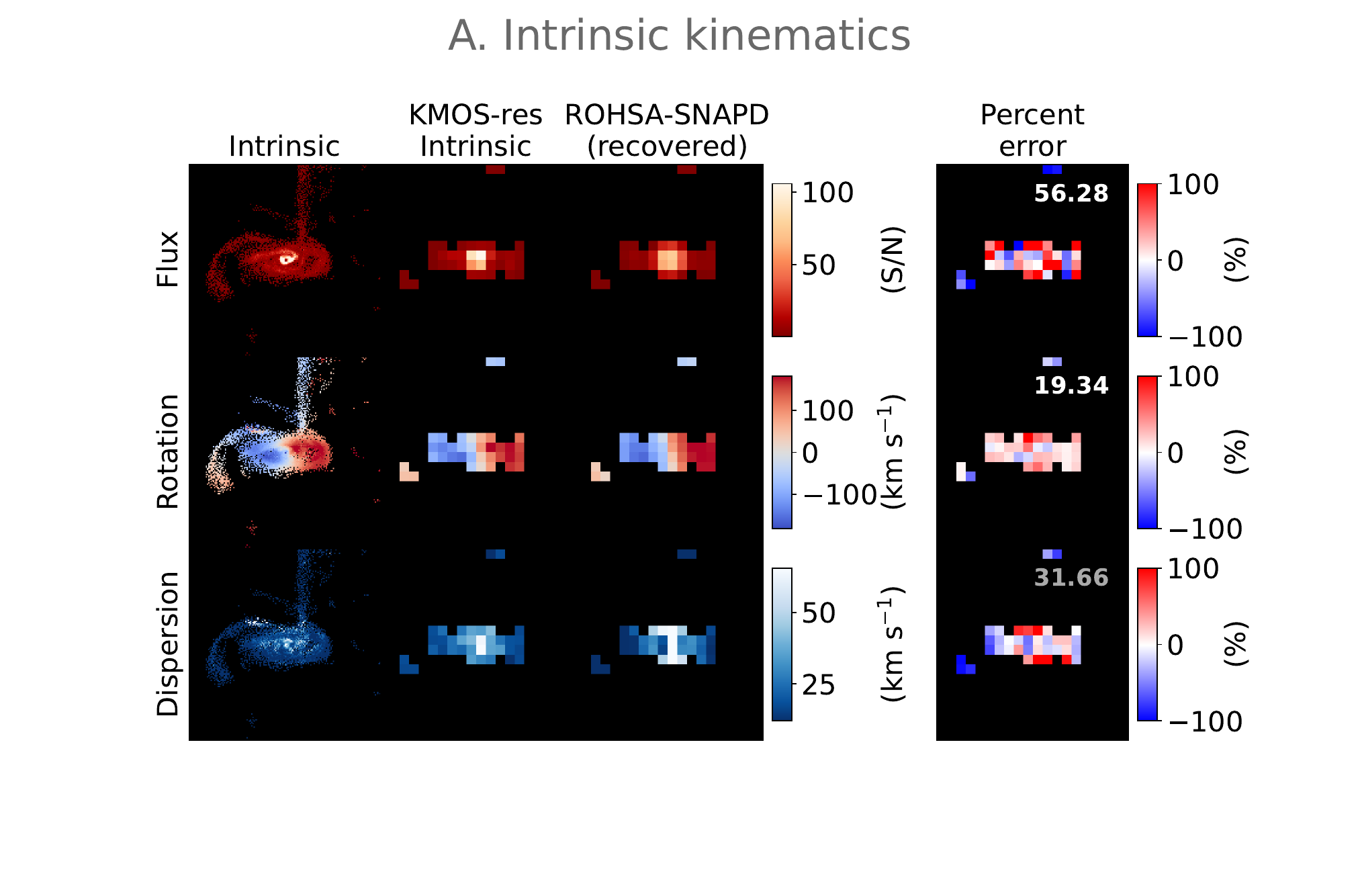}
    \includegraphics[width=0.49\linewidth,trim={3cm 3cm 2cm 0cm},clip]{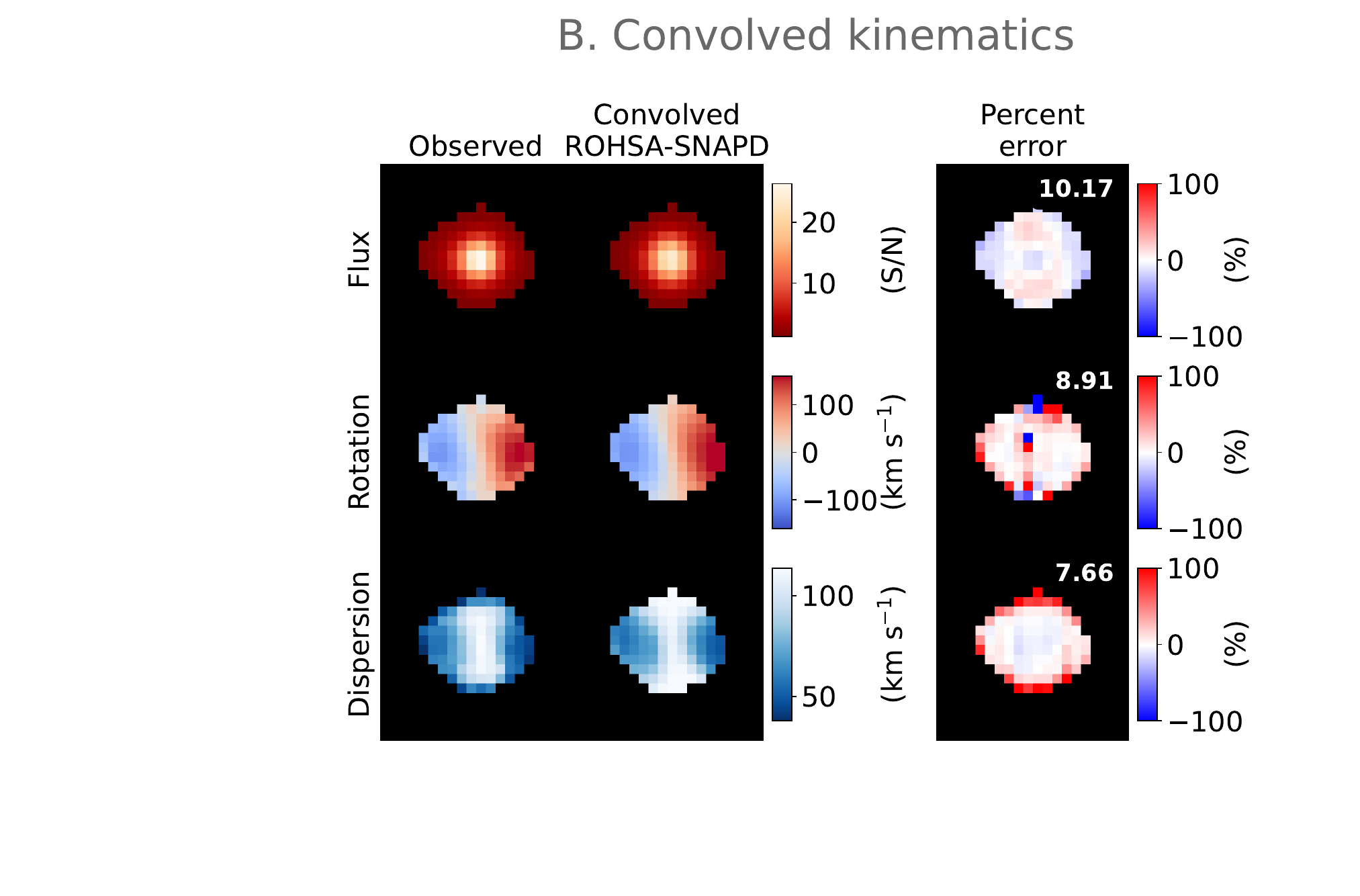}
    \caption{The results of the {\tt ROHSA-SNAPD} fits to the pre-merger snapshot using $\lambda_f = 0.001$, $\lambda_\mu = 5$ and $\lambda_\sigma = 10$, fitting with a varying velocity dispersion profile. The {\tt ROHSA-SNAPD} maps shown are the median fits to the $200$ noise realisations. This figure follows the same layout as Figure \ref{fig:KinMS_kinematic_map_plots}.}
    \label{fig:pre_merger_varying_dispersion}
\end{figure*}

\begin{figure*}
    \centering
    \includegraphics[width=0.49\linewidth,trim={3cm 3cm 2cm 0cm},clip]{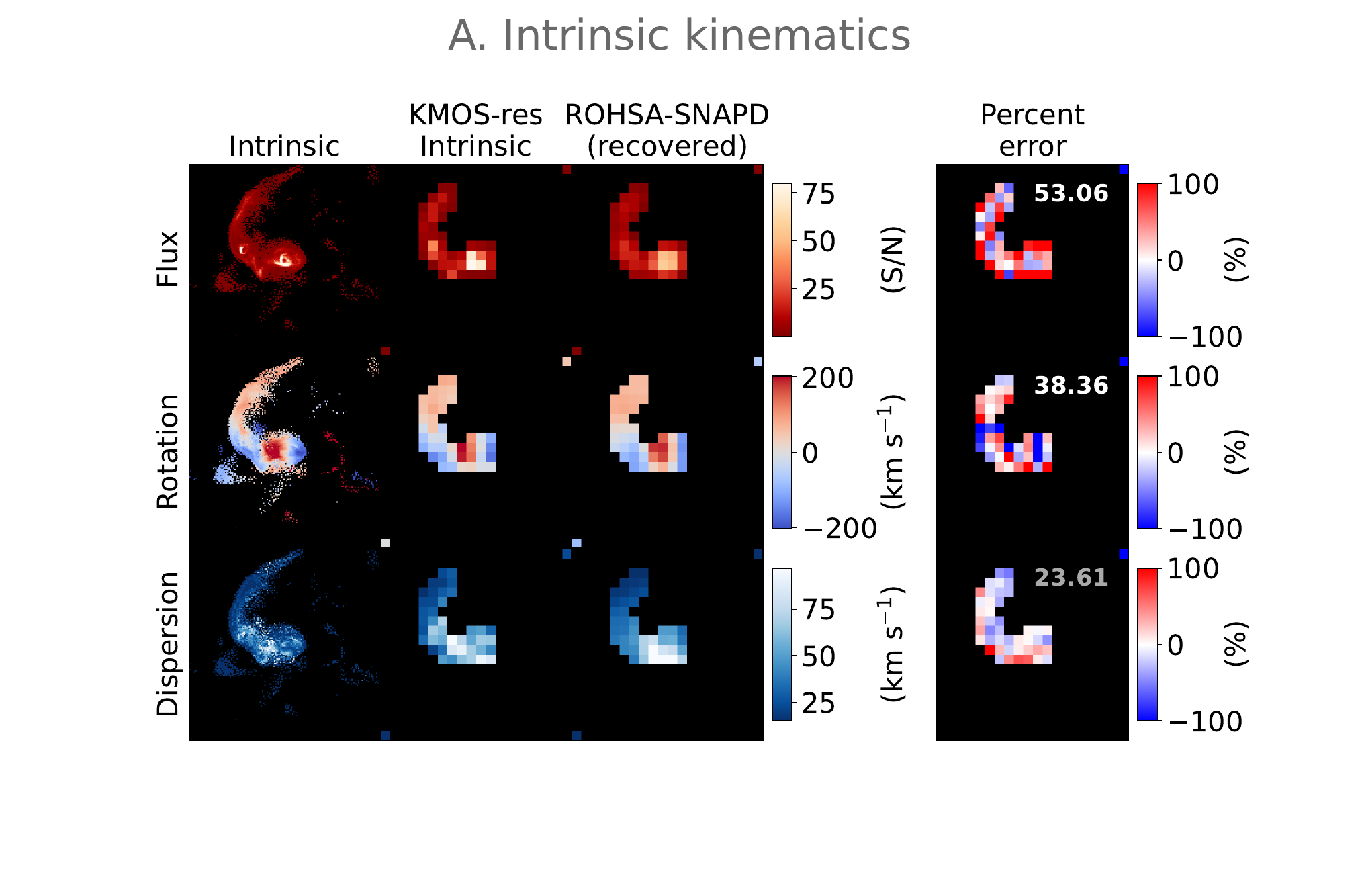}
    \includegraphics[width=0.49\linewidth,trim={3cm 3cm 2cm 0cm},clip]{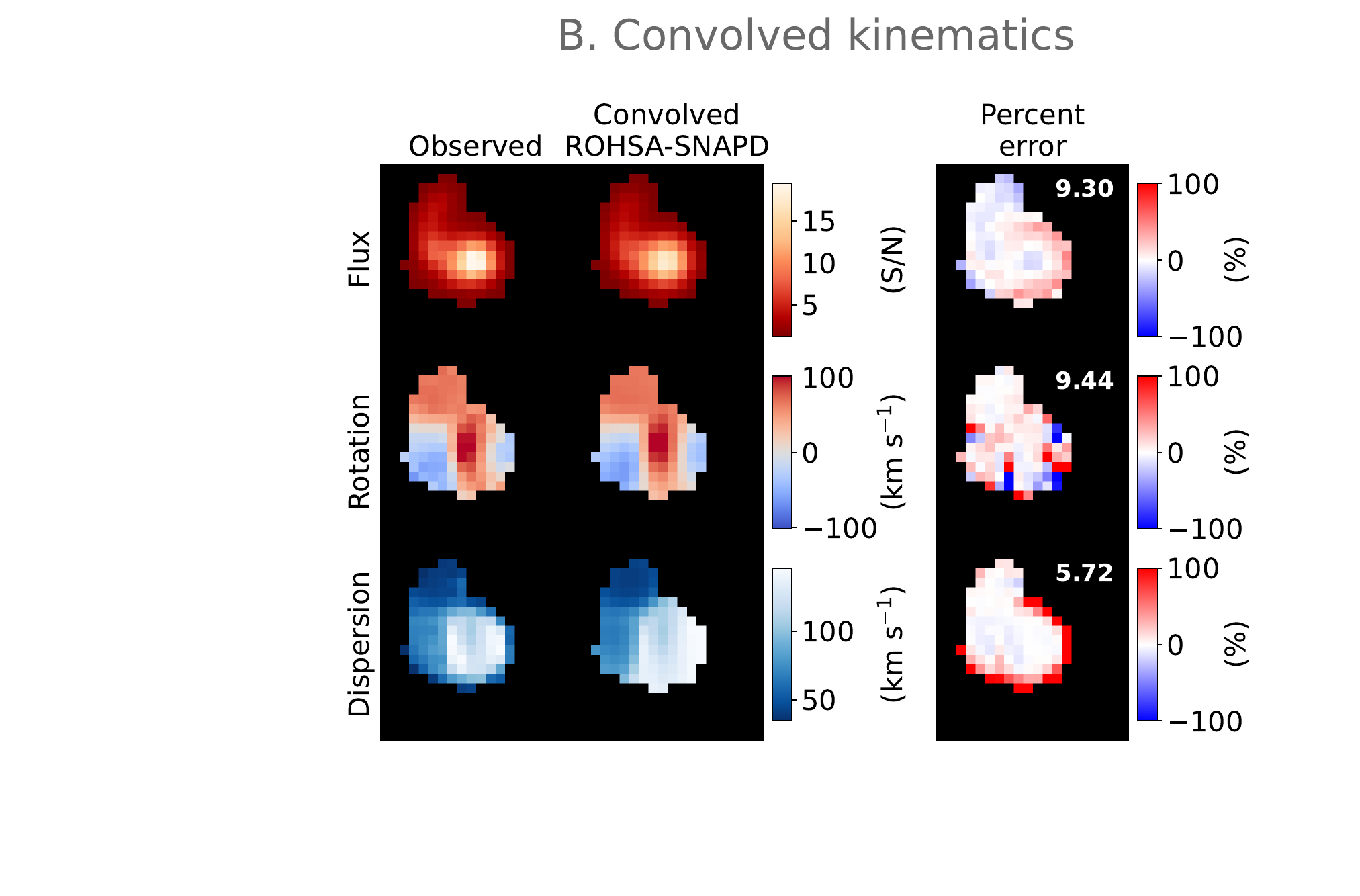}
    \caption{The results of the {\tt ROHSA-SNAPD} fits to the ongoing merger snapshot using $\lambda_f = 0.001$, $\lambda_\mu = 3$ and $\lambda_\sigma = 10$, fitting with a varying velocity dispersion profile. The {\tt ROHSA-SNAPD} maps shown are the median fits to the $200$ noise realisations. This figure follows the same layout as Figure \ref{fig:KinMS_kinematic_map_plots}.}
    \label{fig:ongoing_merger_varying_dispersion}
\end{figure*}

\begin{figure*}
    \centering
    \includegraphics[width=0.49\linewidth,trim={3cm 3cm 2cm 0cm},clip]{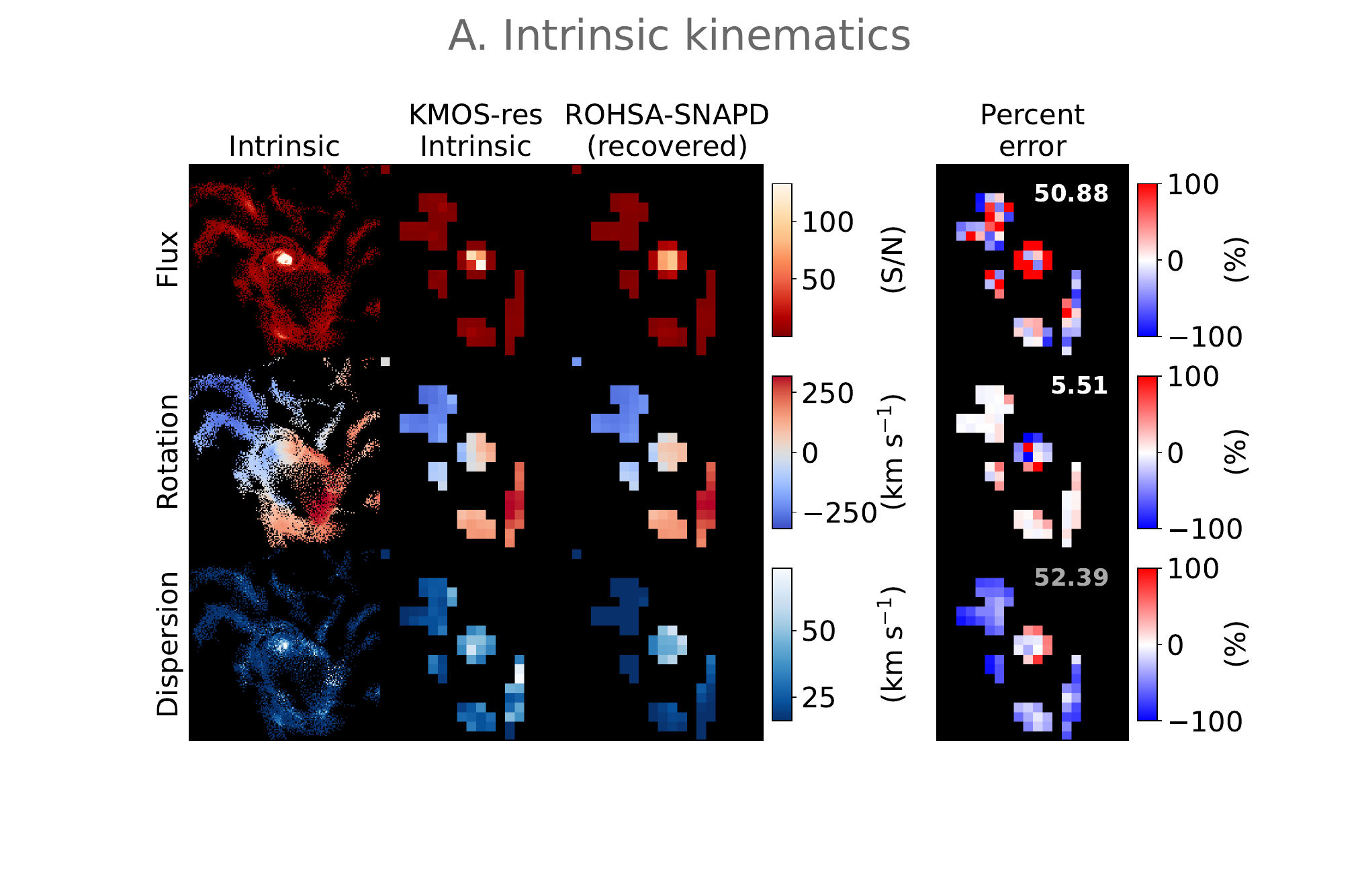}
    \includegraphics[width=0.49\linewidth,trim={3cm 3cm 2cm 0cm},clip]{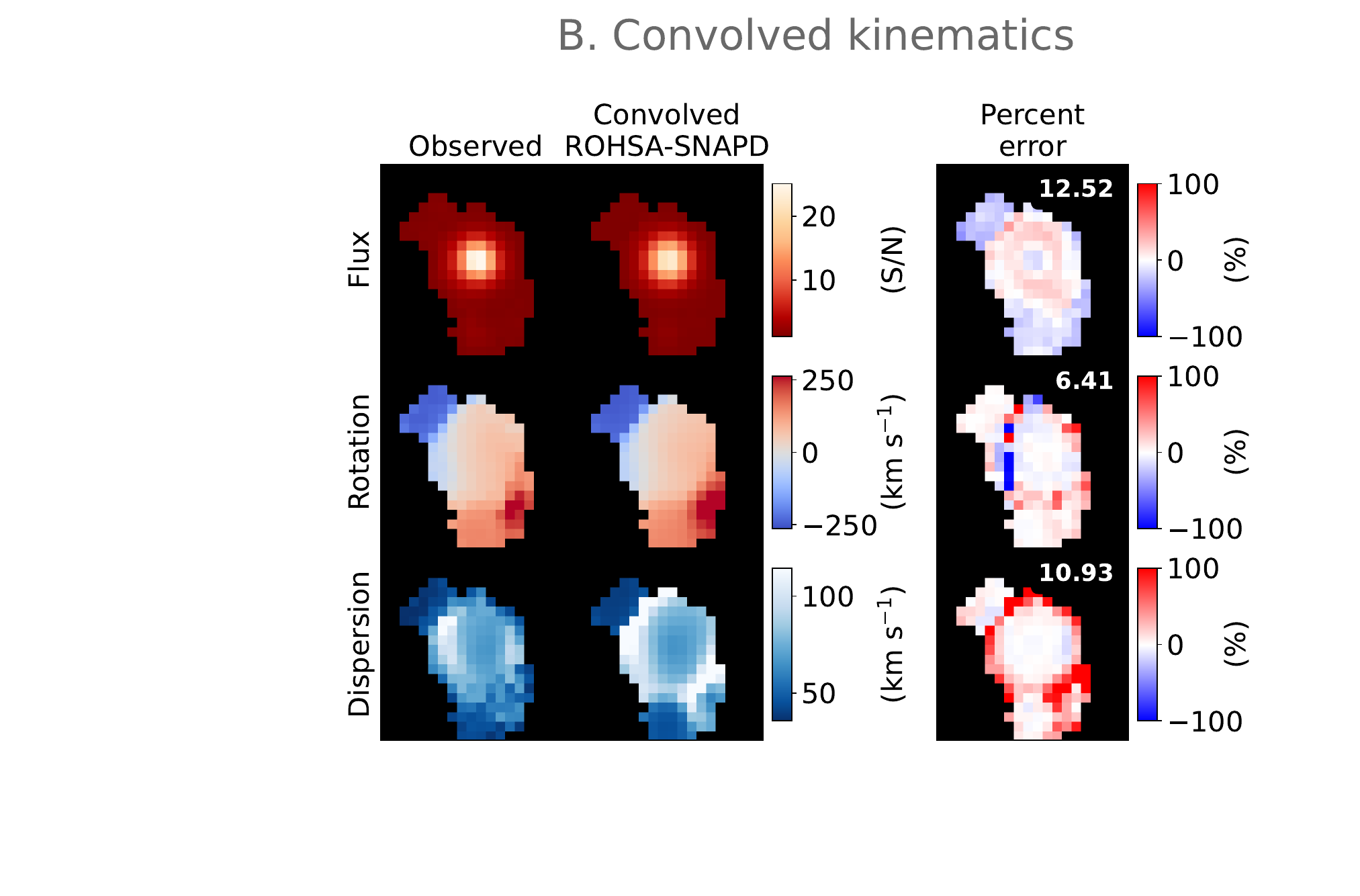}
    \caption{The results of the {\tt ROHSA-SNAPD} fits to the post-merger snapshot using $\lambda_f = 0.001$, $\lambda_\mu = 2$ and $\lambda_\sigma = 10$, fitting with a varying velocity dispersion profile. The {\tt ROHSA-SNAPD} maps shown are the median fits to the $200$ noise realisations. This figure follows the same layout as Figure \ref{fig:KinMS_kinematic_map_plots}.}
    \label{fig:post_merger_varying_dispersion}
\end{figure*}

To demonstrate the recovery of varying velocity dispersion profiles with {\tt ROHSA-SNAPD}, the idealised disc galaxy and TNG100 snapshots were also fit with a non-constant velocity dispersion, using $\lambda_\sigma = 10$ and the same flux and rotation regularisation values used in Sections \ref{sec:evaluation_idealised_disc_galaxy} and \ref{sec:evaluation_tng_galaxies}. Figures \ref{fig:idealised_disc_varying_dispersion}-\ref{fig:post_merger_varying_dispersion} compare the 2D kinematic of each fit, and show that, whilst the velocity dispersion profiles vary, the deconvolved flux and rotational velocity maps are very similar to when velocity dispersion was held constant.

The deconvolved velocity dispersion maps of the ongoing and post-merger snapshots agree reasonably well with the KMOS-resolution intrinsic, considering the additional observational effects introduced when binning the high-resolution intrinsic maps. However, the recovery of the idealised disc and pre-merger velocity dispersions are not as accurate. The most significant differences occur in the central few pixels of the two observations, where the deconvolved dispersion is significantly larger than the intrinsic. This is likely due to varying velocity gradients on sub-pixel scales which {\tt ROHSA-SNAPD} cannot account for and is discussed further in Appendix \ref{appendix:limitations_bilinear_interpolation}.

\begin{figure*}
    \centering
\centering
    \includegraphics[width=0.245\linewidth]{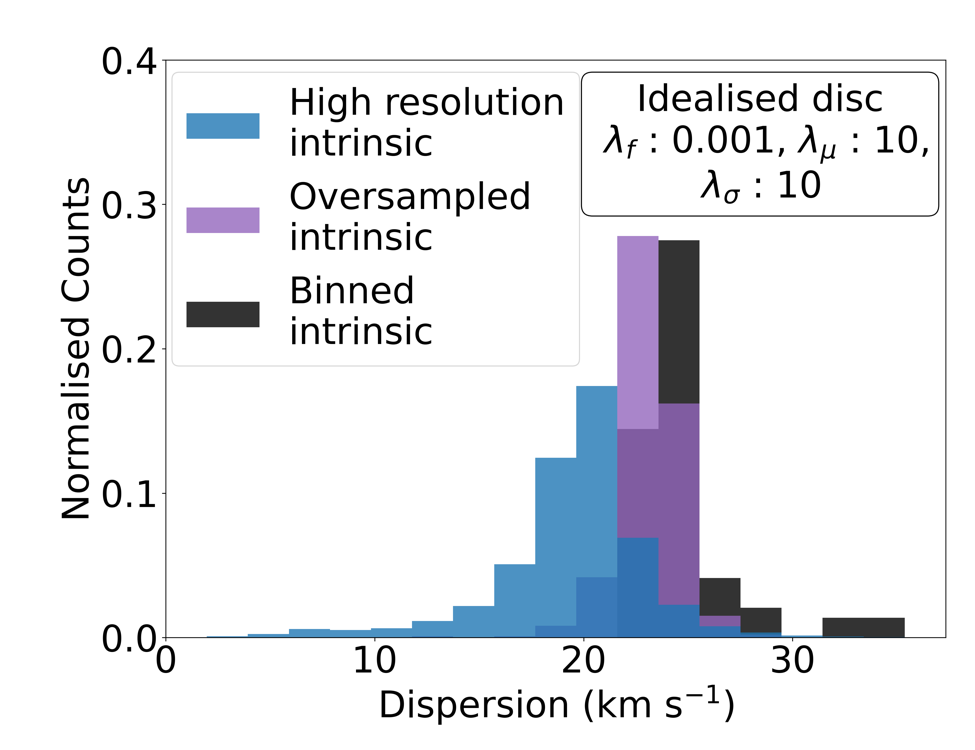}
    \includegraphics[width=0.245\linewidth]{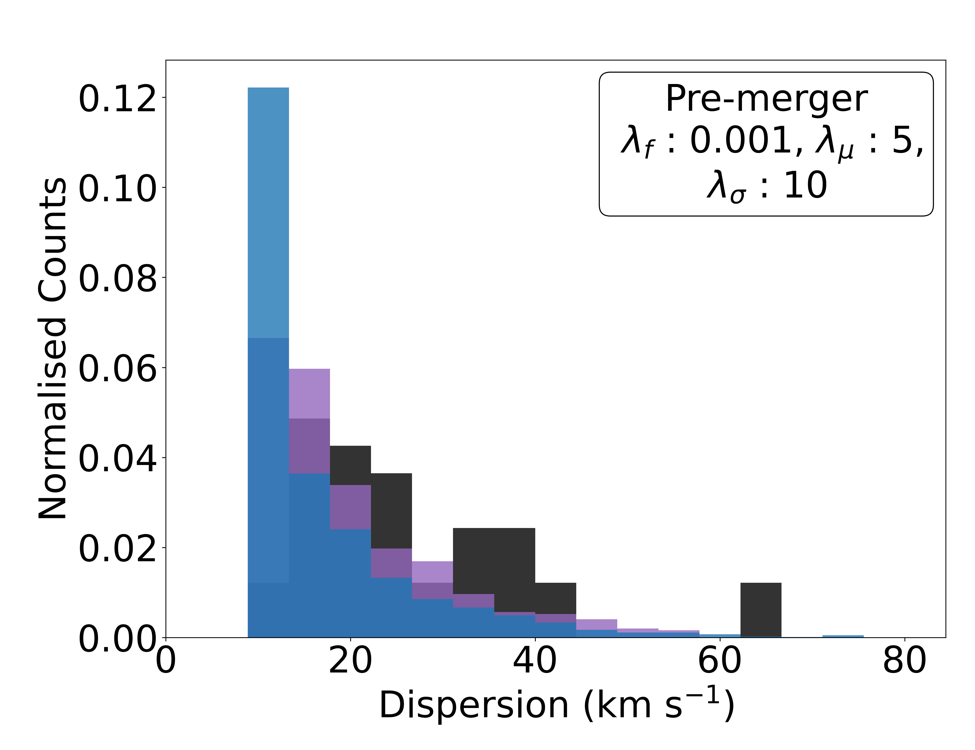}
    \includegraphics[width=0.245\linewidth]{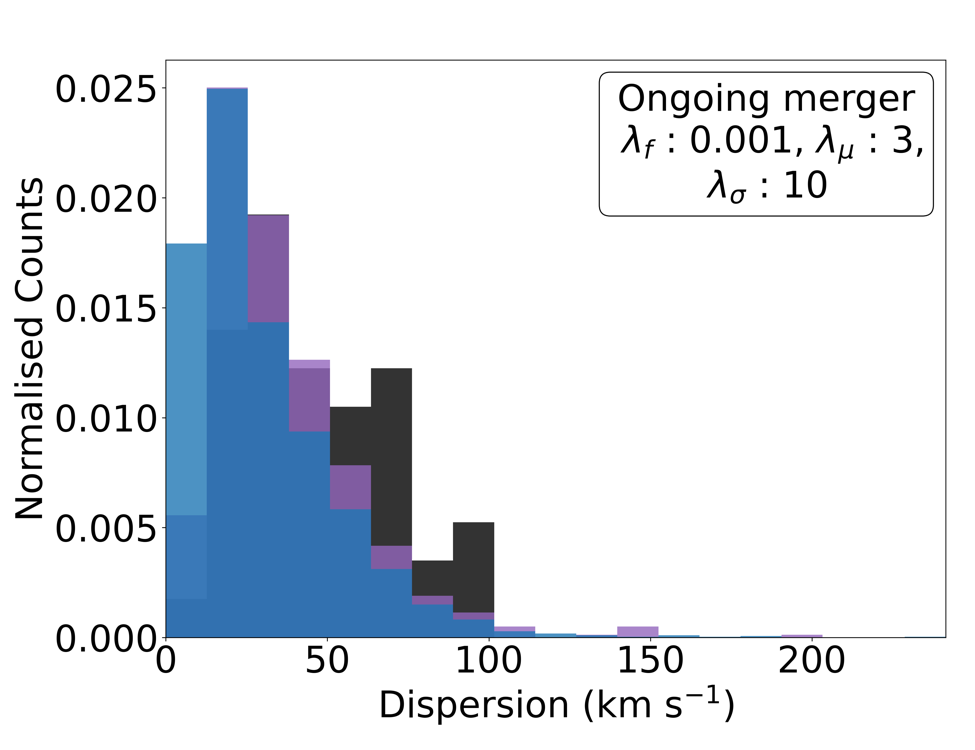}
    \includegraphics[width=0.245\linewidth]{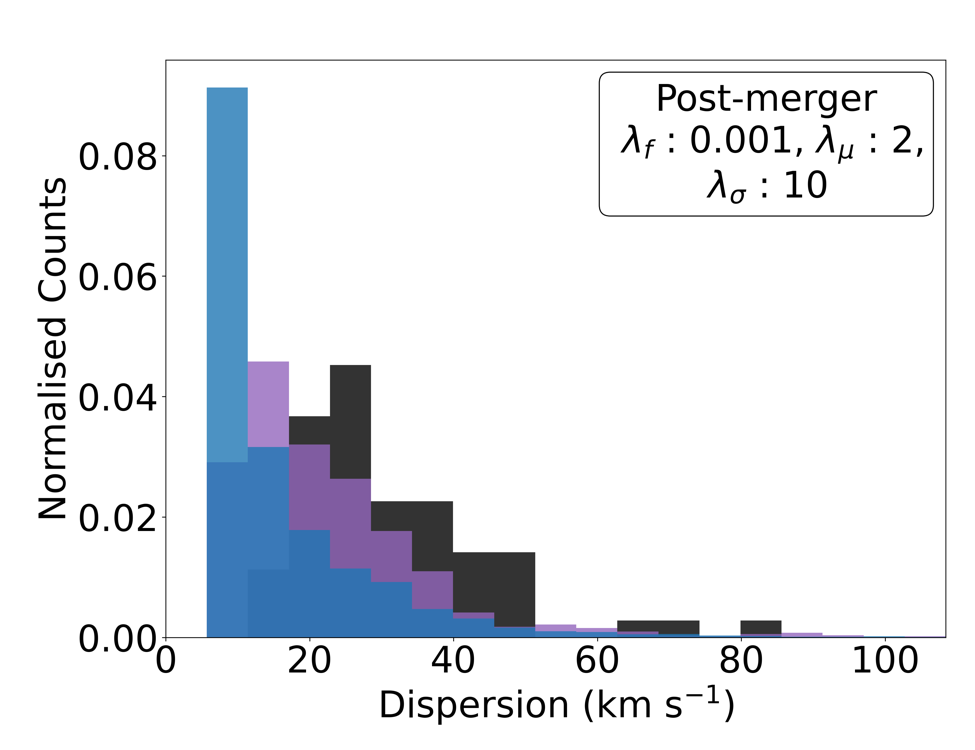}
    \includegraphics[width=0.245\linewidth]{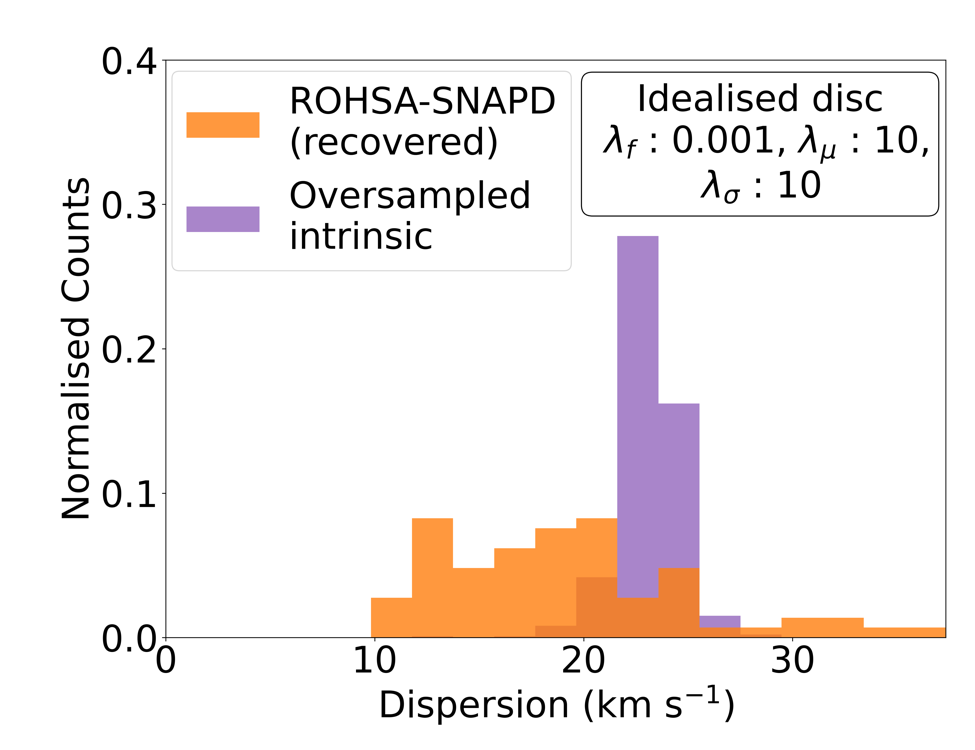}
    \includegraphics[width=0.245\linewidth]{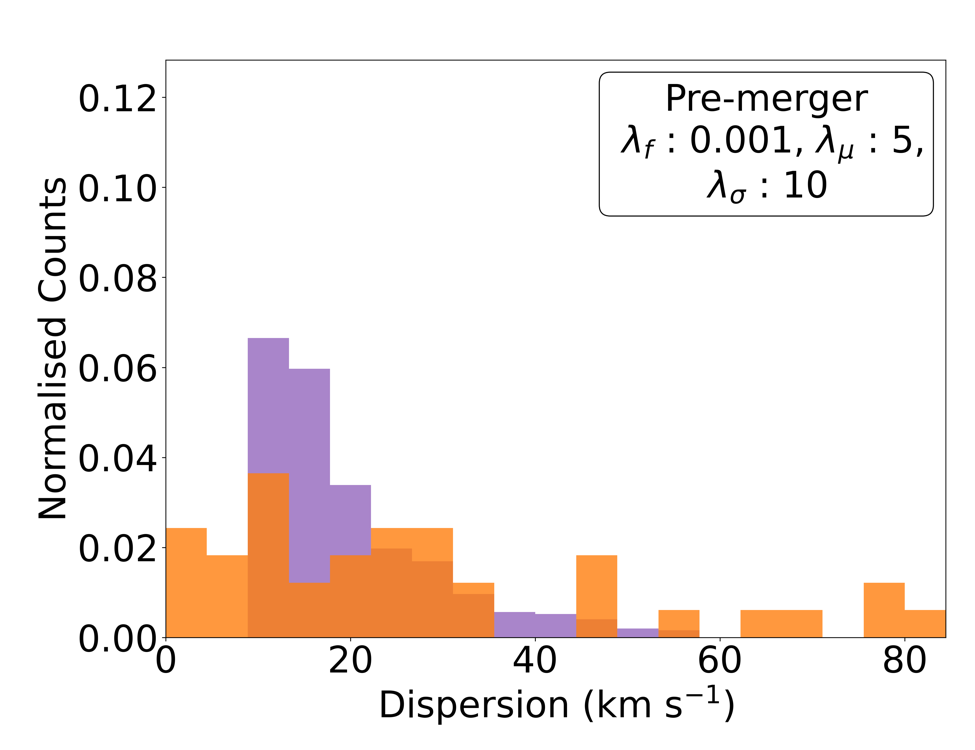}
    \includegraphics[width=0.245\linewidth]{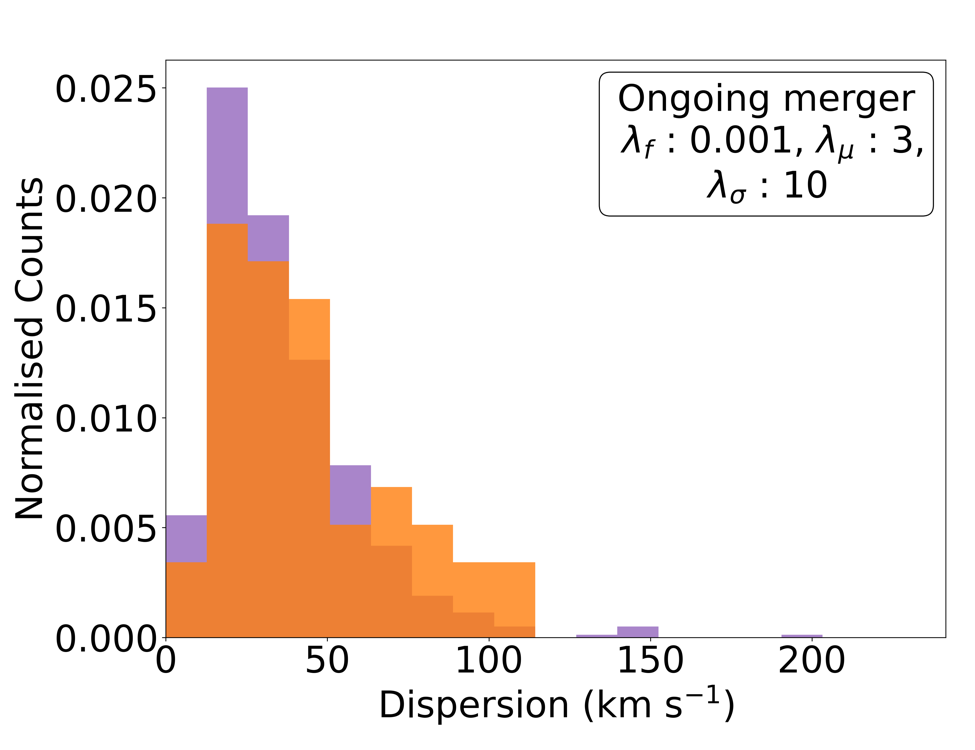}
    \includegraphics[width=0.245\linewidth]{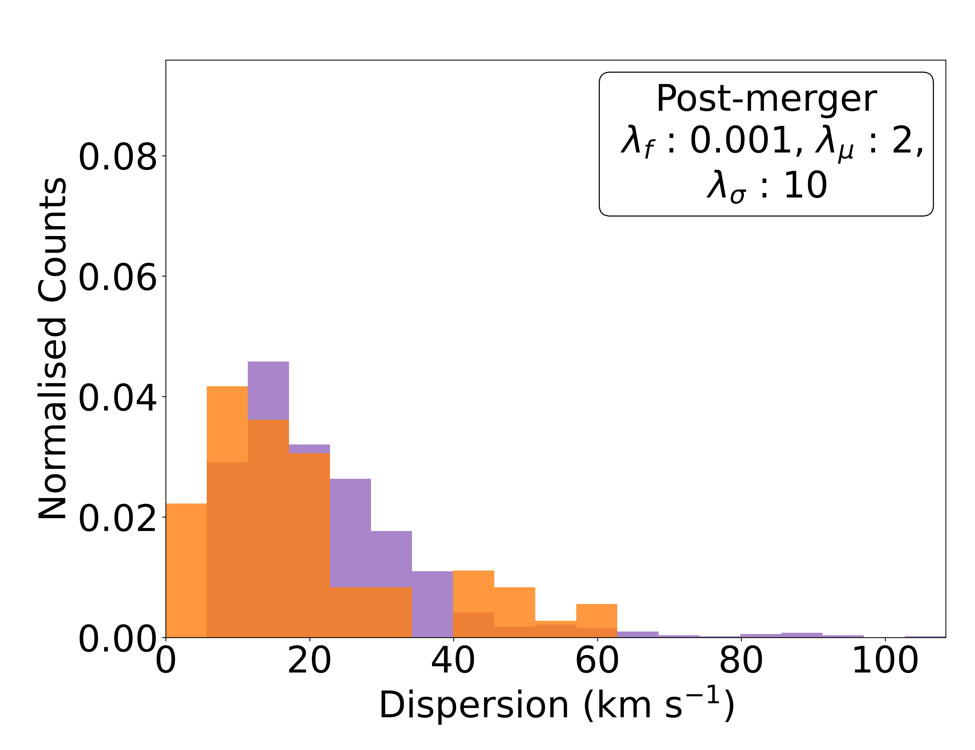}
    \caption{ This figure follows the same layout as Figure \ref{fig:disp_hist_TNG} and compares the deconvolved {\tt ROHSA-SNAPD} velocity dispersion values (orange) with a histogram of the intrinsic velocity dispersion of the respective galaxy at high-resolution ($n=8$) (blue), at oversampled resolution ($n=4$) (purple) and at observed resolution ($n=1$) (black). For a clear comparison, the area under each distribution is normalised to $1$. From the $200$ noise realisations, the median deconvolved {\tt ROHSA-SNAPD} velocity dispersion for each spatial pixel is shown as an orange histogram. The \revonechanged{x-axis} of each panel \revonechanged{spans} from $0$ to the maximum value of any resolution.}
    \label{fig:varying_dispersion_disp_hist_appendix}
\end{figure*}

Apart from a small number of pixels that have overestimated velocity dispersions, Figure \ref{fig:varying_dispersion_disp_hist_appendix} also shows that a large number of deconvolved velocity dispersions also skew lower than the oversampled intrinsic in most systems. This is partially because we are trying to recover velocity dispersion values below the level of the instrument resolution, but also likely because non-parametric codes can be less accurate in low S/N regions \citep{Lee_2024}.

\section{Limitations in using bilinear interpolation to account for finite spatial resolution}
\label{appendix:limitations_bilinear_interpolation}
\begin{figure}
    \centering
    \includegraphics[width=\linewidth]{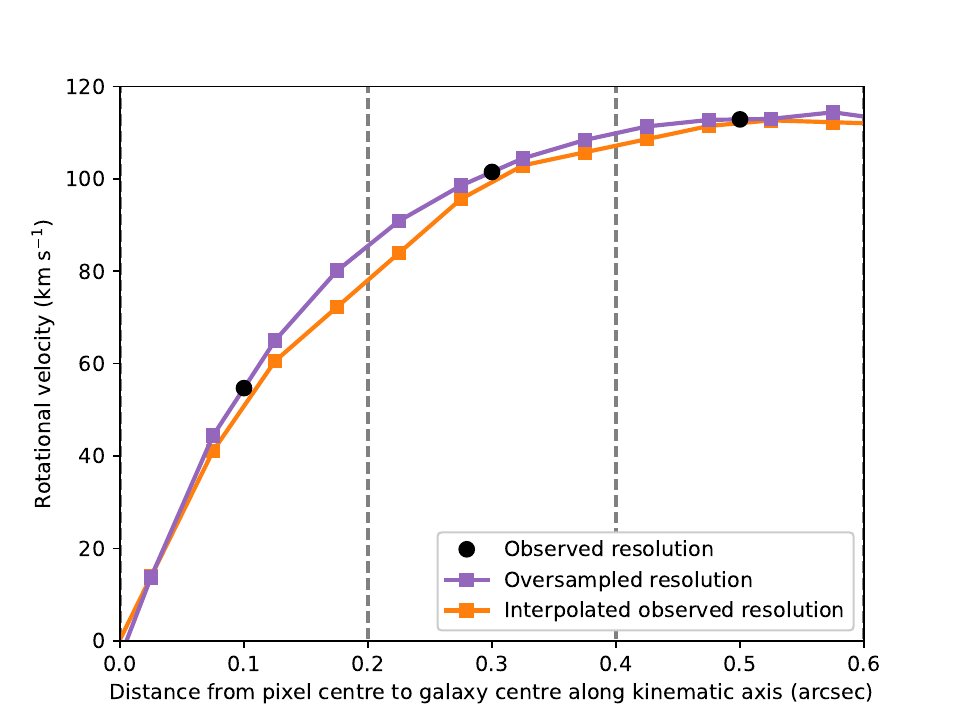}
    \caption{This figure compares a 1D rotation curve sampled at $0".05$ oversampled resolution (purple squares), $0".2$ observed resolution (black points) and observed resolution linearly interpolated to oversampled resolution (orange squares). $0".2$ pixel edges are shown as grey dashed lines.}
    \label{fig:appendix_bilinear_interpolation_example}
\end{figure}

In this work we have used bilinear interpolation to oversample the {\tt ROHSA-SNAPD} kinematics maps within the cost function, aiming to replicate the effects of sub-pixel velocity gradients that are present in the intrinsic data but cannot be resolved observationally due to finite spatial resolution.
In Figure \ref{fig:appendix_bilinear_interpolation_example} we show a simplistic 1D example using linear interpolation to demonstrate the limitations of the method used. 
We take the oversampled resolution rotation map of the idealised disc galaxy introduced in Section \ref{sec:evaluation_idealised_disc_galaxy} and extract the central row of the map to create an oversampled resolution rotation curve sampled at $0".05$ increments (purple squares), focusing on the central region of the galaxy. By binning this curve in $0".2$ increments and taking the median value, we create an observed resolution rotation curve which represents an ideal recovery of the oversampled resolution rotation curve by {\tt ROHSA-SNAPD} (black points). The observed resolution points are then linearly interpolated using $n=4$ to $0".05$ resolution (orange squares) to demonstrate parameter map interpolation within the cost function. All points in Figure \ref{fig:appendix_bilinear_interpolation_example} are shown as a function of the distance from their pixel centre to the galaxy centre. Grey dashed lines denote the observed resolution pixel boundaries.

It can be seen that whilst the interpolation method can account for the average velocity gradient across the pixel, sub-pixel changes in gradient cannot be replicated and the oversampled resolution intrinsic rotation curve therefore includes steeper velocity gradients than {\tt ROHSA-SNAPD} can account for. This effect is exacerbated in the 2D case using bilinear interpolation as there are velocity gradients in all spatial directions. As a result, during the forward modelling process, {\tt ROHSA-SNAPD} must increase the deconvolved velocity dispersion in regions with varying sub-pixel gradients in order to replicate observed line-widths.

Therefore, the oversampled resolution intrinsic velocity and dispersion profiles cannot be perfectly recovered with {\tt ROHSA-SNAPD}, regardless of the oversampling factor used. The most significant deviations occur in regions with low intrinsic velocity dispersion and large sub-pixel velocity gradients. This is most clearly demonstrated in fits with varying velocity dispersions, Appendix \ref{appendix:varying_dispersion_profiles}, as the deconvolved dispersions of the idealised disc and pre-merger snapshot are too high in the central few pixels where sub-pixel gradients vary substantially. In contrast, the ongoing and post-merger deconvolved dispersion values are more appropriate, likely because the dispersion is large in the ongoing merger snapshot and the sub-pixel velocity gradients are less significant in the post-merger snapshot.

\revonechanged{\section{The impact of S/N on deconvolved kinematics}
\label{appendix:SN_impact_regularisation}
To demonstrate the effect of the regularisation hyper-parameters and the impact of S/N, we further analyse the fit results from Appendix \ref{appendixsubsec:decreasing_data_SN_and_regularisation}, where the idealised disc galaxy described in Section \ref{sec:evaluation_idealised_disc_galaxy} was simulated at $4$ different S/N levels (S/N $= 10, 20, 30$ and $40$ within a $1\textnormal{"}.5$ aperture) and fit with a range of regularisation values. To clearly demonstrate the impact of regularisation, we show fit for a single noise realisation and show the full deconvolved flux and velocity maps in Figure \ref{fig:appendix_nxn_grid_plots}. Focusing first on flux regularisation at fixed S/N, the unregularised flux maps vary sharply pixel-to-pixel and are only coherent in the very central pixels, even for the highest S/N mock galaxy. This occurs because isolated bright pixels at the galaxy edges can match the extended observed flux structure when convolved with the PSF. As flux regularisation is increased, such unrealistic pixels are negatively weighted and the deconvolved flux becomes coherent to larger radii. The deconvolved rotation map is mostly unaffected by the choice of flux regularisation, though higher velocity regularisation values cause the map to vary more smoothly. In regions with zero deconvolved flux, the deconvolved rotation map values are driven purely by the rotational regularisation.
As S/N is increased, both the deconvolved flux and rotation maps are better constrained. This is clearly demonstrated by the kinematic axis profiles in Figure \ref{fig:appendix_SN_kin_ax_profiles}, where the results from fitting $20$ noise realisations of each S/N galaxy with fixed regularisation values are shown in Figure \ref{fig:appendix_nxn_grid_plots}. At S/N $= 10$ the error on the observed velocity and dispersion is large, causing the median deconvolved rotation to deviate from the intrinsic by around $\pm 10$ km s$^{-1}$ and the deconvolved dispersion value to fit as $0$ km s$^{-1}$. A zero intrinsic dispersion occurs here because the low S/N observed line width can be completely explained by beam smearing and instrument resolution. As S/N increases, the kinematics are better constrained and the deconvolved {\tt ROHSA-SNAPD} kinematics better match the intrinsic. This highlights the difficulty in recovering accurate velocity dispersion values below the instrument resolution in low S/N galaxies. A S/N greater than around $20$  would be required to accurately recover the intrinsic kinematics of this mock disc galaxy.}

\begin{figure*}
    \centering
    \includegraphics[width=\linewidth,trim={2cm 0cm 5cm 0cm},clip]{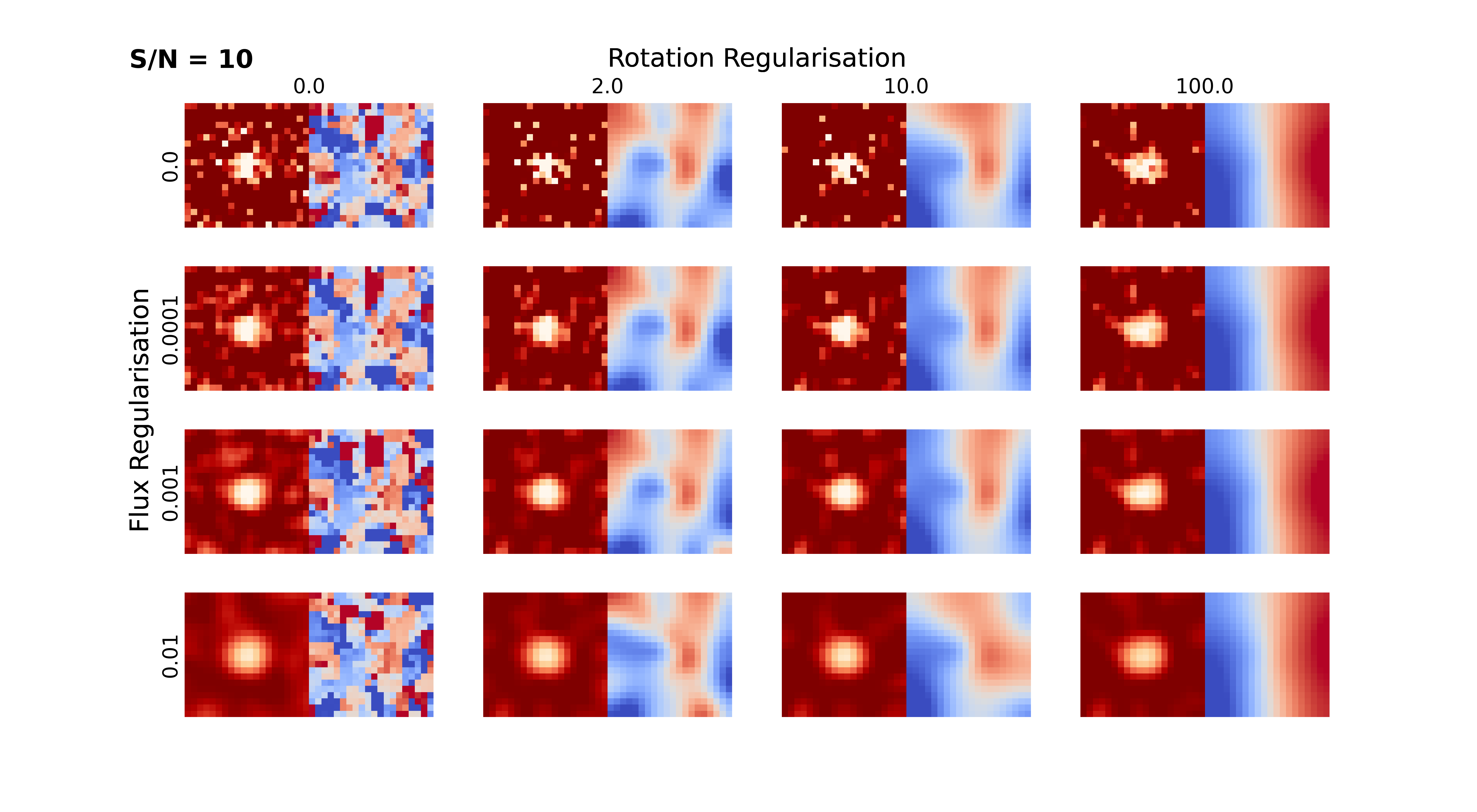}
    \includegraphics[width=\linewidth,trim={2cm 0cm 5cm 0cm},clip]{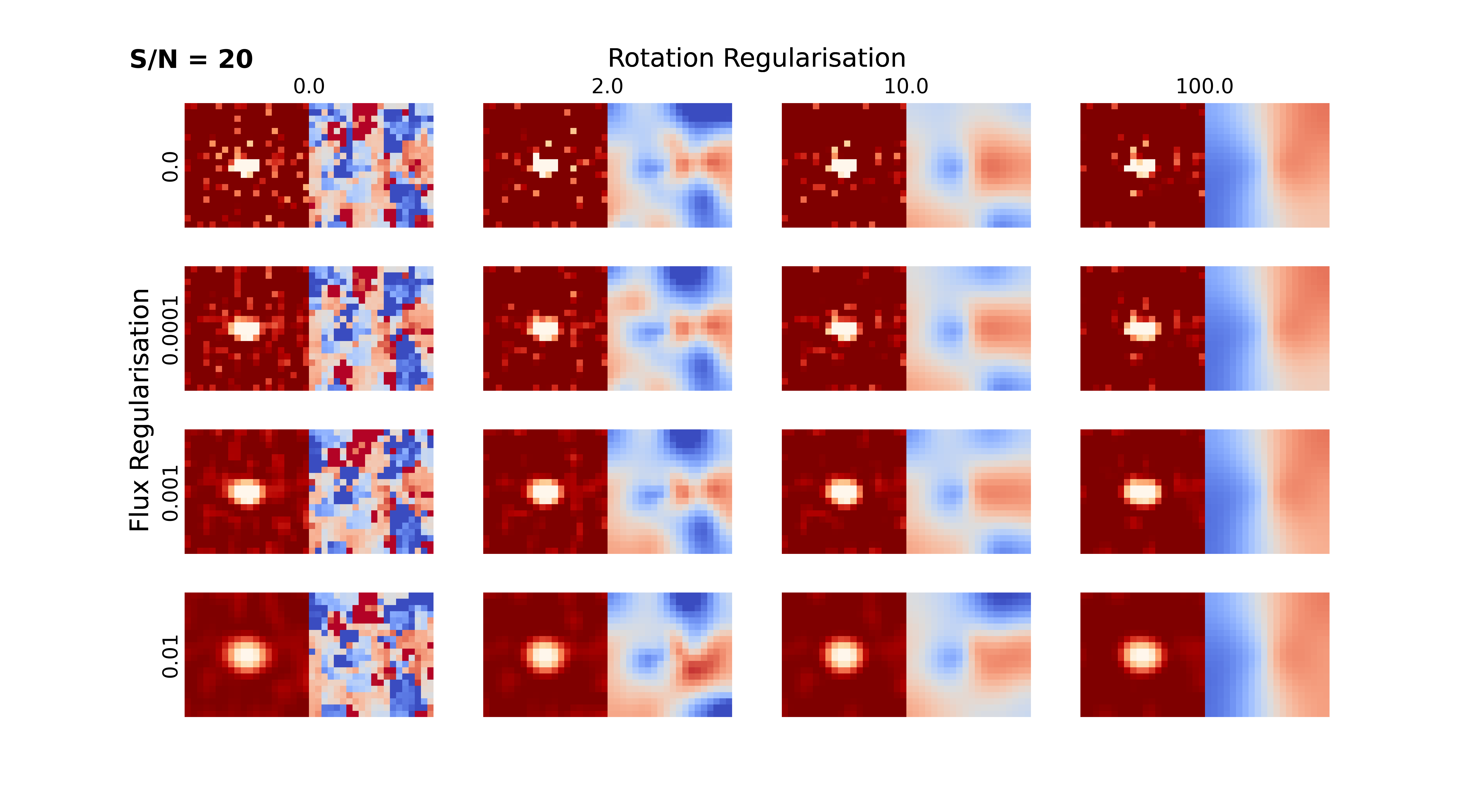}
    \caption{Each of the four panels show the deconvolved {\tt ROHSA-SNAPD} flux and rotation maps from fitting a single noise realisation of the idealised disc galaxy at a given S/N with a range of flux and rotation regularisation values. The specific S/N and regularisation values used are specified along the panel edges. Every flux colourbar has bounds $[0,150]$ in units of intrinsic flux and every rotation colourbar has bounds $[-200,200]$ km s$^{-1}$ around the systemic velocity.}
    \label{fig:appendix_nxn_grid_plots}
\end{figure*}

\begin{figure*}\ContinuedFloat
    \includegraphics[width=\linewidth,trim={2cm 0cm 5cm 0cm},clip]{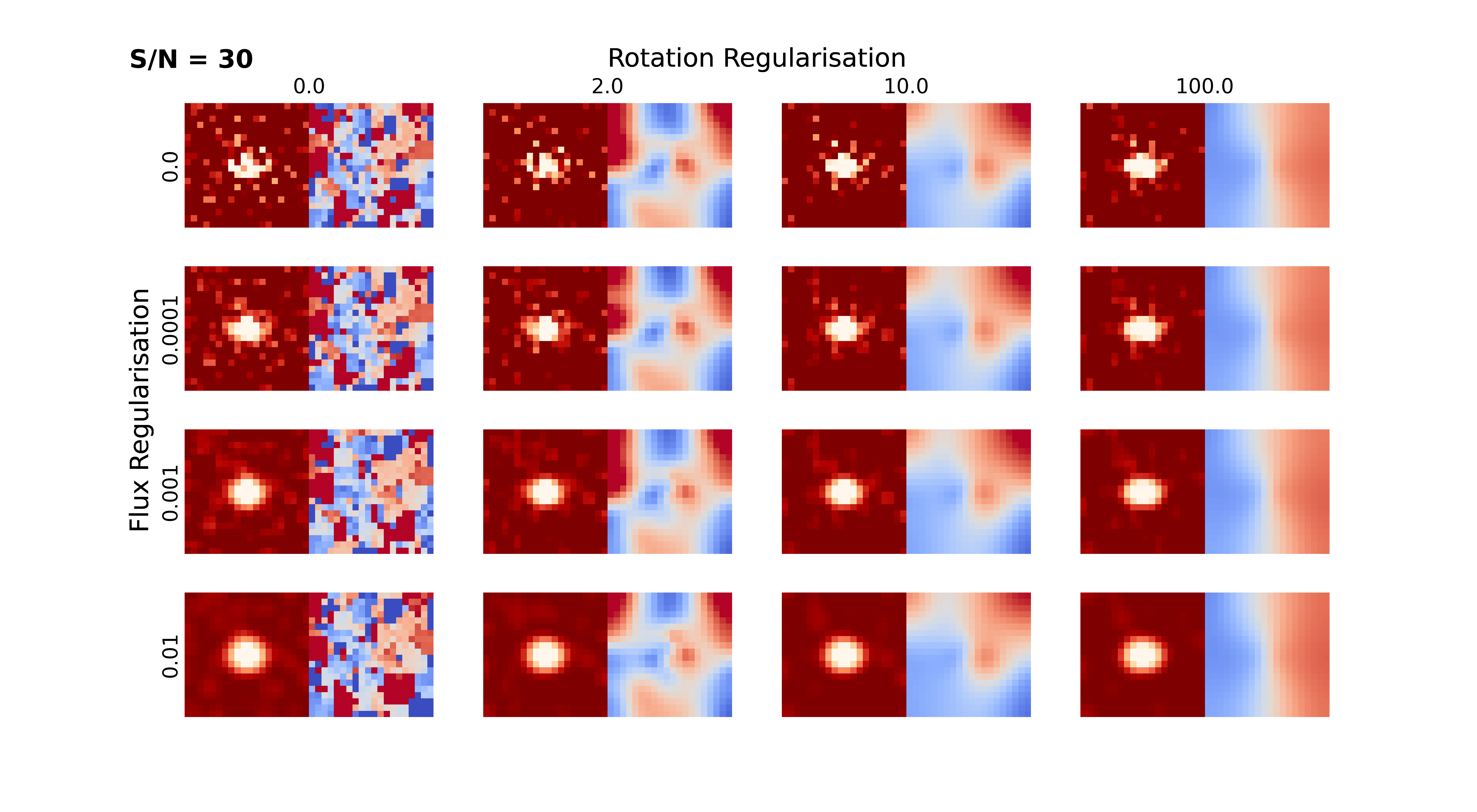}
    \includegraphics[width=\linewidth,trim={2cm 0cm 5cm 0cm},clip]{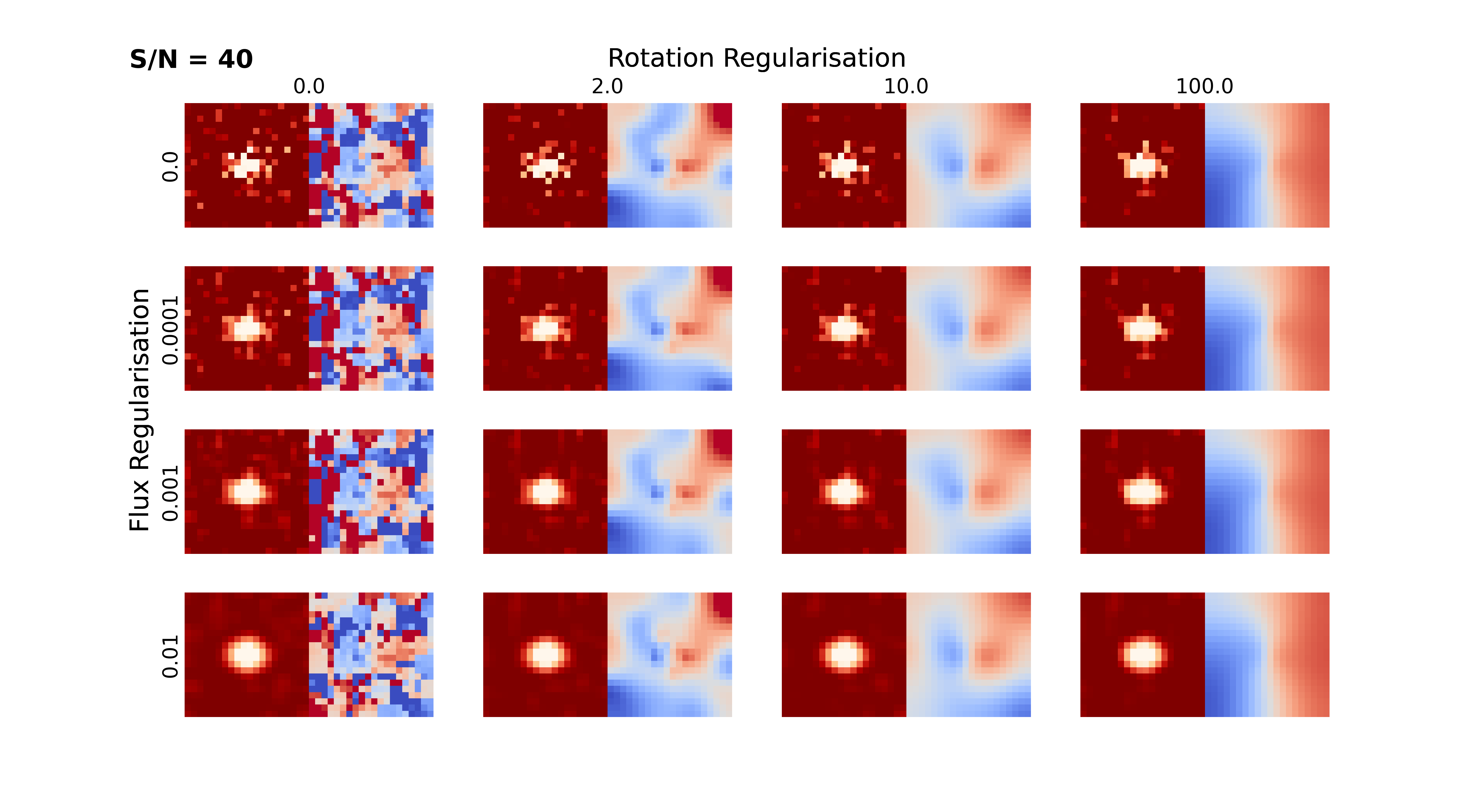}
    \caption{(Continued)}
\end{figure*}

\begin{figure*}
    \centering
    \includegraphics[width=0.9\linewidth]{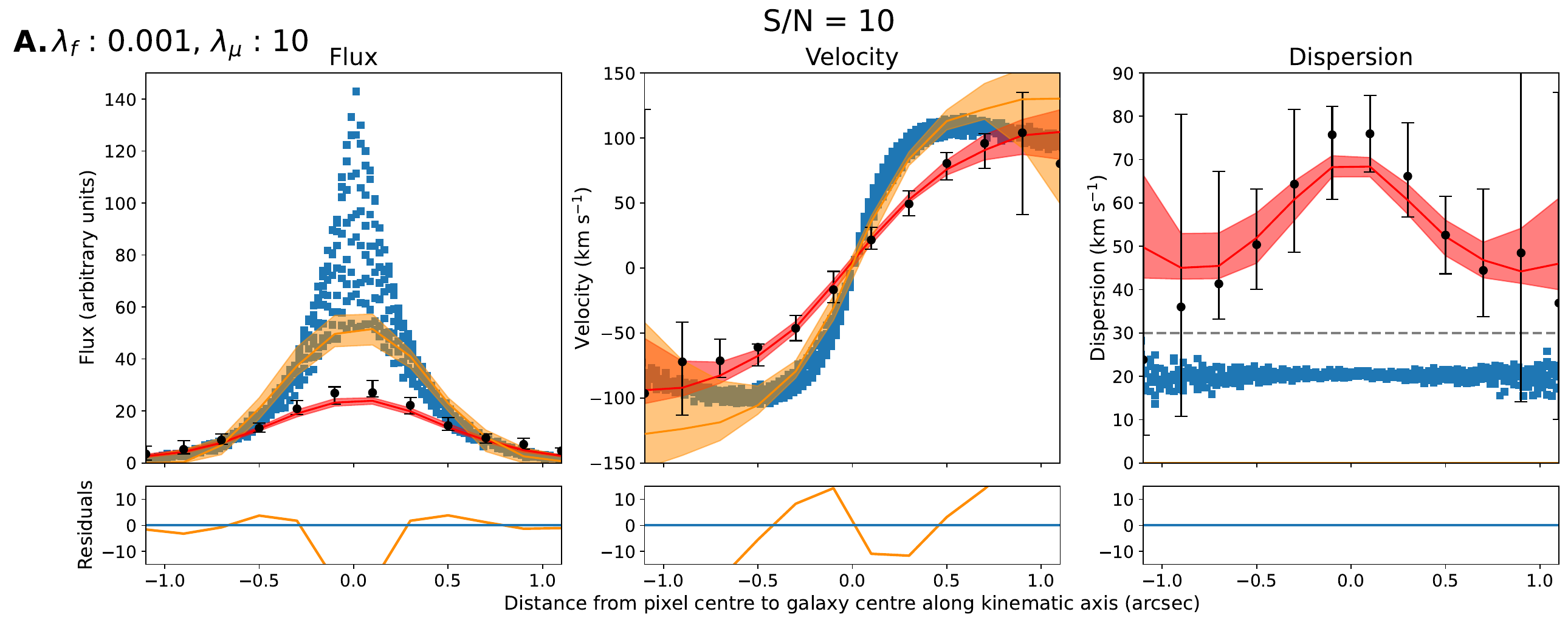} \hspace{5cm}
    
    \includegraphics[width=0.9\linewidth]{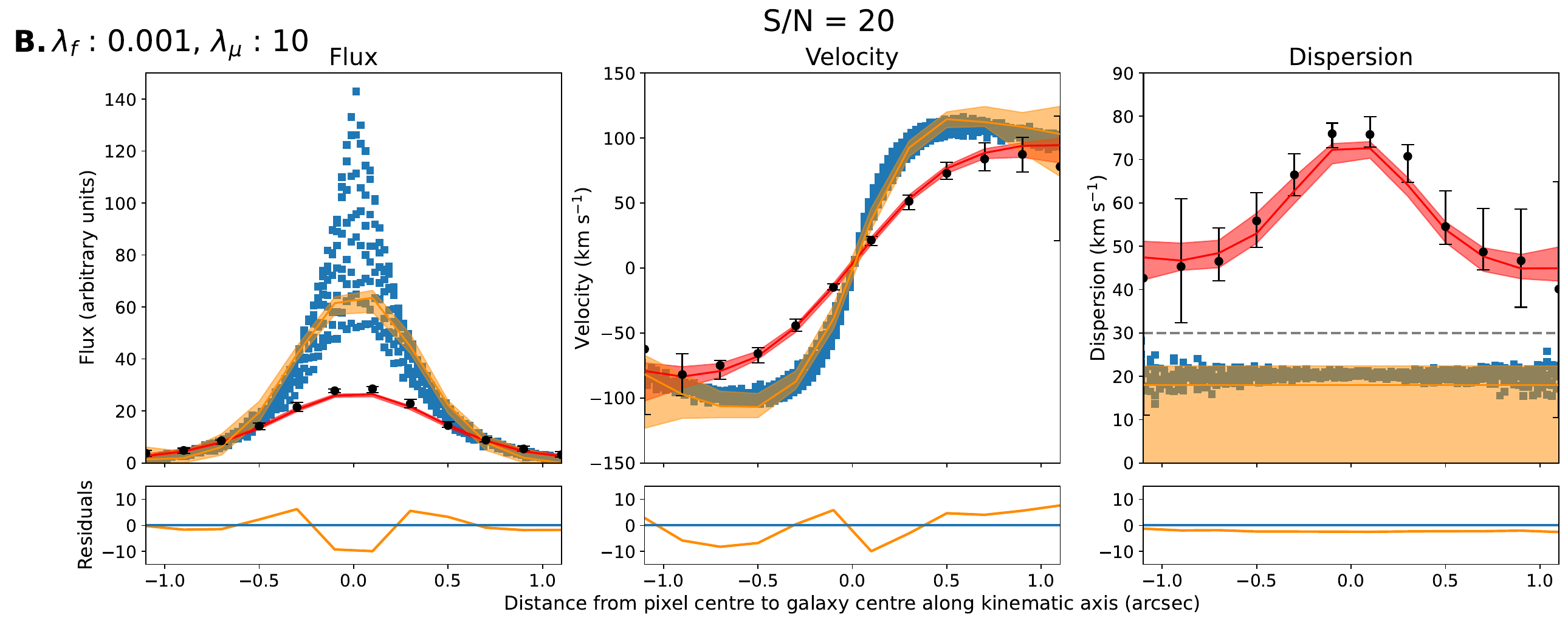} \hspace{5cm}
    
    \caption{Each panel of this figure follows the same general layout as Figure \ref{fig:KinMS_kinematic_axis_profiles} and describes the kinematic axis profiles of four idealised disc galaxies with S/N between $10$ and $40$, as shown in Figure \ref{fig:appendix_nxn_grid_plots}. The median, $16$th and $84$th percentiles in each panel are derived from fitting $20$ noise realisations of each galaxy. Axis profiles extend out to $\pm 1\textnormal{"}.2$, regardless of the S/N level.}
    \label{fig:appendix_SN_kin_ax_profiles}
\end{figure*}

\begin{figure*}\ContinuedFloat
        \includegraphics[width=0.9\linewidth]{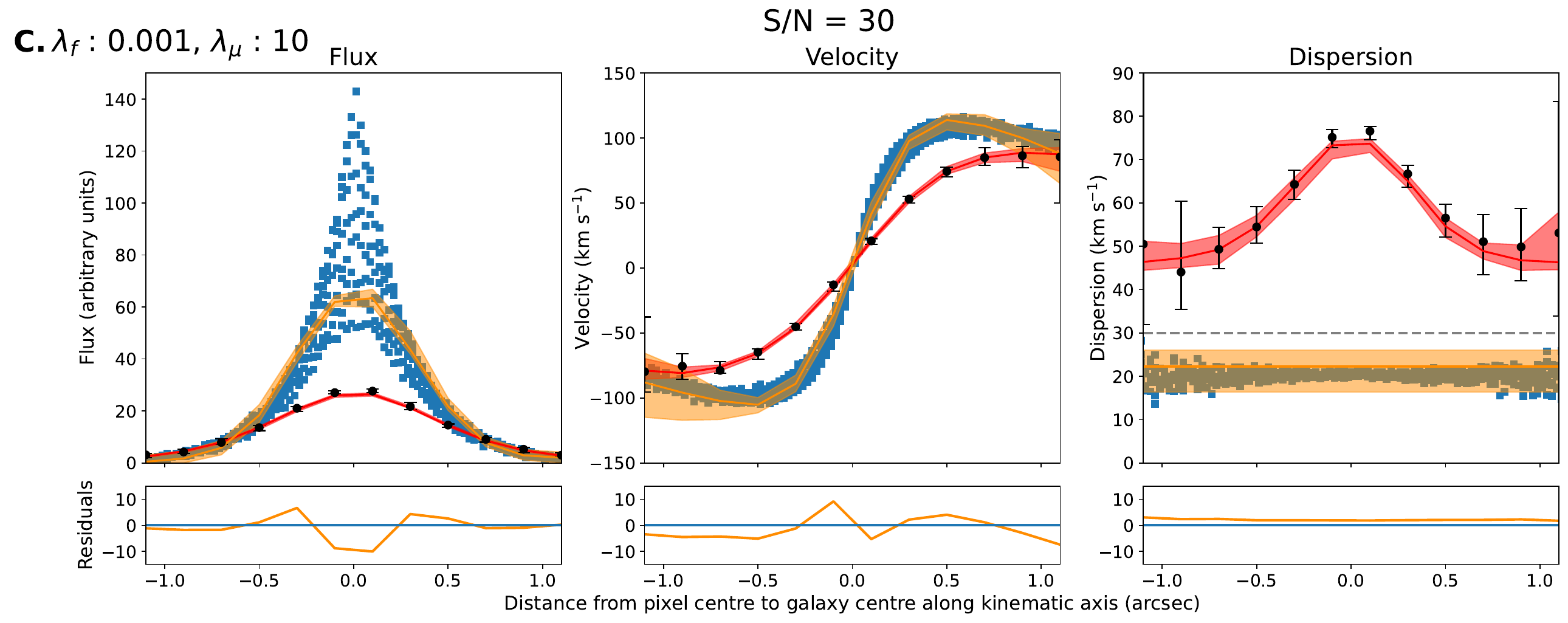}\hspace{5cm}
    
    \includegraphics[width=0.9\linewidth]{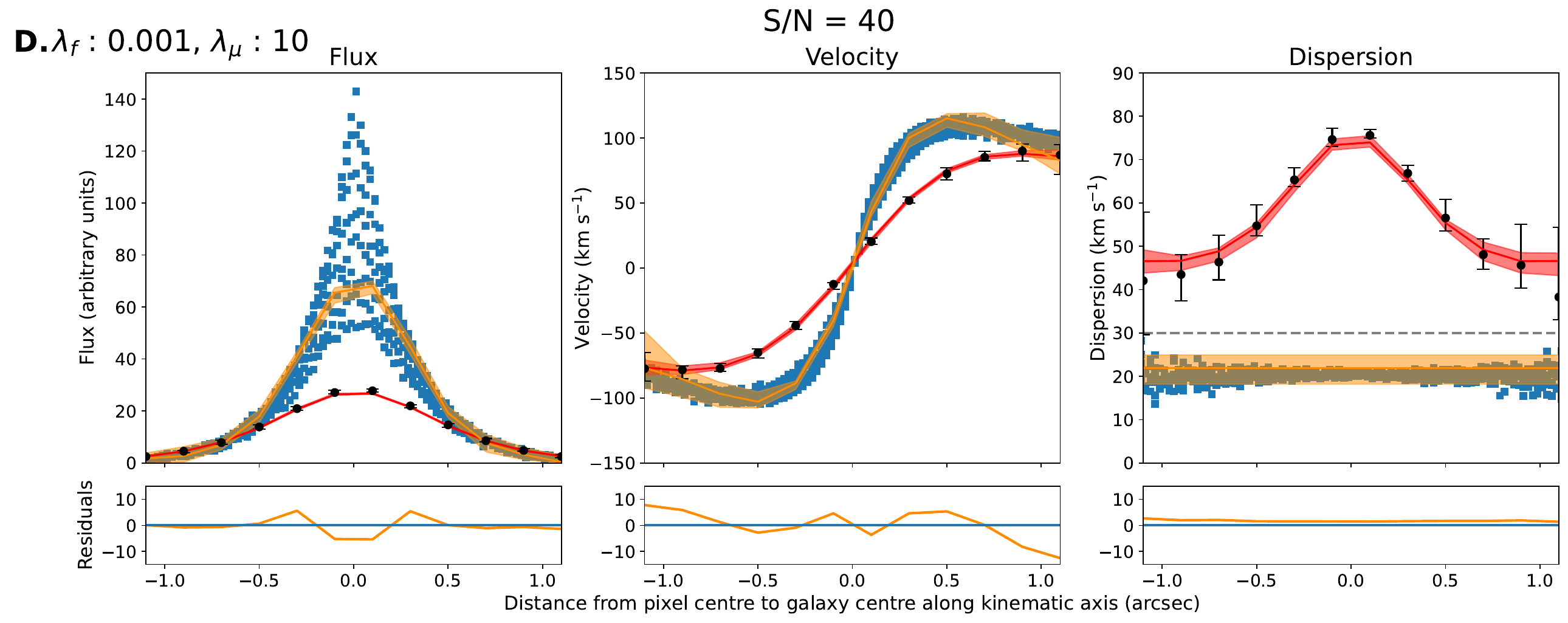}
    \caption{(Continued)}
\end{figure*}



\bsp	
\label{lastpage}
\end{document}